\documentclass[10pt,times,letter]{article}

\usepackage{amsmath, amsfonts,amsthm,amssymb,bm}
\usepackage{dsfont}

\usepackage[english]{babel}

\usepackage{latexsym,amssymb,amsfonts,graphicx}
\usepackage{amsmath,dsfont}
\usepackage{verbatim}
\usepackage{mathrsfs}
\usepackage{bm}
\usepackage{color}
\usepackage{natbib}
\usepackage{epsfig}

\usepackage{url}

\usepackage{bm}
\usepackage{natbib}

\newcommand{\Px}{ \mathbb{P} }
\newcommand{\N}{ \mathds{N} }
\newcommand{\Ex}{ \mathbb{E} }
\newcommand{\gt}{\mathcal{G}_t}
\newcommand{\Ht}{\mathcal{H}_t}

\newcommand{\too}{-\!\!\!\to}
\newcommand{\A}{{\cal A}}

\newcommand{\cA}{\mathcal{A}}

\newcommand{\cO}{\mathcal{O}}
\newcommand{\cL}{\mathcal{L}}
\newcommand{\D}{\mathrm{d}}
\def\cP{\mathcal{P}}

\newcommand{\F}{\mathcal{F}}
\newcommand{\G}{\mathcal{G}}
\newcommand{\R}{\mathds{R}}

\newtheorem{theorem}{Theorem}[section]

\newtheorem{proposition}[theorem]{Proposition}
\newtheorem{remark}[theorem]{Remark}
\newtheorem{lemma}[theorem]{Lemma}

\definecolor{Red}{rgb}{0.00, 0.00, 0.00}
\newcommand{\Red}{\color{Red}}
\definecolor{DRed}{rgb}{0.5, 0.00, 0.00}

\definecolor{Blue}{rgb}{0.00, 0.00, 1.00}

\definecolor{Green}{rgb}{0.0, 0.4, 0.0}

\definecolor{Magenta}{rgb}{1.0, 0, 1.0}

\allowdisplaybreaks

\usepackage{setspace}

\begin{document}

\begin{spacing}{2}
\title{Bilateral Credit Valuation Adjustment for Large Credit Derivatives Portfolios}
\end{spacing}

\vskip 1 cm

\begin{spacing}{1}
\author{
Lijun Bo \thanks{Email: lijunbo@xidian.edu.cn, Department of Mathematics, Xidian University, Xi'an, 710071, China.} \and
Agostino Capponi \thanks{E-mail: capponi@purdue.edu, School of Industrial Engineering, Purdue University, West Lafayette, 47906, IN, USA. }}
\end{spacing}

\begin{spacing}{1}

\maketitle

\begin{center}


\strut

\vskip 1 cm

\textbf{Abstract}
\end{center}

\noindent {\small We obtain an explicit formula for the bilateral counterparty valuation adjustment of a
credit default swaps portfolio referencing an asymptotically large number of entities.
We perform the analysis under a doubly stochastic intensity framework, allowing for default correlation through a common jump process.
The key insight behind our approach is an explicit characterization of the
portfolio exposure as the weak limit of measure-valued processes associated to
survival indicators of portfolio names. We validate our theoretical predictions by means of a numerical analysis,
showing that counterparty adjustments are highly sensitive to portfolio credit risk volatility as well as to default correlation.}

\vspace{0.3 cm}


\vspace{0.3 cm}


\end{spacing}

\newpage
\clearpage
\pagenumbering{arabic}

\section{Introduction}
The recent financial crisis has highlighted the importance of
counterparty risk valuation in over-the-counter derivatives
markets. Indeed, as noted by the Basel Committee on Banking
supervision, see \cite{BaselIII}, under Basel II the risk of
counterparty default and credit migration were addressed, but
mark-to-market losses due to credit valuation adjustments (CVA) were
not. Nevertheless, during the financial crisis, roughly two-thirds
of losses attributed to counterparty credit risk were due to CVA
losses and only about one-third to actual defaults. Credit default swaps (CDSs) have been at the heart of
debates between regulator and supervising authorities. They have been claimed to be responsible for increasing significantly the systemic
risk in the economy, due to large amounts of traded notional (about
\$41 trillion), and consequently high mark-to-market exposures, see \cite{ECB}.

The market price of counterparty risk is usually referred to as \textit{bilateral credit valuation adjustment}, abbreviated with BCVA throughout the paper. This is obtained as the difference between the price of a portfolio transaction, traded between two counterparties assumed default free, and the price of the same portfolio where the default risk of both counterparties is accounted for.
Precise estimates of such adjustments are notoriously difficult to obtain, given that they have to be computed at an aggregate portfolio level and are model dependent;
as such they depend on the volatility of the underlying portfolio, credit spreads of the counterparties, as well as default correlation. On the other hand, accurate assessments of counterparty risk
are essential, given that financial institutions need to mitigate and hedge their counterparty credit exposure. This has originated a significant amount of research, some of which surveyed next.

\cite{BrigoCapPal} develop an arbitrage free valuation framework for bilateral counterparty risk, inclusive of collateralization. Building on \cite{BrigoCapPal}, \cite{assefa} provide a representation formula for BCVA adjustments for a fully netted and collateralized portfolio of contracts. In a series of two papers, \cite{Crepeyfuna} and  \cite{Crepeyfunb} propose a reduced-form backward stochastic differential equation approach to the problem of pricing and hedging CVA, taking into account funding constraints. \cite{BielCrephedge} develop an analytical framework for dynamic hedging unilateral counterparty risk, without excluding simultaneous defaults. \cite{BieleckiJeanblanc} propose a reduced form credit model for dynamically hedging credit default swaptions using credit default swaps.
We refer the reader to \cite{Capponisurv} for a survey on counterparty risk valuation and mitigation.

None of the above mentioned studies provides explicit pricing formulas for counterparty valuation adjustments. We provide a rigorous analysis, which culminates into an analytical representation of
the BCVA adjustment when the traded portfolio consists of an asymptotically large number of CDS contracts. This extends significantly previous literature, which typically resorts to Monte-Carlo simulation methods to evaluate counterparty risk of credit derivatives portfolios, see \cite{CannDuff} and, more recently \cite{Cesari}. Even in the case of single name CDSs portfolios, most of the attempts rely on numerical methods. \cite{BielCrep} employ a Markov chain copula model for pricing counterparty risk embedded in a CDS contract. \cite{lipton} introduce a structural model with jumps and recover the BCVA adjustment on a CDS contract as the numerical solution of a partial differential equation.

We follow a bottom up approach to default and employ doubly stochastic processes to model the default times of the individual names in the portfolio,
as well as of the counterparties, see \cite{bielecki01}. The intensity process of each name in the portfolio, as well as of the counterparties, consists of both diffusive and jump components. The diffusive components follow independent mean-reverting CEV type processes. 
In order to introduce correlation across all portfolio names and counterparties, we assume that the jump process of each name consists of two types of jumps, idiosyncratic and systematic.
Idiosyncratic jumps govern the default risk specific to each reference entity, while common jumps model the occurrence of economic events affecting all parties. In summary, our default correlation model falls within the class of the so called conditionally independent default models.

The key innovation in our approach is a fully explicit characterization of the asymptotic portfolio exposure. Besides
representing a significant departure from currently available techniques to approximate portfolio exposure, this allows us
obtaining explicit representations of the bilateral CVA adjustment.
Our ``law of large numbers'' approximate exposure is recovered as the weak limit of a sequence of weighted empirical measure-valued
process associated with the survival indicators of the portfolio entities.

We employ the heavy weak convergence machinery to martingale
problems related to measure-valued processes driven by
jump-diffusion type processes, as described in \cite{EthKurtz}.
Although such a machinery is well established in the literature and
goes back to the work of \cite{Dawson}, its application to
counterparty risk is novel and requires a detailed mathematical
analysis. As already discussed, our default behavior is captured
through default intensity processes subject to both systematic and
idiosyncratic jumps. This requires to develop a rigorous analysis to
understand the roles that both types of jumps play in the limiting
martingale problem.

Following the weak convergence analysis procedures in \cite{EthKurtz} (see Chapter 4 therein), \cite{GieseckeSow} analyze the default behavior in a large portfolio of interacting firms, while \cite{Gieseckenum} further develop a numerical method for solving the SPDE describing the law of large numbers limiting behavior. Despite our paper exhibits an overlap with \cite{GieseckeSow},
resulting from both employing the same weak-convergence scheme,
there are important differences in the way the various steps are
carried out, which are worth noticing.
{\Red First, in our model
correlation among defaults is introduced through a common jump
process, whereas in \cite{GieseckeSow} default correlation is
modeled by means of a common diffusion process influencing the
intensities of all names, and self-excitation to capture feedback
from default.
Secondly, we are able to recover a fully explicit expression for the limit
measure-valued process (see our Theorem \ref{thm:mart-app}), whereas
\cite{GieseckeSow} only provide an implicit characterization of the
limit.  Thirdly, 
while \cite{GieseckeSow} only have a unique source of jumps coming from self excitation, we
need to capture the behavior of both systematic and idiosyncratic
jumps. By means of a delicate
study (see the analysis leading to Lemma~\ref{lem:limit-geneator}), we can identify how both jump components are incorporated into the generator of the limiting martingale problem. Interestingly, we find that
the limiting killing rate process is purely diffusive with a drift correction given by the sum of the rates of systematic and idiosyncratic jump components.
}

\cite{CMZ} consider a model similar to \cite{GieseckeSow}, assuming that intensities are driven by factors following a diffusion process. Using a fixed point argument, they prove a law of large numbers showing that the limiting process tracking the average number of defaults solves an ordinary differential equation. Using a mean-field interaction model, \cite{Rungga} analyze financial contagion in large networks, and provide characterizations of the entire portfolio loss distribution.

Our default correlation mechanism differs from the ones in the above studies, because we allow for simultaneous jumps of all intensity processes. Our goal is to provide a model which can be calibrated to credit data and at the same time able to match empirical observations. Since the
default intensities are estimated firm-by-firm, our model is consistent with an interpretation based on conditional independence where default correlation is built into the common variation
of the individual default intensities. Moreover, it has been empirical shown by \cite{Yucorr} that if the common factor is properly calibrated, the model is able to reproduce the levels of default correlations historically observed.

The rest of the paper is organized as follows. Section \ref{sec:model} introduces the default model. Section \ref{sec:cdslarge} gives the general expression for bilateral counterparty valuation adjustments in CDS portfolios. Section \ref{sec:portempmeas} develops a weak convergence analysis of the empirical measure associated with the large CDS portfolio. Section \ref{sec:BCVAformula} provides a law of large numbers approximation formula for the bilateral CVA, under the assumption that all default intensity processes follow CEV dynamics. A fully explicit formula is derived in Section \ref{sec:explexpr} under the empirically relevant specialization of CEV to CIR. Section \ref{sec:numerics} presents a numerical and economical analysis of our formulas. Section \ref{sec:conclus} concludes the paper. Technical proofs are delegated to the Appendix.

\section{The Model} \label{sec:model}

Let $(\Omega,\F,\Px)$ be a complete probability space, where $\Px$ denotes the risk-neutral probability measure. The space is endowed with a $K+2$-dimensional Brownian motion $(W^{(1)},\dots, W^{(K)},W^{(A)},W^{(B)})$ and $K+3$ independent Poisson processes  $(\widehat{N}^{(1)},\dots,\widehat{N}^{(K)},\widehat{N}^{(A)},\widehat{N}^{(B)},\widehat{N}^{(c)})$,
with $K\in\N$ being the number of reference names in the CDS portfolio.
The Poisson process $\widehat{N}^{(j)}$ has a constant intensity $\widehat{\lambda}_j>0$ for each $j\in\{1,\dots,K,A,B,c\}$.

We assume that the $K+2$-dimensional Brownian motion is independent
of the $K+3$ Poisson processes above. We use a standard construction  of the default times, see \cite{Lando}, based on doubly stochastic point processes, using given strictly positive $\mathbb F$-adapted intensity processes $\xi^{(k)}=(\xi_t^{(k)}; t \geq 0)$ with $k \in \{1,\ldots,K,A,B \}$. The precise details are provided in the following sections.

\subsection{Intensity Models for Portfolio Names and Counterparties}

The default intensity processes of all names in the portfolio, as well as of the two counterparties $A$ and $B$ are given by
mean-reverting constant elasticity of variance (CEV) processes with jumps: for $k \in \{1,\dots,K\}$,
\begin{eqnarray}\label{eq:intensity-sde}
\xi_t^{(k)} = \xi_0^{(k)} + \int_0^t (\alpha_k -\kappa_k
\xi_s^{(k)})\D s + \int_0^t \sigma_k (\xi_s^{(k)})^{\rho}\D W_s^{(k)} +
c_k\sum_{i=1}^{\widehat{N}_t^{(c)}}Y_i^{(k)}+ d_k\sum_{\ell=1}^{\widehat{N}_t^{(k)}}\widetilde{Y}_\ell^{(k)},
\end{eqnarray}
and for $l \in \{A,B\}$,
\begin{eqnarray}\label{eq:default-intensity-counterparty}
\xi_t^{(l)} = \xi_0^{(l)} + \int_0^t (\alpha_l-\kappa_l\xi_s^{(l)})\D s + \int_0^t \sigma_l (\xi_{s}^{(l)})^{\widehat{\rho}}\D W_s^{{(l)}}
+  c_l\sum_{i=1}^{\widehat{N}_t^{(c)}}Y_i^{(l)} + d_l\sum_{\ell=1}^{\widehat{N}_t^{(l)}}\widetilde{Y}_\ell^{(l)}.
\end{eqnarray}
For each $j \in \{1,\ldots,K,A,B\}$,  $\xi_0^{(j)} >0$. The parameter set $(\alpha_j,\kappa_j,\sigma_j,c_j,
d_j)\in \R_+^5$, while $(Y_1^{(j)},\ldots,Y_i^{(j)},\ldots)$ and
$(\widetilde{Y}_1^{(j)},\ldots,\widetilde{Y}_{\ell}^{(j)},\ldots,)$ are two
independent sequences, each consisting of i.i.d. random variables, and with possibly different distribution functions. The power parameters
$\frac{1}{2} \leq \rho,\ \widehat{\rho} < 1$ are, possibly different, elasticity factors. Obviously the SDEs
\eqref{eq:intensity-sde} and~\eqref{eq:default-intensity-counterparty} will reduce to a CIR type process (with jumps) if the elasticity factor $\rho=\widehat{\rho}=\frac{1}{2}$.

It is well known that, when $\rho=\widehat{\rho}=\frac{1}{2}$, the default intensity processes $\xi_t^{(j)}$, $j \in \{1,\ldots,K,A,B\}$, are strictly positive $\Px$-a.e. if $2\alpha_j\geq\sigma_j^2$, see \cite{Feller}.
Next, we deal with the positivity of the default intensity processes when the elasticity factor belongs to $(\frac{1}{2},1)$. Appendix \ref{app:positive-cve} presents the proof for the CEV process of the $k$-th name given by~\eqref{eq:intensity-sde}. Obviously, the same proof holds for the CEV process in~\eqref{eq:default-intensity-counterparty}.

\begin{lemma}\label{lem:positive-cve}
For each $k\in\{1,\ldots,K\}$, there exists
a unique nonnegative (non-explosive) strong solution
$\xi^{(k)}=(\xi_t^{(k)};\ t\geq0)$ to the stochastic differential
equation (SDE) \eqref{eq:intensity-sde}. Moreover, we have that
$\xi_t^{(k)}>0$, $\Px$-a.e. for all $t\geq0$.
\end{lemma}

\subsection{Default Times and Market Information}

We assume the existence of a sequence of mutually independent unit mean exponential random variables $({\it\Theta}_j;\ j\in\{1,\dots,K,A,B\})$, defined on the probability space $(\Omega,\F, \Px)$, independent of the Brownian and Poisson processes. The default times of the $K$-reference names, as well as of the counterparties, are defined by
\begin{eqnarray}\label{eq:default-time}
\tau_{j} = \inf\left\{t\geq0;\ \int_0^t\xi_s^{(j)}\D s\geq{\it\Theta}_j\right\},\ \ \ \ \ \ \ j\in\{1,\dots,K,A,B\}.
\end{eqnarray}
The corresponding default indicator processes are given by
\begin{eqnarray}\label{eq:default-indicator}
H_t^{(j)}={\bf1}_{\{\tau_j\leq t\}}, \ \ \ \ \ \ t\geq0,
\end{eqnarray}
and the survival indicator processes are denoted by $\overline{H}_t^{(j)}=1-H_t^{(j)}={\bf1}_{\{\tau_j>t\}}$ for each $j\in\{1,\dots,K,A,B\}$.

Given $t\geq0$, let $\Ht^{(j)}=\sigma(H_s^{(j)};\ s\leq t)$, after completion and regularization on the right, see \cite{Belanger}.
Let $\G_t^{(j)}=\F_t^{(j)}\vee\Ht^{(j)}$ with $j\in\{1,\dots,K,A,B\}$, where $\F_t^{(j)}=\sigma(\xi_s^{(j)};\ s\leq t)$. For $t\geq0$,
the market {filtration} is given by
\begin{eqnarray}\label{eq:filtration}
\gt^{(K,A,B)}=\F_t^{(K,A,B)}\vee\Ht^{(K,A,B)},\ \ \ \ \ t\geq0,
\end{eqnarray}
where $\F_t^{(K,A,B)}=\bigvee_{j\in\{1,\dots,K,A,B\}}\F_t^{(j)}$ and $\Ht^{(K,A,B)}=\bigvee_{j\in\{1,\dots,K,A,B\}}\Ht^{(j)}$ with $t\geq0$. The filtration $\F_t^{(K,A,B)}$ is also
referred to as the reference filtration, and models all observable market quantities except default events.

A straightforward application of results in \cite{bielecki01} shows that the default times of the reference names and of the counterparties, are conditionally independent given the reference filtration, see Appendix \ref{app:basics} for details.

\begin{remark}
The modeling choices adopted above lead to a different stochastic analysis than in \cite{GieseckeSow} for verifying the various steps of the weak-convergence procedure. As we demonstrate in Section \ref{sec:portempmeas}, we develop a different method to prove the validity of a stronger compact containment condition of the underlying sequence of measure-valued processes. Our technique consists in transforming such a sequence into a sequence of real-valued stochastic processes defined on the Skorokhod space through test functions (see Li (2010)). \cite{GieseckeSow} instead prove the compact containment condition for the sequence of their empirical measure-valued process by establishing a compact set of the corresponding functional space.
\end{remark}

\section{Pricing of Bilateral Counterparty Risk in CDS Portfolio} \label{sec:cdslarge}

We provide the general arbitrage-free valuation formula for bilateral CVA in portfolios of credit default swaps. Such a formula generalizes the one provided in
\cite{BrigoCapPal}, who focus on a portfolio consisting of one credit default swap. We denote by $A$ and $B$ the two counterparties of the trade,
and by $T$ the maturity of all contracts. For $0\leq t\leq T$, we denote by $D(t,T) = e^{-r(T-t)}$ the discount factor from $t$ to $T$,
where $r>0$ is the constant interest rate. Define the conditional survival function of the $k$-th name as: for $0\leq t\leq T$,
\begin{eqnarray}\label{eq:survival-names}
G^{(k)}(t,T)=\Px\left(\tau_{k}>T| \G_t^{(K,A,B)}\right),\ \ \ \ \
k\in\{1,\dots,K\}.
\end{eqnarray}
Then, on the event $\{\widetilde{\tau}_K>t\}$, where $\widetilde{\tau}_K=\min_{k\in\{1,\dots,K\}}\tau_k$, the CDS price process of the $k$-th reference name is given by
\begin{eqnarray}\label{eq:cds-i}
{CDS}^{(k)}(t,T)=\int_t^T S_kD(t,s)G^{(k)}(t,s)\D s +
\int_t^TL_k D(t,s) \D G^{{(k)}}(t,s),
\end{eqnarray}
where $S_k$ and $L_k$ are constants for each $k\in\{1,\dots,K\}$. The \emph{exposure} of $A$ to $B$ is given by
\begin{equation}
\varepsilon^{(K)}(t,T)=\sum_{k=1}^K z_k {CDS}^{(k)}(t,T),\ \ \ \ \ 0\leq t\leq T,
\label{eq:expos}
\end{equation}
where $z_k = 1$ if the counterparty $A$ is long on the $k$-th CDS, i.e. $A$ sold the $k$-th CDS to $B$ ($A$ is receiving the spread payments
$S_k$ from $B$), and $z_k = -1$ if the counterparty $A$ is short on the $k$-th CDS, i.e. $A$ bought the $k$-th CDS from $B$ (i.e. $A$ is paying the spread premium $S_k$ to
$B$).

On the event $\{\widetilde{\tau}_K>t\}$, such exposure may be rewritten as
\begin{eqnarray}\label{eq:cds-i1}
\frac{\varepsilon^{(K)}(t,T)}{K} &=& \int_t^T D(t,s)\Ex\left[\frac{1}{K}\sum_{k=1}^Kz_kS_k\overline{H}^{(k)}_s\bigg| \G_t^{(K,A,B)}\right]\D s\nonumber\\
&&+\int_t^T D(t,s) \D \left( \Ex\left[\frac{1}{K}\sum_{k=1}^Kz_kL_k\overline{H}^{(k)}_s\bigg| \G_t^{(K,A,B)}\right] \right),\ \ \ K\in\N.
\end{eqnarray}
where $\D f(t,s)$ denotes the differential of the
function $f$ w.r.t. the time variable $s$. On
$\{\widetilde{\tau}>t\}$, where $\widetilde{\tau}= \min_{j \in
\{1,\ldots,K,A,B\}} \tau_j$, the bilateral counterparty valuation
adjustment, denoted by BCVA, on the portfolio of $K$ CDSs is given
by
\begin{eqnarray}\label{eq:cva}
BCVA^{(K)}(t,T) &=& L_A\Ex\left[{\bf1}_{\{t<\tau_A\leq \min(\tau_B,T)\}}D(t,\tau_A)\varepsilon_{-}^{(K)}(\tau_A,T)\Big|\G_t^{(K,A,B)}\right] \\
& & - L_B\Ex\left[{\bf1}_{\{t<\tau_B\leq \min(\tau_A,T)\}}D(t,\tau_B)\varepsilon_{+}^{(K)}(\tau_B,T)\Big|\G_t^{(K,A,B)}\right],\  \ \ 0\leq t\leq T \nonumber,
\end{eqnarray}
where $x_+=x \vee 0$ and $x_{-}=(-x) \vee 0$ for any real number $x\in\R$. Here $L_A$ and $L_B$ denote the percentage losses incurred when counterparty $A$, respectively $B$, defaults on its obligations.

Notice that the above formula is fully consistent with industry practise, where netting is applied before computing the exposure. Our goal is to provide a rigorous law of large numbers approximation formula for the bilateral CVA above defined.

\section{Weak Convergence of Portfolio Empirical Measures} \label{sec:portempmeas}
\label{sec:weak-convergence}

We analyze the weak convergence of a sequence of empirical measure-valued processes. The latter are associated with the survival
indicators of the large number of reference entities in the CDS portfolio (i.e., $K\too\infty$). We follow a classical martingale approach.

From Section \ref{sec:model}, the intensity processes of the $K$-reference entities follow CEV processes extended with jumps given by~\eqref{eq:intensity-sde}.
As in \cite{GieseckeSow}, we first define the `type' parameter set related to the $K$-intensity processes:
\begin{eqnarray}\label{eq:type-parameters}
p_k = \left(\alpha_k,\kappa_k,\sigma_k,c_k,
d_k,\widehat{\lambda}_k\right),\ \ \ \ k\in\{1,\dots,K\},
\end{eqnarray}
taking values in the space $\cO_p:=\R_+^6$. Here
$\widehat{\lambda}_k$ is the intensity of the idiosyncratic
component $\widehat{N}^{(k)}$. Throughout the paper, we make the
following assumption
\begin{itemize}
\item[({\bf A1})] Let $q^K=\frac{1}{K}\sum_{k=1}^K\delta_{p_k}$,
$\eta^K=\frac{1}{K}\sum_{k=1}^K
\delta_{(Y_1^{(k)},\widetilde{Y}_1^{(k)})}$ and
$\phi_0^K=\frac{1}{K}\sum_{k=1}^K\delta_{\xi_0^{(k)}}$. Then
$q=\lim_{K\too\infty}q^K$, $\eta=\lim_{K\too\infty}\eta^K$ and
$\phi_0=\lim_{K\too\infty}\phi_0^K$ exist in $\cP(\cO_p)$,
$\cP( \R_+^2)$ and $\cP(\R_+)$ respectively, where $\cP(A)$
denotes all Borel measures $\nu(\cdot)$ defined on
${\mathcal{B}}(A)$ such that $\nu(A)\leq1$ for a given space $A$. We
use $\delta(\cdot)$ to denote the Dirac-delta measure.
\end{itemize}

\begin{remark}\label{rem:assum-A1}
If there exists $q^*=(\alpha^*,\kappa^*,\sigma^*,c^*,
d^*,\widehat{\lambda}^*)\in\cO_p$ such that
$\lim_{k\too\infty}p_k=q^*$, then $q=\delta_{q^*}$.
\end{remark}

Let the space $\cO=\cO_p\times\R_+^2$. Define a sequence of
measure-valued processes by
\begin{eqnarray}\label{eq:measure-process}
\nu_t^{(K)}=
\frac{1}{K}\sum_{k=1}^K\delta_{(p_k,(Y_1^{(k)},\widetilde{Y}_1^{(k)}),\xi_t^{(k)})}\
\overline{H}_t^{(k)},\ \ \ \ \ \ \ \ t\geq0,
\end{eqnarray}
on ${\mathcal{B}}(\cO)$. Let $S={\cP}(\cO)$ (i.e., the set of all
Borel measures $\nu$ on ${\mathcal{B}}(\cO)$ such that
$\nu(\cO)\leq1$). For any smooth function $f\in C^{\infty}(\cO)$ and
$\nu\in S$, define
\[
\nu(f):= \int_{\cO}f(p,y,x)\nu(\D p\times\D y\times\D x).
\]
Obviously, it holds that
\begin{eqnarray}\label{eq:inner-product}
\nu_t^{(K)}(f)= \int_{\cO}f(p,y,x)\nu_t^{(K)}(\D p\times\D y\times\D
x)=\frac{1}{K}\sum_{k=1}^K
f(p_k,(Y_1^{(k)},\widetilde{Y}_1^{(k)}),\xi_t^{(k)})\
\overline{H}_t^{(k)},\ \ \ \ \ t\geq0.
\end{eqnarray}
Application of It\^o's formula yields that, for all $t\geq0$,
\begin{eqnarray}\label{eq:ito}
\nu_t^{(K)}(f)&=&\nu_0^{(K)}(f) +
\int_0^t\nu_s^{(K)}(\cL_{11}f)\D s+
\frac{1}{K}\int_0^t\sum_{k=1}^Kf(p_k,(Y_1^{(k)},\widetilde{Y}_1^{(k)}),\xi_{s-}^{(k)})\D\overline{H}_s^{(k)}\nonumber\\
&&+\frac{1}{K}\int_0^t\sum_{k=1}^K\left[\sigma_k\frac{\partial
f}{\partial
x}(p_k,(Y_1^{(k)},\widetilde{Y}_1^{(k)}),\xi_s^{(k)})\overline{H}_s^{(k)}
(\xi_s^{(k)})^{\rho} \D W_s^{(k)}\right]\\
&&+\frac{1}{K}\int_0^t\sum_{k=1}^K\left([f(p_k,(Y_1^{(k)},\widetilde{Y}_1^{(k)}),\xi_{s-}^{(k)}+
d_k\widetilde{Y}_1^{(k)})
-f(p_k,(Y_1^{(k)},\widetilde{Y}_1^{(k)}),\xi_{s-}^{(k)})]\overline{H}_s^{(k)}\D
\widehat{N}_s^{(k)}\right)\nonumber\\
&&+\frac{1}{K}\int_0^t \left(\sum_{k=1}^K[f(p_k,(Y_1^{(k)},\widetilde{Y}_1^{(k)}),\xi_{s-}^{(k)}+
c_kY_1^{(k)})
-f(p_k,(Y_1^{(k)},\widetilde{Y}_1^{(k)}),\xi_{s-}^{(k)})]\overline{H}_s^{(k)}\right)\D
\widehat{N}_s^{(c)},\nonumber
\end{eqnarray}
{\Red where, for $(p,y,x)=(\alpha,\kappa,\sigma,c,
d,\widehat{\lambda},y,x)\in\cO$,}
\begin{eqnarray}\label{eq:operatorL11}
\cL_{11}f(p,y,x) =  \frac{1}{2}\sigma^2x^{2\rho}\frac{\partial^2
f}{\partial x^2}(p,y,x) + (\alpha-\kappa x)\frac{\partial
f}{\partial x}(p,y,x).
\end{eqnarray}

Next, we analyze the jump behavior of the process
$\nu^{(K)}(f)=(\nu_t^{(K)}(f);\ t\geq0)$ using~\eqref{eq:ito}.
Consider $M$ smooth functions $f_m\in C^{\infty}(\cO)$, where
$m=1,\dots,M$. For each $K\in\N$, we can define the $M$-dimensional
stochastic process
\[
\nu_t^{(K)}({\bm f})
:=\left(\nu_t^{(K)}(f_1),\dots,\nu_t^{(K)}(f_M)\right),\ \ \ \
\ \ t\geq0.
\]
Since the Poisson processes
$(\widehat{N}^{(1)},\dots,\widehat{N}^{(K)},\widehat{N}^{(c)})$ are
mutually independent, they cannot experience simultaneous jumps.
Hence, the above jump decomposition \eqref{eq:ito} shows that
all components of the $M$-dimensional process $\nu^{(K)}({\bm f})$
must jump together with positive probability. More precisely, if the
$k$-th ($k\in\{1,\dots,K\}$) Poisson process $\widehat{N}^{(k)}$
jumps at time $t>0$, the corresponding jump amplitude of the
$M$ components of the process $\nu^{(K)}({\bm f})$ would be
\begin{eqnarray*}
{\bm J}_t^{(K,k)}(
Y_1^{(k)},\widetilde{Y}_1^{(k)})&:=&\frac{1}{K}\Big(\left[f_1(p_k,(Y_1^{(k)},\widetilde{Y}_1^{(k)}),\xi_{t-}^{(k)}+
d_k\widetilde{Y}_1^{(k)})
-f_1(p_k,(Y_1^{(k)},\widetilde{Y}_1^{(k)}),\xi_{t-}^{(k)})\right],
\dots,\nonumber\\
&&\quad\dots,\left[f_M(p_k,(Y_1^{(k)},\widetilde{Y}_1^{(k)}),\xi_{t-}^{(k)}+
d_k\widetilde{Y}_1^{(k)})
-f_M(p_k,(Y_1^{(k)},\widetilde{Y}_1^{(k)}),\xi_{t-}^{(k)})\right]\Big)\
\overline{H}_t^{(k)}.
\end{eqnarray*}
The probability that the above event occurs is given by
\[
\frac{\widehat{\lambda}_k}{\widehat{\lambda}_c +
\sum_{k=1}^K\widehat{\lambda}_k}.
\]
The other possibility is that the Poisson process
$\widehat{N}^{(c)}$, common to the $K$ intensity processes,
jumps at time $t>0$. Then the corresponding common jump amplitudes are given
by
\begin{eqnarray*}
{\bm J}_t^{(K,c)}({\bm
Y}_1, {\bm \tilde{Y}}_1)&:=&\frac{1}{K}\bigg(\sum_{k=1}^K[f_1(p_k,(Y_1^{(k)},\widetilde{Y}_1^{(k)}),\xi_{t-}^{(k)}+c_kY_1^{(k)})
-f_1(p_k,(Y_1^{(k)},\widetilde{Y}_1^{(k)}),\xi_{t-}^{(k)})]\
\overline{H}_{t}^{(k)},
\dots,\\
&&\qquad\quad\dots,\sum_{k=1}^K[f_M(p_k,(Y_1^{(k)},\widetilde{Y}_1^{(k)}),\xi_{t-}^{(k)}+c_kY_1^{(k)})
-f_M(p_k,(Y_1^{(k)},\widetilde{Y}_1^{(k)}),\xi_{t-}^{(k)})]\
\overline{H}_{t}^{(k)}\bigg),
\end{eqnarray*}
where ${\bm Y}_1:=(Y_1^{(1)},\dots,Y_1^{(K)})$, and ${\widetilde{\bm Y}}_1:=(\widetilde{Y}_1^{(1)},\dots,\widetilde{Y}_1^{(K)})$.
The probability of a common jump is given by
\[
\frac{\widehat{\lambda}_c}{\widehat{\lambda}_c +\sum_{k=1}^K\widehat{\lambda}_k}.
\]

Based on the above analysis of the behavior of systematic and idiosyncratic jumps in the sequence of empirical measure-valued processes, we define, for any smooth function $\varphi\in C^{\infty}(\R^M)$ and $\nu\in S$,
\begin{eqnarray}\label{eq:phi}
{\it\Phi}(\nu) = \varphi\left(\nu({\bm f})\right),
\end{eqnarray}
where $\nu({\bm f})=(\nu(f_1),\dots,\nu(f_M))\in\R^M$. Moreover, for all functions ${\it\Phi}$ of the form
\eqref{eq:phi}, define the operator
\begin{eqnarray}\label{eq:limit-generator}
\A{\it\Phi}(\nu):=\sum_{m=1}^M\frac{\partial\varphi}{\partial
x_m}(\nu({\bm f}))\left(\nu(\cL_1f_m)+\nu(\cL_{21}f_m)
+\widehat{\lambda}_c\nu(\cL_{22}f_m)\right),\ \ \ \ \ \nu\in S,
\end{eqnarray}
where for $p=(\alpha,\kappa,\sigma,c,
d,\widehat{\lambda})\in\cO_p$, $x\in\R_+$ and $y=(y_1,y_2)\in\R_+^2$,
\begin{eqnarray}\label{eq:operators-weak-convergence}
\cL_1f(p,y,x) &=& \cL_{11}f(p,y,x) - x f(p,y,x),\nonumber\\
\cL_{21}f(p,y,x)&=&\widehat{\lambda} d y_2 \frac{\partial f}{\partial x}(p,y,x),\\
\cL_{22}f(p,y,x)&=&c y_1 \frac{\partial f}{\partial x}(p,y,x).\nonumber
\end{eqnarray}
We recall that the operator $\cL_{11}$ has been defined in~\eqref{eq:operatorL11}. Let us analyze the operator in \eqref{eq:limit-generator}.
The component $\nu(\cL_1 f_m)$ corresponds to the diffusive part of
the limit process with a killing rate $x$, while the component
$\nu(\cL_{21}f_m)$ is related to the individual jumps of the $K$
default intensity processes. Furthermore, the component
$\widehat{\lambda}_c\nu(\cL_{22}f_m)$ corresponds to the common
jump of the $K$ default intensity processes with arrival intensity
$\widehat{\lambda}_c$. This observation is consistent with our
original model setup, inclusive of systematic and idiosyncratic jumps.

Then we have the following
\begin{lemma}\label{lem:limit-geneator}
The operator $\A$ given by \eqref{eq:limit-generator} is the generator
of our limit martingale problem in the sense of
\begin{eqnarray}\label{eq:limit-mart}
\lim_{K\to\infty}\Ex\left[\left({\it\Phi}(\nu_{t_{n+1}}^{(K)})-{\it\Phi}(\nu_{t_n}^{(K)})
-\int_{t_n}^{t_{n+1}}\A{\it\Phi}(\nu_s^{(K)})\D
s\right)\prod_{j=1}^n{\it\Psi}_{j}(\nu_{t_j}^{(K)})\right]=0,
\end{eqnarray}
whenever $0\leq t_1<\cdots<t_{n+1}<+\infty$ and
${\it\Psi}_1,\dots,{\it\Psi}_n\in B(S)$ (all bounded functions on the space $S$).
\end{lemma}
The proof of the above lemma is reported in Appendix \ref{app:lem-generator}.

To prove the weak convergence of the measure-valued process $\nu^{(K)}=(\nu_t^{(K)};\ t\geq0)$
defined by \eqref{eq:measure-process} in $D_{S}([0,\infty))$ as $K\too\infty$, the following condition is assumed to hold throughout
the paper.
\begin{itemize}
  \item[({\bf A2})] Let $m_k^{Y}:=\Ex[|Y_1^{(k)}|^4]<+\infty$, and $m_k^{\tilde{Y}}:=\Ex[|\tilde{Y}_1^{(k)}|^4]<+\infty$. Assume that the
  parameters set
  $(\alpha_k,\sigma_k,c_k, d_k,\widehat{\lambda}_k,m_k^{Y},m_k^{\tilde{Y}}, \xi_0^{(k)})$ are bounded by a
  common constant $C_p>0$ for all $k\in\{1,2,\dots,K\}$.
\end{itemize}

\subsection{Relative Compactness of $(\nu^{(K)};\ K\in\N)$}
In order to prove the weak convergence of the family of measure-valued processes $(\nu^{(K)};\ K\in\N)$ defined by
\eqref{eq:measure-process}, we need to check the relative compactness of $(\nu^{(K)};\ K\in\N)$.
Following the standard procedure specified in Chapter 3 of \cite{EthKurtz}, it is enough to verify
{\bf(a)} the compact containment condition and {\bf(b)} the condition (ii) of Theorem 8.6 of Chapter 3 in \cite{EthKurtz}.


We first check the compact containment condition of the
family of stochastic processes $(\nu^{(K)};\ K\in\N)$ whose sample
paths are in $D_{S}[0,\infty)$.
Notice that we consider the convergence of the martingale problem
for functions of the form:
\[
{\it\Phi}(\nu)=\varphi(\nu(f_1),\dots,\nu(f_M)),
\]
where $\varphi\in C^{\infty}(\R^M)$ and $f_1,\dots,f_M\in
C^\infty(\cO)$. {\Red It, thus, suffices} to prove the relatively compactness
for $(\nu^{(K)}(f);\ K\in\N)$ as a stochastic process with sample
path in $D_{\R}([0,\infty))$, where $f\in C^{\infty}(\cO)$ (see,
e.g. \cite{Li}). We have the following
\begin{lemma}\label{lem:ccc}
Let the assumption {\rm({\bf A2})} hold. For every $T>0$, it holds that for any $f\in C^{\infty}(\cO)$,
\begin{eqnarray}\label{lem:ccc-eq}
\sup_{K\in\N}\Px\left(\sup_{0\leq t\leq T}\left|\nu_t^{(K)}(f)\right|\geq m\right)\too 0,
\end{eqnarray}
as $m\too+\infty$.
\end{lemma}
The proof of Lemma \ref{lem:ccc} is postponed to Appendix \ref{app:weak-convergene}. Eq.~\eqref{lem:ccc-eq}
implies the following compact containment condition: for every
$\eta>0$ and $T>0$, there exists a compact set
${{\it\Gamma}_{\eta,T}}\subset E$ such that
{\Red
\begin{eqnarray}\label{eq:strong-ccc}
\inf_{K\in\N}\Px\left(\nu_t^{(K)}\in{\it\Gamma}_{\eta,T},\ \forall\ 0\leq t\leq T\right)>1-\eta
\end{eqnarray}
}
holds for the stochastic process $(\nu^{(K)}(f);\ K\in\N)$ for any $f\in C^\infty(\cO)$ (see Remark 7.3 on Page 129 in \cite{EthKurtz}).

Next we prove that {\bf (b)} the condition (ii) of Theorem 8.6 of
Chapter 3 in \cite{EthKurtz} holds. We have the following lemma, whose proof is reported in Appendix
\ref{app:weak-convergene}.
\begin{lemma}\label{lem:cond-b}
Let $h(u,v)=|u-v|\wedge 1$ for any $u,v\in\R$. Then there exists a positive r.v. $H_K(\delta)$ with $\lim_{\delta\too 0}\sup_{K\in\N}\Ex[H_K(\delta)]=0$
such that for all $0\leq t\leq T$, $0\leq u\leq \delta$ and $0\leq v\leq\delta\wedge t$, it holds that
\begin{eqnarray}\label{eq:lem-b}
\Ex\left[h^2(\nu_{t+u}^{(K)}(f),\nu_{t}^{(K)}(f))\cdot h^2(\nu_{t}^{(K)}(f),\nu_{t-v}^{(K)}(f))\Big|\bigvee_{k=1}^K\G_t^{(k)}\right]\leq
\Ex\left[H_K(\delta)\Big|\bigvee_{k=1}^K\G_t^{(k)}\right],
\end{eqnarray}
for each $K\in\N$.
\end{lemma}

Finally, we need the uniqueness of the martingale
problem for the generator $(\cA,\delta_{q\times\eta\times\phi_0})$
given by \eqref{eq:limit-generator}. This result will be used in the next subsection to
identify the limit measure-valued process.
\begin{lemma}\label{lem:uniqueness-mart}
The uniqueness of the martingale problem of the generator
$(\cA,\delta_{q\times\eta\times\phi_0})$ given by
\eqref{eq:limit-generator} holds.
\end{lemma}
The proof of the above lemma is based on the standard dual argument (see Theorem 4.4.2, pag. 184 and Proposition 4.4.7, pag. 189 in \cite{EthKurtz}). Hence, the details are omitted here.

\subsection{Limit Measure-Valued Process}

\begin{theorem}\label{thm:mart-app}
Let the measure-valued process $\nu^{(K)}=(\nu_t^{(K)};\ t\geq0)$ be defined as in \eqref{eq:measure-process} for each $K\in\N$.
Then $\nu^{(K)}=\!\!=\!\!\Rightarrow\nu$ as
$K\too\infty$ (i.e. in a large portfolio), where the limit
measure-valued process $\nu=(\nu_t;\ t\geq0)$ is given by
\begin{eqnarray}\label{eq:limit-measure-pro}
\nu_t(A\times B\times C)=\int_{\cO}{\bf1}_{A\times
B}(p,y)\Ex\left[\exp\left(-\int_0^tX_s({\bm p})\D
s\right){\bf1}_{\{X_t({\bm p})\in C\}}\right]q(\D p)\eta(\D
y)\phi_0(\D x),
\end{eqnarray}
with the sets $A\in{\cal B}(\cO_p)$, $B\in{\cal B}(\R_+)$ and
$C\in{\cal B}(\R_+)$. The (random) measures $q(\D p),\eta(\D
y),\phi_0(\D x)$ are given in Assumption {\rm({\bf A1})}. Here, the process $X({\bm p})=(X_t({\bm p});\ t\geq0)$
satisfies the following SDE:
\begin{eqnarray}\label{eq:lim-pro}
X_t({\bm p})=x + \int_0^t\left(D({\bm p}) + \alpha -\kappa
X_s({\bm p})\right)\D s + \sigma\int_0^t\left(X_s({\bm
p})\right)^{\rho}\D W_s
\end{eqnarray}
where ${\bm p}=(p,y,x)\in\cO$ is a parameter set, with
$y=(y_1,y_2)\in\R_+^2$ and $p=(\alpha,\kappa,\sigma,c,d,\widehat{\lambda})\in\cO_p$. Here the elasticity factor
$\frac{1}{2}\leq\rho<1$, and $W=(W_t;\ t\geq0)$ is a Brownian
motion. Moreover, the drift rate is given by
\begin{eqnarray}\label{eq:drift}
D({\bm p})=d\widehat{\lambda}y_2+c\widehat{\lambda}_cy_1.
\end{eqnarray}
\end{theorem}

{\Red
\begin{remark}
We stress that~\eqref{eq:limit-measure-pro} provides an explicit representation of the limit measure valued process.
This contrasts with the limit intensity process provided in \cite{GieseckeSow}, where
the recursive dependence on a term coming from self-excitation (see their equations (4.3) and (4.4)) prevents an explicit computation of the
limiting measure. Considering that our objective is to obtain closed-form expressions for bilateral CVA adjustments, the explicit
form of the limit measure-valued process is of crucial importance. As we demonstrate in Section \ref{sec:BCVAformula}, this allows us obtaining
semi-closed form representations for the portfolio exposure under general CEV specifications, and closed-form expressions in case of square root diffusion
intensities.
\end{remark}
}
\noindent{\it Proof of Theorem \ref{thm:mart-app}.} \quad As in standard weak convergence analysis
procedures (see Chapter 3 of \cite{EthKurtz}), the weak convergence of
$\nu^{(K)}=\!\!=\!\!\Rightarrow\nu$ for some measure-valued process
$\nu=(\nu_t;\ t\geq0)$ as $K\too\infty$ can be obtained by using
Lemma \ref{lem:limit-geneator}, Lemma \ref{lem:ccc}, and Lemma
\ref{lem:cond-b}. Next, we prove that the limit measure-valued
process $\nu(\cdot)$ is given by \eqref{eq:limit-measure-pro}.

First, we note that $D({\bm p})>0$ for all ${\bm p}\in\cO$. Using
Lemma \ref{lem:positive-cve}, we can immediately show that $X_t({\bm
p})\geq0$ for all $t\geq0$. For any smooth function $f\in C^{\infty}(\cO)$, it holds that
\begin{eqnarray*}
\nu_t(f)=\int_{\cO}\Ex\left[e^{-\int_0^tX_s({\bm p})\D s}\
f\left(p,y,X_{t}({\bf p})\right)\right]q(\D p){\eta}(\D y)\phi_0(\D
x),\ \ \ \forall\ t\geq0.
\end{eqnarray*}
Using It\^{o}'s formula, we have
\begin{eqnarray*}
&&\frac{\partial}{\partial t}\Ex\left[e^{-\int_0^tX_s({\bm p})\D s}\
f\left(p,y,X_{t}({\bf
p})\right)\right]\nonumber\\
&&\quad =\frac{\partial}{\partial t}
\Ex\left[\int_0^te^{-\int_0^sX_u({\bm p})\D u}\ \Big(\cL_1f(p,y,X_{s}({\bm p})) + D({\bf p})\ \cL_2f\left(p,y,X_{s}({\bm p})\right)\Big)\D s\right]\nonumber\\
&&\quad =\Ex\left[e^{-\int_0^tX_s({\bm p})\D s}\
\Big(\cL_1f(p,y,X_{t}({\bm p})) + D({\bm p})\
\cL_2f\left(p,y,X_{t}({\bm p})\right)\Big)\right].
\end{eqnarray*}
Then, we have the equality
\begin{eqnarray*}
\frac{\D\nu_t(f)}{\D t}=\nu_t(\cL_{1}f)+\nu_t(\cL_{21}f)+
\widehat{\lambda}_c\nu_t(\cL_{22}f).
\end{eqnarray*}
Using~\eqref{eq:phi} and \eqref{eq:limit-generator}, we {\Red have that}
\begin{eqnarray*}
\frac{\D {\it\Phi}(\nu_t)}{\D t}&=&\sum_{m=1}^M\frac{\partial\varphi}{\partial x_m}(\nu_t({\bm f}))\frac{\D\nu_t(f_k)}{\D t}\nonumber\\
&=&\sum_{m=1}^M\frac{\partial\varphi}{\partial x_m}(\nu_t({\bm
f}))\left(\nu_t(\cL_{1}f_m)+\nu_t(\cL_{21}f_m)+
\widehat{\lambda}_c\nu_t(\cL_{22}f_m)\right)\nonumber\\
&=&\A{\it\Phi}(\nu_t).
\end{eqnarray*}
The above equality implies that, for all functions ${\it\Phi}$ of
the form \eqref{eq:phi},
\[
{\it\Phi}(\nu_t) = {\it\Phi}(\nu_s) + \int_s^t {\it\Phi}(\nu_u)\D
u,\ \ \ \forall\ 0\leq s<t<+\infty.
\]
Hence, the measure $\delta_{\nu}(\cdot)$ satisfies the martingale
problem for $(\A,\delta_{q\times{\eta}\times\phi_0})$, which is given by
\eqref{eq:limit-generator} due to the uniqueness of the martingale
problem for the operator $(\A,\delta_{q\times{\eta}\times\phi_0})$
(see Lemma \ref{lem:uniqueness-mart}). \hfill$\Box$

\subsection{Approximating Formula of Exposure in Large Portfolio} \label{sec:approx}

The computation of the exposure in Eq.~\eqref{eq:cds-i1} requires evaluating sums of the form:
$$
\frac{1}{K}\sum_{k=1}^Ka_k\overline{H}_{t}^{(k)}.
$$
According to Theorem \ref{thm:mart-app}, we have the following weak convergence as $K\too\infty$,
\begin{eqnarray}\label{eq:app00}
\frac{1}{K}\sum_{k=1}^Ka_k\overline{H}_{t}^{(k)}  =\!\!=\!\!\Rightarrow a^*\nu_t(\cO)
\ \ \ \ \ 0\leq t\leq T,
\end{eqnarray}
where
\begin{eqnarray}\label{eq:vutO}
\nu_t(\cO)=\int_{\cO}\Ex\left[\exp\left(-\int_0^{t}X_s({\bm p})\D s\right)\right]q(\D p)\eta(\D y)\phi_0(\D
x), \ \ \ \ \ 0\leq t\leq T,
\end{eqnarray}
with the process $X({\bm p})=(X_t({\bm p});\ t\geq0)$ satisfying the SDE \eqref{eq:lim-pro},
and $(a_k;\ k=1,\ldots,K)$ is a sequence of real numbers satisfying
$a^*=\lim_{K\too\infty}\frac{1}{K}\sum_{k=1}^Ka_k$, with $a^*$ being finite.
The killing rate process $X({\bm p})$ acts as the limit intensity process of the portfolio of $K$-names, as $K\to\infty$.
Accordingly, we use $\tau_X^*$ to denote the limit default time of the large portfolio associated to the killing rate $X({\bm p})$. Hence in a large pool, $K \too \infty$, the default indicator ${\bf1}_{\{\tau_X^*>t\}}$ plays the role of the average default indicator $\frac{1}{K}\sum_{k=1}^K \overline{H}_{t}^{(k)}$.

Using the weak convergence result \eqref{eq:app00} and noticing that weak convergence implies convergence of expectations, we obtain that as $K-\!\!\!\!\to\infty$, for each $t\geq0$, the conditional expectation on the event $\{\tau_X^*>t\}$,
$$\Ex\left[\frac{1}{K}\sum_{k=1}^K \overline{H}_{s}^{(k)} \Big| \G_t^{(K,A,B)} \right], \qquad s > t, $$
approaches the function $\widehat{F}(t,s)$ given by, for $0\leq t<s\leq T$,
\begin{eqnarray}\label{eq:FLS1}
\widehat{F}(t,s):=\Ex\left[\int_{\cO}{\Ex}\left[\exp\left(-\int_t^{s}X_u({\bm
p})\D u\right)\right]q(\D p)\eta(\D y)\phi_0(\D x)\right].
\end{eqnarray}
Hence, on $\{\tau_X^*>t\}$, we can characterize the default time of the limiting portfolio in terms of
\begin{eqnarray}\label{eq:csp-tauX}
\Px\left(\tau_X^*>s| \G_t^{(K,A,B)}\right)=\widehat{F}(t,s),\ \ \ \
\ 0\leq s\leq t\leq T,
\end{eqnarray}
which represents the conditional survival probability of the
portfolio when $K \too \infty$.

Moreover, recall the formula for the actual exposure given in
Eq.~\eqref{eq:cds-i1}. Using Eq.~\eqref{eq:csp-tauX}, on the event
$\{\tau_X^*>t\}$ as $K\too\infty$, we obtain
\begin{eqnarray}\label{eq:expo}
\frac{\varepsilon^{(K)}(t,T)}{K}\too\overline{\varepsilon}^{(*)}(t,T),\
\ \ \ \ 0\leq t\leq T,
\end{eqnarray}
where
\begin{eqnarray}\label{eq:approx-exposure0}
\overline{\varepsilon}^{(*)}(t,T)&=&S_z^*\int_t^T D(t,s)\widehat{F}(t,s)\D s+L_z^*\int_t^TD(t,s) \D \widehat{F}(t,s) \nonumber\\
&=&L_z^*\left[D(t,T)\widehat{F}(t,T)-1\right]+\left(S_z^*+rL_z^*\right)\int_t^T
D(t,s)\widehat{F}(t,s)\D s.
\end{eqnarray}
Here $L_z^*:=\lim_{K\too\infty}\frac{1}{K}\sum_{k=1}^Kz_kL_k$ and
$S_z^*:=\lim_{K\too\infty}\frac{1}{K}\sum_{k=1}^Kz_kS_k$, both
assumed to be finite. We recall the reader that $r$ denotes the constant risk-free rate as defined in Section \ref{sec:cdslarge}.
Eq.~\eqref{eq:approx-exposure0} is obtained using integration by parts, along with the trivial equality $\widehat{F}(t,t)=1$.
For future purposes, we introduce the following quantity
\begin{eqnarray}\label{eq:solu-express-Bexp}
B(\kappa,\sigma;u) = -\frac{2(e^{\varpi u}-1)}{2\varpi +
(\kappa+\varpi)(e^{\varpi u}-1)},\ \ \  \ \ \ 0\leq u\leq T,
\end{eqnarray}
where $\kappa,\sigma>0$ and
$\varpi=\sqrt{\kappa^2+2\sigma^2}$.

\begin{remark}\label{rem:limit}
Consider the special case where $z_k=1$, $L_k=L$, $S_k=S$ for all $k\in\{1,2,\dots,K\}$, and
\begin{itemize}
   \item $\rho = \frac{1}{2}$, i.e. the intensities of the portfolio names are CIR processes.
  \item $c_k=d_k=0$ for all $k\in\{1,2,\dots,K\}$, i.e., the intensities of the CIR processes do not experience
  jumps.
  \item there exists $p^*=(\alpha^*,\kappa^*,\sigma^*,\widehat{\lambda}^*)\in\cO_p$ and $x^*\in\R_+$ such that
        $\lim_{k\to\infty}p_k=p^*$ and $\lim_{k\to\infty}\xi_0^{(k)}=x^*$.
\end{itemize}
Under the above assumption, we write the intensity of the $k$-th name as $\xi^{(k)}\sim{\rm CIR}(\alpha_k,\kappa_k,\sigma_k)$ with
$k\in\{1,2,\dots,K\}$. From~\eqref{eq:cds-i}, we have, on the event $\{\tau_X^*>t\}$,
\begin{eqnarray}\label{eq:rem-cds}
\frac{\varepsilon^{(K)}(t,T)}{K} = L\left[D(t,T)F^{(K)}(t,T)-1\right]
+ (S + rL) \int_t^T D(t,s)F^{(K)}(t,s)\D s,
\end{eqnarray}
where for $s>t$,
\begin{eqnarray}\label{eq:FK}
F^{(K)}(t,s)=\frac{1}{K}\sum_{k=1}^K\Ex\left[e^{-\int_t^s\xi_v^{(k)}\D v}\Big|\F_t^{(k)}\right].
\end{eqnarray}
Moreover, using that $\xi^{(k)}\sim{\rm CIR}(\alpha_k,\kappa_k,\sigma_k)$, we obtain
\begin{eqnarray}\label{FKSolution}
\Ex\left[F^{(K)}(t,s)\right]=\frac{1}{K}\sum_{k=1}^K e^{A^{(k)}(s-t) + B^{(k)}(s-t)\xi_{0}^{(k)}},
\end{eqnarray}
where $B^{(k)}(u)=B(\kappa_k,\sigma_k;u)$ has been given in
\eqref{eq:solu-express-Bexp} and
$A^{(k)}(u)=\alpha_k\int_0^uB(\kappa_k,\sigma_k;v)\D v$. Thus
$B^{(k)}(u)\too B(\kappa^*,\sigma^*;u)$ and $A^{(k)}(u)\too
\alpha^*\int_0^uB(\kappa^*,\sigma^*;v)\D v$ as $k\too\infty$. Hence
\begin{eqnarray}\label{FKSolution-Con}
\lim_{K\too\infty}\Ex\left[F^{(K)}(t,s)\right]&=&\exp\left(\alpha^*\int_0^{s-t} B(\kappa^*,\sigma^*;v)\D v + B(\kappa^*,\sigma^*;s-t)x^*\right)\nonumber\\
&=&\Ex\left[\exp\left(-\int_t^sX_v(p^*,x^*)\D v\right)\right]\nonumber\\
&=&\Ex\left[\int_{\cO}\exp\left(-\int_t^sX_v(p,x)\D v\right)\delta(p-p^*)\delta(x-x^*)\D p\D x\right]\nonumber\\
&=&\widehat{F}(t,s),
\end{eqnarray}
where we have used that $D({\bm p})=0$, following by the assumption
$c^*=\lim_{k\too\infty}c_k=0$ and
$d^*=\lim_{k\too\infty}d_k=0$. It follows from \eqref{eq:rem-cds}
that
\begin{equation}
\lim_{K\too\infty}\Ex\left[\frac{\varepsilon^{(K)}(t,T)}{K}\right]=\overline{\varepsilon}^{(*)}(t,T),\
\ \ 0\leq t\leq T, \label{eq:unbiasest}
\end{equation}
where $\overline{\varepsilon}^{(*)}(t,T)$ is given in Eq.~\eqref{eq:approx-exposure0}.

Eq.~\eqref{eq:unbiasest} shows that, in case when the default
intensities do not jump, the actual $K$-names CDS exposure given
by~\eqref{eq:rem-cds} is an asymptotically unbiased estimator of the
approximate limiting exposure.

\end{remark}

\section{The Bilateral Credit Valuation Adjustment Formula} \label{sec:BCVAformula}
This section develops a semi-closed form expression for the ``law of large numbers'' BCVA on a portfolio of credit default swaps contracts. This is obtained by
using formula~\eqref{eq:expo} to approximate the actual exposure.

\subsection{Joint Survival Probability of Counterparties}

Recall that the default intensity processes
$\xi^{(A)}=(\xi^{(A)}_t;\ t\geq0)$ and $\xi^{(B)}=(\xi^{(B)}_t;\
t\geq0)$ of counterparties $A$ and $B$  are given by
\eqref{eq:default-intensity-counterparty}. The corresponding default
times $\tau_A$ and $\tau_B$ are defined by \eqref{eq:default-time}.

Denote by $G(t,t_A,t_B)$ the conditional joint survival probability
that the counterparties $A$ and $B$ do not default before time
$t_A\geq t$ and $t_B\geq t$ respectively, given that the $K$ names
and the counterparties $A$ and $B$ in the portfolio survive up to
time $t\geq0$.
\[
G(t,t_A,t_B)=\Px\left(\tau_A>t_A,\tau_B>t_B\Big |\G_t^{(K,A,B)}\right).
\]
\begin{lemma}\label{lemma:joint-survive}
Let $t\geq0$. Then for any $t_A,t_B\geq t$, we have
\begin{eqnarray}\label{eq:joint-survival}
G(t,t_A,t_B)=\left(\prod_{j\in\{1,\dots,K,A,B\}}H_t^{(j)}\right)\Ex\left[\exp\left(-\int_{t}^{t_{A}}\xi_s^{(A)}\D
s - \int_{t}^{t_{B}}\xi_s^{(B)} \D s
\right)\Big|\F_t^{(K,A,B)}\right].
\end{eqnarray}
Moreover, the corresponding conditional joint density of
$(\tau_A,\tau_B)$ is given by
\begin{eqnarray}\label{eq:joint-density}
\frac{\partial^2 G}{\partial t_A\partial
t_B}(t,t_A,t_B)=\left(\prod_{j\in\{1,\dots,K,A,B\}}H_t^{(j)}\right)\Ex\left[\exp\left(-\int_{t}^{t_{A}}\xi_s^{(A)}\D
s - \int_{t}^{t_{B}}\xi_s^{(B)} \D s
\right)\xi_{t_A}^{(A)}\xi_{t_B}^{(B)}\Big|\F_t^{(K,A,B)}\right].
\end{eqnarray}
\end{lemma}

\noindent{\it Proof.}\quad
{On $0\leq t\leq \widetilde{\tau}:=\min_{j\in\{1,\dots,K,A,B\}}\tau_j$, define
the survival probability function for all $K$ names and the
counterparties $A,B$ as }
\[
\widehat{G}(t,t_1,\dots,t_K,t_A,t_B)=\Px\left(\cap_{j=1}^K\{\tau_j>t_j\}\cap\{\tau_A>t_A,\tau_B>t_B\}\Big|\G_t^{(K,A,B)}\right).
\]
By Lemma 9.1.2 in \cite{bielecki01} and Lemma \ref{lem:cond-indepedence} above, we have that, on the event $\{\widetilde{\tau}>t\}$,
\begin{eqnarray}\label{eq:N+2-survival1}
\widehat{G}(t,t,\dots,t,t_A,t_B)=\left(\prod_{j\in\{1,\dots,K,A,B\}}H_t^{(j)}\right)\Ex\left[\exp\left(-\int_t^{t_A}\xi_s^{(A)}\D
s-\int_t^{t_B}\xi_s^{(B)}\D s\right)\Big|\F_t^{(K,A,B)}\right],
\end{eqnarray}
On the other hand, we also have
\begin{eqnarray*}
\widehat{G}(t,t,\dots,t,t_A,t_B)=\left(\prod_{j=1}^KH_t^{(j)}\right)G(t,t_A,t_B).
\end{eqnarray*}
Hence it holds that for all $t_A,t_B\geq t$,
\begin{eqnarray*}
\left(\prod_{j=1}^KH_t^{(j)}\right)G(t,t_A,t_B)=\left(\prod_{j\in\{1,\dots,K,A,B\}}H_t^{(j)}\right)\Ex\left[\exp\left(-\int_{t}^{t_A}\xi_s^{(A)}\D
s -\int_{t}^{t_B}\xi_s^{(B)}\D s\right)\Big|\F_t^{(K,A,B)}\right],
\end{eqnarray*}
which yields \eqref{eq:joint-survival}. \hfill$\Box$

\subsection{The BCVA Formula}

On the event $\{\widetilde{\tau}>t\}$, for sufficiently large $K$, using the BCVA formula given in~\eqref{eq:cva} we obtain
\begin{eqnarray}\label{eq:cvagen}
 BCVA^{(K,*)}(t,T) &=& L_A\Ex\left[{\bf1}_{\{t<\tau_A\leq \min(\tau_B,T)\}} {\bf1}_{\{\tau_A<\tau_X^*\}}D(t,\tau_A)\varepsilon_{-}^{(K,*)}(\tau_A,T)\Big|\G_t^{(K,A,B)}\right] \\
& & - L_B\Ex\left[{\bf1}_{\{t<\tau_B \leq \min(\tau_A,T)\}} {\bf1}_{\{\tau_B<\tau_X^*\}}D(t,\tau_B)\varepsilon_{+}^{(K,*)}(\tau_B,T)\Big|\G_t^{(K,A,B)}\right], \ \ 0\leq t\leq T \nonumber,
\end{eqnarray}
where
$\varepsilon^{(K,*)}(t,T):=K\overline{\varepsilon}^{(*)}(t,T)$
is the ``law of large numbers'' approximation to the exposure in the
large CDS portfolio, given by \eqref{eq:approx-exposure0}.
\begin{remark}
In market language, the first term of Eq.~\eqref{eq:cvagen} is often referred to as debit valuation adjustment, and denoted by DVA. The second term in the above formula is often referred to as credit valuation adjustment, and denoted by CVA.
\end{remark}

For the above bilateral CVA formula \eqref{eq:cvagen}, we then have the following semi-closed form representation (where the default intensities
of the counterparties satisfy the CEV processes given by \eqref{eq:default-intensity-counterparty}):

\begin{theorem}\label{thm:cva1}
On the event $\{\tau_X^*\wedge\tau_A\wedge\tau_B>t\}$, the BCVA formula in the large portfolio admits the following semi-closed representation:
\begin{eqnarray}\label{thm:cva-formula}
BCVA^{(K,*)}(t,T)=L_A A^{(K,*)}(t,T) - L_B B^{(K,*)}(t,T),\ \ \ \ 0\leq t\leq T,
\end{eqnarray}
where, on the event $\{\tau_X^*\wedge\tau_A\wedge\tau_B>t\}$:
\begin{eqnarray}\label{eq:cva-rhs1}
B^{(K,*)}(t,T)&:=&\Ex\left[{\bf1}_{\{t<\tau_B \leq \min(\tau_A,T) \}}{\bf1}_{\{\tau_B<\tau_X^*\}}D(t,\tau_B)\varepsilon_{+}^{(K,*)}(\tau_B,T)\Big|\G_t^{(K,A,B)}\right]\nonumber\\
&=&\int_t^TD(t,t_B)\varepsilon_{+}^{(K,*)}(t_B,T)\widehat{F}(t,t_B)
H_1(t_B-t,\xi_t^{(A)},\xi_t^{(B)})\D t_B,
\end{eqnarray}
and
\begin{eqnarray}\label{eq:cva-rhs2}
A^{(K,*)}(t,T)&:=&\Ex\left[{\bf1}_{\{t<\tau_A\leq\min(\tau_B,T)\}}{\bf1}_{\{\tau_A<\tau_X^*\}}D(t,\tau_A)
\varepsilon_{-}^{(K,*)}(\tau_A,T)\Big|\G_t^{(K,A,B)}\right]\nonumber\\
&=&\int_t^TD(t,t_A)\varepsilon_{-}^{(K,*)}(t_A,T)\widehat{F}(t,t_A)
H_2(t_A-t,\xi_t^{(A)},\xi_t^{(B)})\D t_A.
\end{eqnarray}
Here the conditional survival functions $\widehat{F}(t,t_B)$ and
$\widehat{F}(t,t_A)$ are given by \eqref{eq:FLS1} and the functions
$H_1,H_2$ are defined as follows: for $x_A,x_B>0$,
\begin{eqnarray}\label{eq:H-hatH12}
H_1(t_B-t,x_A,x_B)&:=&\Ex\left[\exp\left(-\int_t^{t_B}(\xi_s^{(A)}+\xi_s^{(B)})\D s\right)\xi_{t_B}^{(B)}\bigg|\xi_t^{(A)}=x_A,\xi_t^{(B)}=x_B\right],\nonumber\\
H_2(t_A-t,x_A,x_B)&:=&\Ex\left[\exp\left(-\int_t^{t_A}(\xi_s^{(A)}+\xi_s^{(B)})\D
s\right)\xi_{t_B}^{(A)}\bigg|\xi_t^{(A)}=x_A,\xi_t^{(B)}=x_B\right].
\end{eqnarray}
\end{theorem}
The proof of Theorem \ref{thm:cva1} is reported in Appendix
\ref{app:basics}. Both $B^{(K,*)}$ and $A^{(K,*)}$ have an intuitive
economic meaning. Indeed, $B^{(K,*)}$ is the integral over time $s
\in [t,T]$ of the positive exposure of the investor to the
counterparty at time $s$ weighted by the probability that the
limiting portfolio as well as both counterparties survive up to time
$s$, and the counterparty defaults at time $s$. When multiplied by
$L_B$, this precisely identifies the CVA contribution to the BCVA
adjustment. A symmetric argument holds for the DVA contribution.

From Eq.~\eqref{thm:cva-formula}, we deduce that a
fully explicit formula for \eqref{eq:cvagen} requires a closed-form
representation of {\bf(i)} the expectations $H_1, H_2$
defined by \eqref{eq:H-hatH12}, and {\bf(ii)} the survival function
$\widehat{F}(t,s)$ associated to the limit default time given in
Eq.~\eqref{eq:FLS1}. From the definitions of $H_1$ and $H_2$,
we see immediately that explicit representations can only be
obtained when the elasticity factor $\widehat{\rho}=\frac{1}{2}$,
i.e. the default intensity processes of both counterparties belong
to the affine class. Similarly, to evaluate the expectation
in~\eqref{eq:FLS1}, we need the limit process to be affine.

Nevertheless, in the general case when $\rho,\
\widehat{\rho}\in[\frac{1}{2},1)$ it is possible to accurately
evaluate such expectations numerically upon writing the
corresponding Feynman-Kac representations, and then solving the
resulting PDE. We remark that such a PDE would be two dimensional in
the case of~\eqref{eq:H-hatH12} and one dimensional
for~\eqref{eq:FLS1}. Consequently, the PDE solutions can be
accurately and efficiently computed via standard finite difference
methods. Alternatively, such expectations may be efficiently
estimated using Monte-Carlo methods, see for instance the
discretized Euler scheme introduced in \cite{AndersenAnd}. Notice
also that the two dimensional CEV process is an important ingredient
of the popular SABR model, hence a plethora of methods are available
for computing such expectations, given that they naturally arise
when pricing options under SABR.

\section{Explicit Expression of BCVA} \label{sec:explexpr}
We provide closed-form expressions for $H_1$, $H_2$, and $\widehat{F}$,
which yield an explicit expression for the law of large numbers bilateral CVA formula.
To this purpose, we further specialize the model and specify the empirical measures as well as the distribution of
the jump sizes. Throughout the section, we set the elasticity factor
$\rho=\widehat{\rho}=\frac{1}{2}$, i.e. choose the default
intensities of each name in the portfolio, as well as of investor
and counterparty to be CIR processes. Besides mathematical
tractability such a choice is empirically relevant, considering that
square root diffusion models allow for an automatic calibration of
the term structure of credit default swaps. Moreover, they can also
be used to calibrate option data, such as caps for the interest rate
market and options on CDSs. We refer the reader to \cite{Alfonsi}
for further discussions on this aspect.

\subsection{BCVA Model Specifications}\label{subsec:bcva-model}

Let $p^*$ be such that $\lim_{k \rightarrow \infty} p_k = p^*$,
where $p^*= (\alpha^*, \kappa^*, \sigma^*, c^*,d^*,\widehat{\lambda}^*)\in\cO_p$ for all $k\in\{1,\dots,K\}$. Let
$x^*>0$ be the limit of the initial intensities of all names, i.e.
$\lim_{k \rightarrow \infty} \xi_0^{(k)}=x^*$. Under the above
setting, we have that the limiting measure $q(\D p)$ is a delta
function, i.e. $q(\D p)=\delta(p-p^*)\D p$ and the limit
distribution of the time zero intensity is $\phi_0(\D x)=\delta(x-x^*)\D x$.
Let $\mu_{Y}$ and $\mu_{\widetilde{Y}}$ be the probability measures on $\R_+$ associated
with $Y$ and $\widetilde{Y}$, respectively.
Next, we characterize the limit jump measure $\eta(\D
y)$. To this purpose, we firstly state the following lemma:
\begin{lemma}\label{lem:limit-jump-eta}
Define the empirical measure $\mu_Y^K(x) := \frac{1}{K} \sum_{k=1}^K
\delta(x-Y^{(k)})$ with $x\in\R_+$. Then for each $\varrho\in\R$,
$F_{\mu}^K(\varrho) \too F_Y(\varrho)$, as $K \too \infty$, where
$F_{\mu}^K(\varrho)$ and $F_Y(\varrho)$ denote, respectively, the
characteristic function of $\mu_K$ and $Y^{(1)}$.
\end{lemma}

\proof We have that $F_{\mu}^K(\varrho) = \int_0^{\infty} e^{{\rm i}
\varrho x} \mu_K(\D x) = \frac{1}{K} \sum_{k=1}^K e^{{\rm i} \varrho
Y^{(k)}}$, where ${\rm i}=\sqrt{-1}$. By the strong law of large
numbers, we obtain that $\lim_{K \rightarrow \infty}
F_{\mu}^K(\varrho) = \Ex[e^{{\rm i} \varrho Y^{(1)}}]$, a.s. hence
proving the statement of the lemma.
\endproof

By the previous lemma, using the L\'evy continuity theorem, we
obtain that the measure $\mu_{Y}^{K}$ converges weakly to $\mu_Y$,
the distribution of the r.v. $Y^{(1)}$, i.e. for every $A \in {\cal
B}((0, x])$ for which $\mu_Y(x) = 0$, we have that $\lim_{K
\rightarrow \infty} \mu_{Y}^K(A) = \mu_Y(A)$. Similarly, defining the empirical measure
$\mu_{\widetilde{Y}}^K(x) = \frac{1}{K}
\sum_{k=1}^K \delta(x-\widetilde{Y}^{(k)})$ with $x\in\R_+$, we
obtain that $\mu_{\widetilde{Y}}^{K}$ converges weakly to
$\mu_{\widetilde{Y}}$. Here, we choose the jump measures to be
\begin{equation} \mu_Y(\D y_1)=\gamma_1 e^{-\gamma_1 y_1}, \quad
y_1\in \R_+,\ \ \ \ {\rm and}\ \ \mu_{\widetilde{Y}}(\D
y_2)=\gamma_2 e^{-\gamma_2 y_2}, \quad y_2\in \R_+,
\label{eq:exponjump}
\end{equation}
where $Y$ and $\widetilde{Y}$ are two independent exponential random
variables with parameters $\gamma_1,\gamma_2>0$. Hence the
corresponding limit measure is $\eta(\D y)=\delta(y_1-Y)\delta(y_2-\widetilde{Y})\D y_1\D y_2$.

We take the elasticity factor to be $\rho = \frac{1}{2}$. This
yields intensity processes, which are given by square root diffusion
processes with jumps. We remark that the latter are heavily used in
CVA computations, see for instance \cite{BrigoPalla} and
\cite{BielCrep}.

Next, we specify the distribution of the jump sizes
$(Y_1^{(A)},Y_1^{(B)})$ of the counterparty intensities due to the
common Poisson process. Further, we specify the distribution of the jump sizes
$(\widetilde{Y}_1^{(A)},\widetilde{Y}_1^{(B)})$ of the intensities due to the
idiosyncratic Poisson processes of the two counterparties.
We assume that $(Y_1^{(A)},Y_1^{(B)})$ is given by a
bivariate exponential distribution with parameters
$\gamma_A,\gamma_B,\gamma_{AB}>0$. Hereafter, we write
$(Y_1^{(A)},Y_1^{(B)})\sim {\rm
BVE}(\gamma_A,\gamma_B,\gamma_{AB})$. As in \cite{MOlkin}, we have
$Y^{(i)}\sim\exp(\gamma_i+\gamma_{AB})$ for $i\in\{A,B\}$ and the
correlation of $(Y_1^{(A)},Y_1^{(B)})$ is given by
$\rho_{AB}=\frac{\gamma_{AB}}{\gamma_0}$, where
$\gamma_0=\gamma_{A}+\gamma_B+\gamma_{AB}$. Moreover, the moment
generating function of the bivariate exponential jump
$(Y_1^{(A)},Y_1^{(B)})$ is given by (see Lemma 3.2 in
\cite{MOlkin}):
\begin{eqnarray}\label{eq:mgf-BVE}
{\it\Phi}(\theta_A,\theta_B)=\frac{(\gamma_0 -
\theta_A-\theta_B)(\gamma_{A}+\gamma_{AB})(\gamma_{B}+\gamma_{AB})+\theta_A\theta_B\gamma_{AB}}{(\gamma_0
- \theta_A-\theta_B)(\gamma_A + \gamma_{AB}-\theta_A)(\gamma_B +
\gamma_{AB}-\theta_B)},\ \ \ \ \ \ \theta_A,\ \theta_B\leq0,
\end{eqnarray}
It can be further checked that
\begin{eqnarray}\label{eq:mgf-deriv}
\frac{\partial {\it\Phi}(\theta_A,\theta_B)}{\partial \theta_A}
&=&\frac{(\gamma_0-\theta_B)\theta_B\gamma_{AB}(\gamma_A+\gamma_{AB}-\theta_A)
+{(\gamma_0-\theta_A-\theta_B)^2\gamma_{AB}^*+(\gamma_0-\theta_A-\theta_B)\theta_A\theta_B\gamma_{AB}}}
{{(\gamma_0-\theta_A-\theta_B)^2}(\gamma_A+\gamma_{AB}-\theta_A)^2(\gamma_B+\gamma_{AB}-\theta_B)},\nonumber\\
&&\nonumber\\
&&\nonumber\\
\frac{\partial {\it\Phi}(\theta_A,\theta_B)}{\partial \theta_B}
&=&\frac{(\gamma_0-\theta_A)\theta_A\gamma_{AB}(\gamma_B+\gamma_{AB}-\theta_B)
+{(\gamma_0-\theta_A-\theta_B)^2\gamma_{AB}^*+(\gamma_0-\theta_A-\theta_B)\theta_A\theta_B\gamma_{AB}}}
{{(\gamma_0-\theta_A-\theta_B)^2}(\gamma_B+\gamma_{AB}-\theta_B)^2(\gamma_A+\gamma_{AB}-\theta_A)},\nonumber\\
\end{eqnarray}
where $\gamma_{AB}^*=(\gamma_A+\gamma_{AB})(\gamma_B+\gamma_{AB})$.
Similarly, we assume that
$(\widetilde{Y}_1^{(A)},\widetilde{Y}_1^{(B)})\sim {\rm
BVE}(\widetilde{\gamma}_A,\widetilde{\gamma}_B,\widetilde{\gamma}_{AB})$,
where
$\widetilde{\gamma}_A,\widetilde{\gamma}_B,\widetilde{\gamma}_{AB}>0$.

\subsection{Closed-Form Representation of $\widehat{F}$ } \label{sec:exposcomput}
We provide a closed-form expression for the time $t$ conditional survival probability $\widehat{F}(t,s)$, defined in Eq.~\eqref{eq:FLS1}.
We state the result in the form of the following proposition.

\begin{proposition}\label{lem:Ftsexpl}
Under the model setting specified in Section \ref{subsec:bcva-model}, the function $\widehat{F}$ in the large portfolio admits the following closed-form representation:
\begin{eqnarray*}
\widehat{F}(t,s) &=& \exp\left(x^* B_{p^*}(s-t) + \alpha^*\int_0^{s-t} B_{p^*}(u)\D u
\right)\frac{\gamma_1}{\gamma_1-c^*\widehat{\lambda}_c\int_0^{s-t}
B_{p^*}(u)\D u}\nonumber\\
&&\times\frac{\gamma_2}{\gamma_2-d^*\widehat{\lambda}^*\int_0^{s-t}
B_{p^*}(u)\D u},
\end{eqnarray*}
where $0\leq t<s\leq T$, and $B_{p^*}(u):=B(\kappa^*,\sigma^*;u)$ with $B$ given by Eq.~\eqref{eq:solu-express-Bexp}.
\end{proposition}
The proof of the proposition is reported in Appendix \ref{app:closeforms}.
In the special case where $Y=\widetilde{Y}$, we obtain
\begin{eqnarray}\label{hatF:simulation0}
\widehat{F}(t,s) &=& e^{x^* B_{p^*}(s-t)} \Ex\left[\exp\left(\left[\alpha^*+ (d^*\widehat{\lambda}^* + c^*\widehat{\lambda}_c)\ Y\right]
\int_0^{s-t} B_{p^*}(u)\D u\right)\right]\nonumber\\
&=&\frac{\gamma}{\gamma-(d^*\widehat{\lambda}^*+c^*\widehat{\lambda}_c)\int_0^{s-t}
B_{p^*}(u)\D u}\exp\left(x^* B_{p^*}(s-t) + \alpha^*\int_0^{s-t}
B_{p^*}(u)\D u \right),
\end{eqnarray}
where $\gamma=\gamma_1\ = \gamma_2$.

\subsection{Closed-Form Expressions for $H_1$ and $H_2$}

We provide the closed-form representation of the functions
$H_1$ and $H_2$ defined by \eqref{eq:H-hatH12}. As stated earlier,
we fix $\widehat{\rho}=\frac{1}{2}$ in
\eqref{eq:default-intensity-counterparty}.
\begin{proposition}\label{prop:HhatH12}
Let $\lambda =
\widehat{\lambda}_A+\widehat{\lambda}_B+\widehat{\lambda}_c$. Then,
the functions $H_1$ and $H_2$ defined in \eqref{eq:H-hatH12}
admit the following explicit representations:
\begin{itemize}
\item For $t\leq t_B\leq T$ and $(x_A,x_B)\in\R_+^2$, the function $H_1$ admits the closed-form:
\begin{eqnarray}\label{eq:propH1hatH1}
H_1(t_B-t,x_A,x_B)
&=&\left[h_1(t_B-t)+h_A(t_B-t)x_A+h_B(t_B-t)x_B\right]\nonumber\\
&&\times\exp\left(\widehat{h}_1(t_B-t) +\widehat{h}_A(t_B-t)x_A+\widehat{h}_B(t_B-t)x_B\right),
\end{eqnarray}
where the functions $(\widehat{h}_1(u),\widehat{h}_A(u),\widehat{h}_B(u))$ are given by: for $0\leq u\leq t_B$,
\begin{eqnarray}\label{eq:solu-hat-h}
\widehat{h}_A(u) &=& B(\kappa_A,\sigma_A; u),\nonumber\\
\widehat{h}_B(u) &=& B(\kappa_B,\sigma_B; u),\nonumber\\
\widehat{h}_1(u) &=& \int_0^u\Big[\alpha_A\widehat{h}_A(v)+\alpha_B\widehat{h}_B(v)
+\widehat{\lambda}_c{\it\Phi}(c_A\widehat{h}_A(v),c_B\widehat{h}_B(v)),\\
&&\qquad\
+\widehat{\lambda}_A\widetilde{\it\Phi}(d_A\widehat{h}_A(v),0)
+\widehat{\lambda}_B\widetilde{\it\Phi}(0,d_B\widehat{h}_B(v))\Big]\D
v - \lambda u,\nonumber
\end{eqnarray}
with $B(\cdot)$ specified in~\eqref{eq:solu-express-Bexp} and the functions
$(h_1(u),h_A(u),h_B(u))$ given by: for $0\leq u\leq t_B$,
\begin{eqnarray}\label{eq:solu-h}
{h}_A(u) &\equiv&0,\nonumber\\
h_B(u) &=& \exp\left(-\kappa_B u + \sigma_B^2 \int_0^u \widehat{h}_B(v)\D v\right),\\
h_1(u) &=&
\int_0^u\Big[\alpha_Bh_B(v)+\widehat{\lambda}_cc_Bh_B(v)\frac{\partial{\it\Phi}(c_A\widehat{h}_A(v),c_B\widehat{h}_B(v))}
{\partial\theta_B}\nonumber\\
&&\qquad\ +\widehat{\lambda}_Bd_Bh_B(v)\frac{\partial\widetilde{\it\Phi}(0,d_B\widehat{h}_B(v))}{\partial\theta_B}\Big]\D v.\nonumber
\end{eqnarray}

\item For $t\leq t_A\leq T$ and $(x_A,x_B)\in\R_+^2$, the function $H_2$ admits the closed-form:
\begin{eqnarray}\label{eq:propH2hatH2}
H_2(t_A-t,x_A,x_B)&=&\left[w_1(t_A-t)+w_A(t_A-t)x_A+w_B(t_A-t)x_B\right]\nonumber\\
&&\times\exp\Big(\widehat{w}_1(t_A-t) +\widehat{w}_A(t_A-t)x_A+\widehat{w}_B(t_A-t)x_B\Big),
\end{eqnarray}
where the functions $(\widehat{w}_1(u),\widehat{w}_A(u),\widehat{w}_B(u))$ are given by: for $0\leq u\leq t_A$,
\begin{eqnarray}\label{eq:solu-hat-w}
\widehat{w}_A(u) &=& B(\kappa_A,\sigma_A; u),\nonumber\\
\widehat{w}_B(u) &=& B(\kappa_B,\sigma_B; u),\nonumber\\
\widehat{w}_1(u) &=& \int_0^u\Big[\alpha_A\widehat{w}_A(v)+\alpha_B\widehat{w}_B(v)+\widehat{\lambda}_c{\it\Phi}(c_A\widehat{w}_A(v),c_B\widehat{w}_B(v)),\\
&&\qquad\
+\widehat{\lambda}_A\widetilde{\it\Phi}(d_A\widehat{w}_A(v),0)+\widehat{\lambda}_B\widetilde{\it\Phi}(0,d_B\widehat{w}_B(v))\Big]\D
v - \lambda u,\nonumber
\end{eqnarray}
with the functions $(w_1(u),w_A(u),w_B(u))$ given by: for $0\leq u\leq t_A$,
\begin{eqnarray}\label{eq:solu-w}
{w}_A(u) &=&\exp\left(-\kappa_A u + \sigma_A^2 \int_0^u \widehat{w}_A(v)\D v\right),\nonumber\\
w_B(u) &\equiv& 0,\\
w_1(u) &=& \int_0^u\Big[\alpha_A
w_A(v)+\widehat{\lambda}_cc_Aw_A(v)\frac{\partial{\it\Phi}(c_A\widehat{w}_A(v),c_B\widehat{w}_B(v))}{\partial\theta_A}\nonumber\\
&&\qquad\ +\widehat{\lambda}_Ad_Aw_A(v)\frac{\partial\widetilde{\it\Phi}(d_A\widehat{w}_A(v),0)}{\partial\theta_A}\Big]\D v.\nonumber
\end{eqnarray}
\end{itemize}
\end{proposition}
The proof of the above proposition is postponed to Appendix \ref{app:prop-HhatH12}.

The results derived in propositions~\ref{lem:Ftsexpl} and ~\ref{prop:HhatH12} along with Theorem \ref{thm:cva1} yield
a closed-form expression for the BCVA formula.

\section{Numerical and Economic Analysis} \label{sec:numerics}
We provide a numerical and economic analysis of the BCVA formula. We first analyze the accuracy of
the weak limit approximation to the exposure in Subsection \ref{sec:qualapprox}. We then perform a comparative statics analysis of the BCVA
adjustment in Subsection \ref{sec:BCVAanal}.

\subsection{Quality of Exposure Approximation}\label{sec:qualapprox}
We assess the quality of the exposure approximation given by
Eq.~\eqref{eq:expo} and~\eqref{hatF:simulation0}. We set $K=300$,
i.e. sufficiently large, and analyze how our approximate formula for
the exposure compares versus the corresponding Monte-Carlo estimate.
The latter is obtained by first simulating the multivariate $K+2$
intensity process, including all names in the portfolio and the two
counterparties via the Euler scheme, see \cite{Andersen}. Let us
denote by $\tilde{\xi}^{(k),m}_t$ the value of $\xi^{(k)}_t$ on the
$m$-th simulation path. Then, we compute the Monte-Carlo exposure on
$\{\widetilde{\tau}_K>t\}$, as
\begin{eqnarray*}
\frac{\tilde{\varepsilon}^{(K)}(t,T)}{K}&=& \frac{1}{M} \sum_{m=1}^M \frac{1}{K}\sum_{k=1}^K z_k CDS^{(k)}(t,T)\nonumber\\
&=&\frac{1}{M} \sum_{m=1}^M \left[e^{-r(T-t)}\frac{1}{K}\sum_{k=1}^Kz_kL_kH_G(T;t,\tilde{\xi}_t^{(k),m})-\frac{1}{K}\sum_{k=1}^Kz_kL_k\right]\nonumber\\
 && + \frac{1}{M} \sum_{m=1}^M \int_t^Te^{-r(s-t)}\frac{1}{K}\sum_{k=1}^Kz_k(S_k+rL_k)H_G(s;t,\tilde{\xi}_t^{(k),m})\D s,
\end{eqnarray*}
where the function
\begin{eqnarray*}
H_G(s;t,\tilde{\xi}_t^{(k),m})=\exp\left(A_0^{(k)}(s-t)+B_0^{(k)}(s-t)
\tilde{\xi}_t^{(k),m}\right),\ \ \ \ s\geq t\geq0.
\end{eqnarray*}
Here, $B_0^{(k)}(u)=B(\kappa_k,\sigma_k;u)$ is given by \eqref{eq:solu-express-Bexp}, and the function
\[
A_0^{(k)}(u)=\int_0^u \left[\alpha_kB_0^{(k)}(v)+(\widehat{\lambda}_c+\widehat{\lambda}_k)\left(\frac{\gamma}{\gamma-c_kB_0^{(k)}(v)}-1\right)\right]\D v.
\]
We fix $\alpha^*=x^*  \kappa^*$, $c^*=d^*$, $\lambda^*=0.5$. Further, $\gamma_1 = \gamma_2 = 1.5$, $\widehat{\lambda}_c=2.5$, $r = 0.03$. We set $S_z^* = 0.02$,  $L_z^* =  0.4$. We choose $L_A = L_B = 0.4$.
Next, we define the sequence of credit risk and contractual parameters of the $K$ names in the portfolio as
\begin{eqnarray*}
\xi_0^{(k)} &=& x^* \left(1+ \frac{1}{k} \right) \qquad \alpha_k = \alpha^* \left(1+ \frac{1}{k} \right) \\
\kappa_k &=& \kappa^* \left(1+ \frac{1}{k} \right) \qquad \sigma_k = \sigma^* \left(1+ \frac{1}{k} \right) \\
c_k  &=& c^* \left(1+ \frac{1}{k} \right) \qquad d_k  = d^* \left(1+ \frac{1}{k} \right)\\
S_k &=& S^* \left(1+ \frac{1}{k} \right) \qquad L_k = L^* \left(1- \frac{1}{k} \right) \\
\widehat{\lambda}_k &=& \widehat{\lambda}^* \left(1+ \frac{1}{k} \right) \\
\end{eqnarray*}
Notice that all parameters are decreasing to their limit, except for $L_k$, which increases to its limit (this is done to maintain the realistic assumption that $L_k \leq 1$). We assume that the counterparty $A$ is equally long on each contract, i.e. $z_k = 1$. 

\begin{figure}
\centering
\begin{tabular}{cc}
\epsfig{file={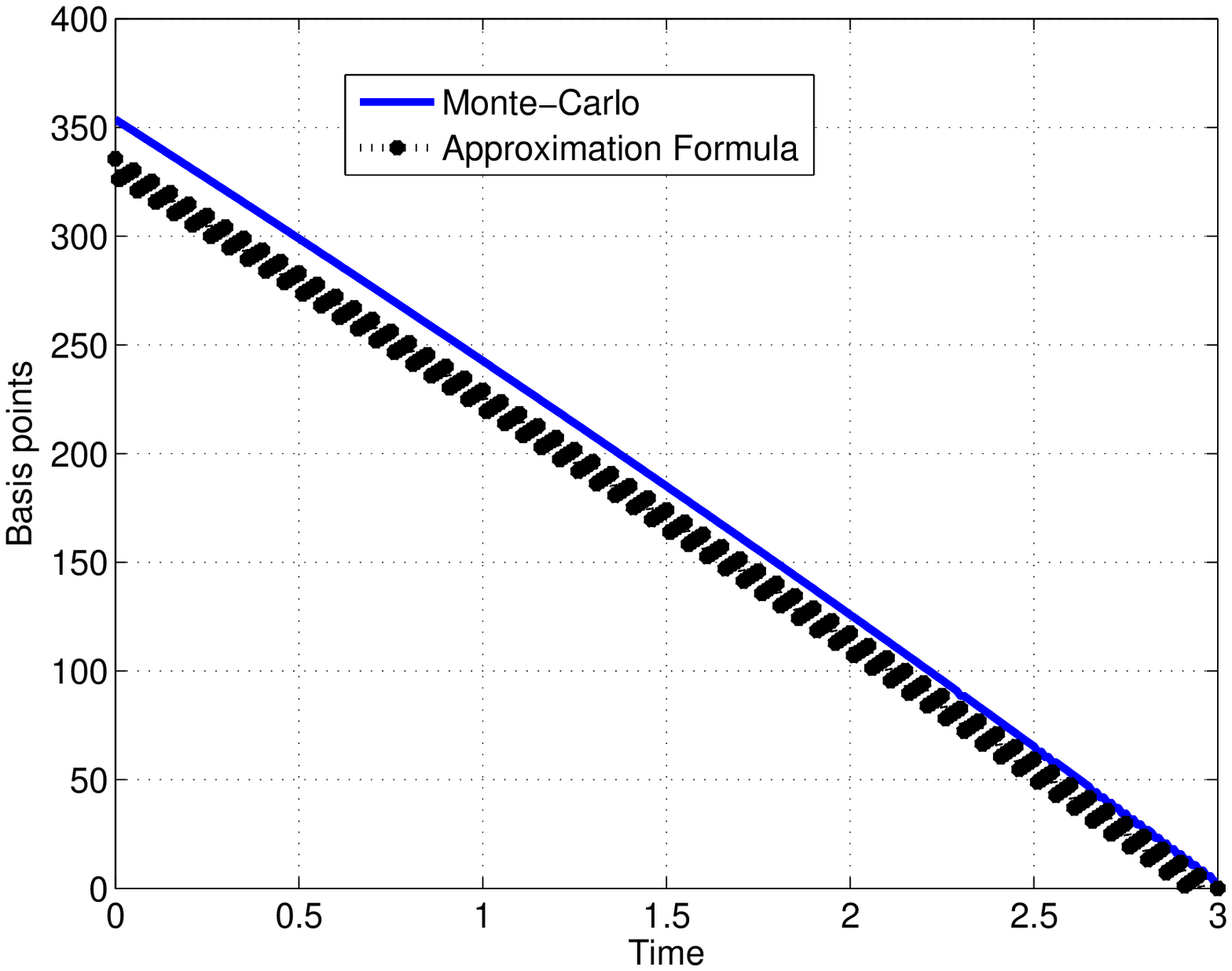},width=0.4\linewidth,clip=}
\epsfig{file={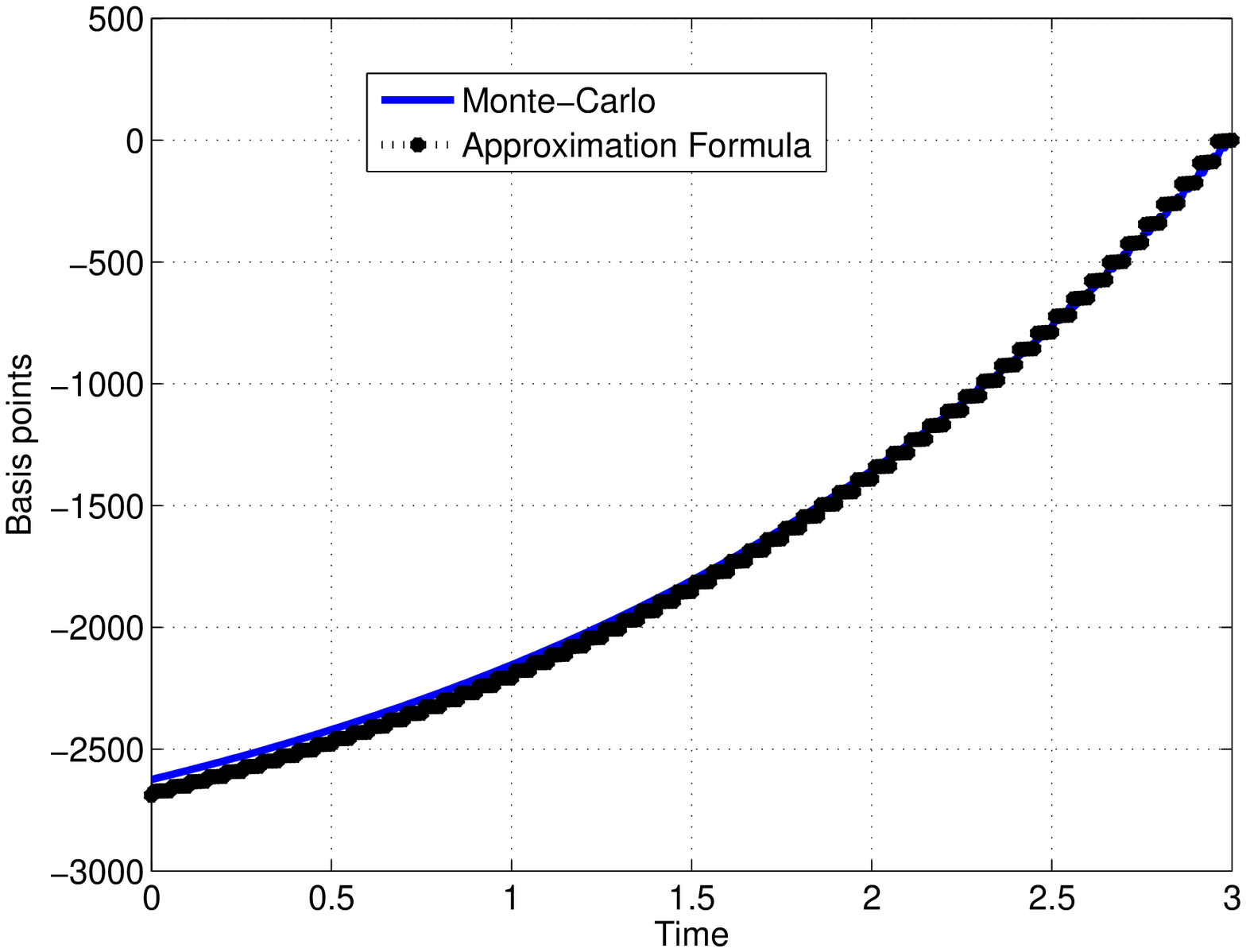},width=0.4\linewidth,clip=} \\
\epsfig{file={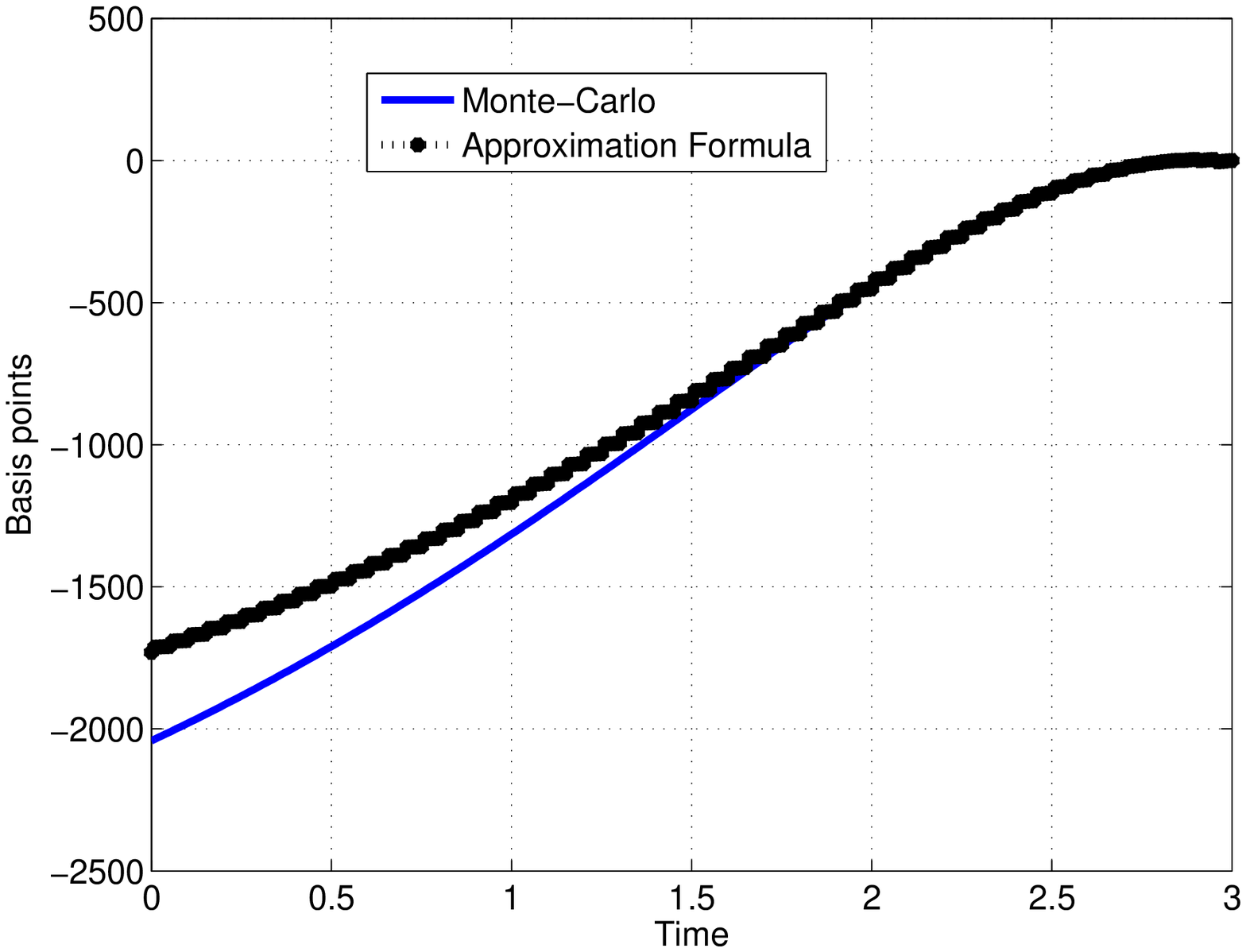},width=0.4\linewidth,clip=}
\epsfig{file={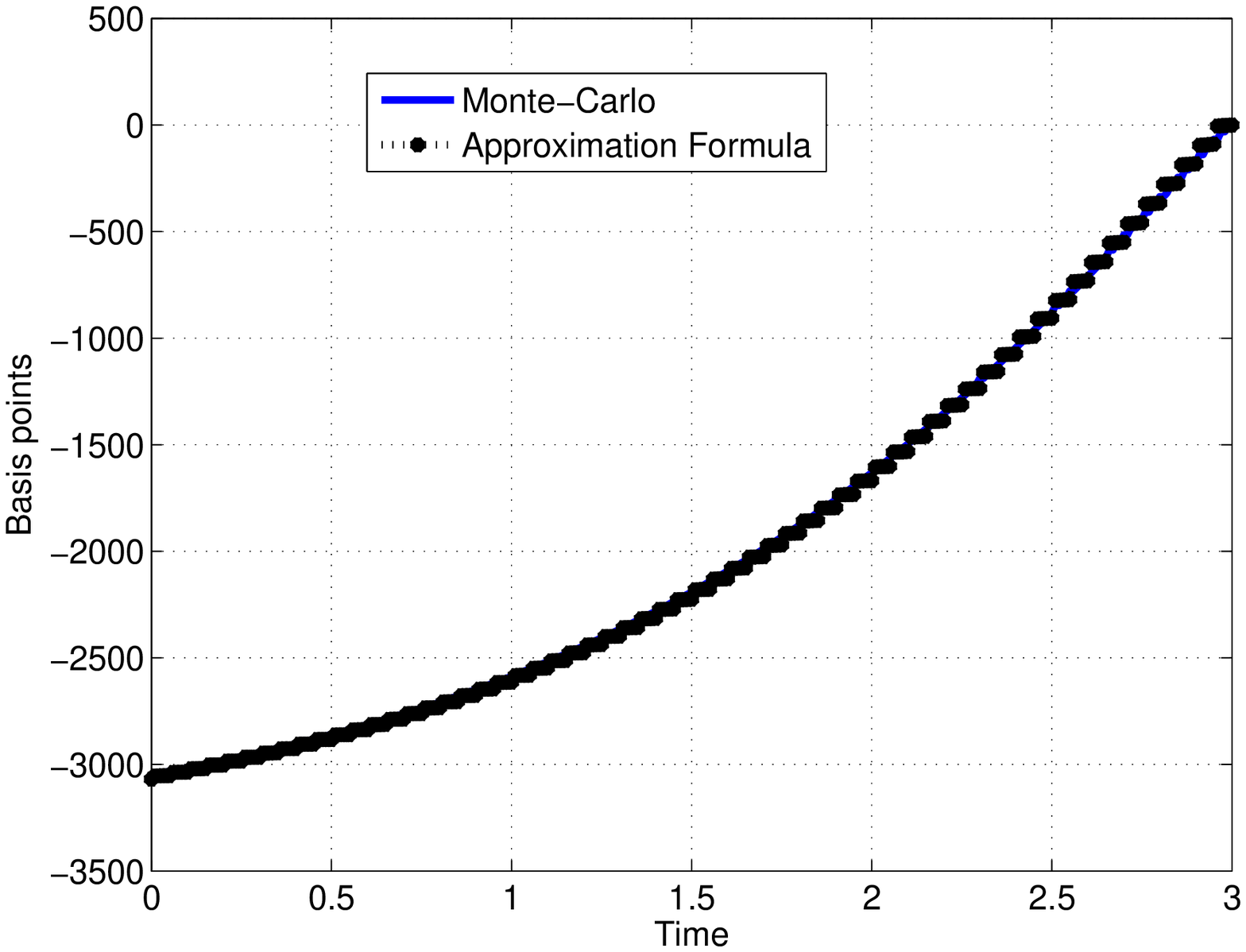},width=0.4\linewidth,clip=}
\end{tabular}
\caption{Monte-Carlo exposure estimate with $K$ names in the portfolio versus ``law of large numbers'' exposure. The left top panel is associated with
$(x^*,\kappa^*,\sigma^*,c^*,d^*) = (0.02, 0.5, 0.01, 0,0)$. The left right panel is associated with
$(x^*,\kappa^*,\sigma^*,c^*,d^*) = (0.5, 1.5, 0.2, 0,0)$.  The bottom top panel is associated with
$(x^*,\kappa^*,\sigma^*,c^*,d^*) = (0.02, 0.5, 0.2,0.2,0.2)$. The bottom right panel is associated with
$(x^*,\kappa^*,\sigma^*,c^*,d^*) = (0.5, 1.5, 0.2, 0.2,0.2)$.}
\label{fig:convergence}
\end{figure}

Figure \ref{fig:convergence} shows how the approximation behaves under different configuration settings, where we vary the level of default risk, and allow or not for the presence of jumps in the intensity processes.
Figure \ref{fig:convergence} clearly shows that as time approaches expiration, the exposure decreases in magnitude since both counterparties are exposed to default risk for a shorter time horizon.
The top panels show that both in the case when the exposure is positive or negative for the investor, the approximation formula yields very accurate results when compared to the Monte-Carlo estimate.
The bottom panels indicate that if the default risk is low (bottom left), then the presence of jumps introduces a small approximation error if the time to horizon is large. However, the mismatch decreases fast and disappears as the time to the horizon decreases. If the portfolio is instead very risky (bottom right panel), then our approximation formula exbibits a perfect match with the Monte-Carlo estimate.

\subsection{BCVA Economic Analysis} \label{sec:BCVAanal}

We analyze the behavior of CVA and DVA adjustments computed using
formula~\eqref{eq:cvagen}. We use the following parameters for
counterparties $A$ and $B$: $\gamma_A=\gamma_B = \tilde{\gamma}_A = \tilde{\gamma}_B = 1.5$, $\gamma_{AB}=\tilde{\gamma}_{AB} = 0$,
$\widehat{\lambda}_A = \widehat{\lambda}_B = 0.4,\ c_A = d_A = c_B =d_B = 0.3$, $L_A =
L_B = 0.4$, $\xi_0^{(A)} =  \xi_0^{(B)} = 0.2$, $\kappa_A =
\kappa_B=0.6$, $\sigma_A = \sigma_B=0.3$, $\alpha_A = \alpha_B =
0.4$, $\gamma = 2$. We fix the time horizon $T = 3$, and assume the following limit parameters for the intensity process:
$c^*=d^*=0.1, \kappa^*=0.5, \alpha^*=0.01, \sigma^* = 0.3, \widehat{\lambda}^* = 0.2, x^*=0.02$. The limiting contractual parameters of the CDS are set to $L^*=0.4, S^*= 0.02$. Further, we assume $z_k = 1$, i.e
$A$ is long and $B$ is short on each CDS contract.

\begin{figure}
\centering
\begin{tabular}{cc}
\epsfig{file={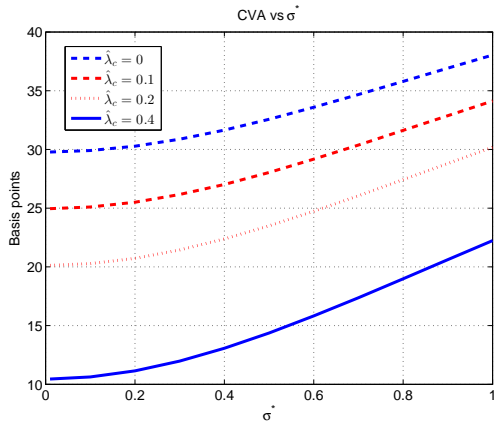},width=0.4\linewidth,clip=}
\end{tabular}
\caption{CVA adjustments with respect to $\sigma^*$.}
\label{fig:CVAvol}
\end{figure}

Under the above scenario, the trading counterparties have a symmetric credit risk profile, and are riskier than the names in the underlying portfolio. Hence we expect the on-default exposure to be non-negligible. Figure \ref{fig:CVAvol}
shows that CVA adjustments are increasing in the volatility $\sigma^*$. This is expected because larger volatility increases the credit risk of the names in the portfolio, hence increasing the exposure of $A$ to $B$, and resulting in larger CVA adjustments. As the CDS portfolio is always in the money for $A$, and out of money for $B$, DVA adjustments are zero and hence not reported here. When the intensity of the common Poisson process increases, the CVA adjustments decrease. This happens because, when common jumps are more likely to occur, the default likelihood of both portfolio names and counterparties increase. Hence, a smaller number of names would default after either counterparty, implying reduced market exposures and smaller CVA adjustments.

\begin{figure}
\centering
\begin{tabular}{cc}
\epsfig{file={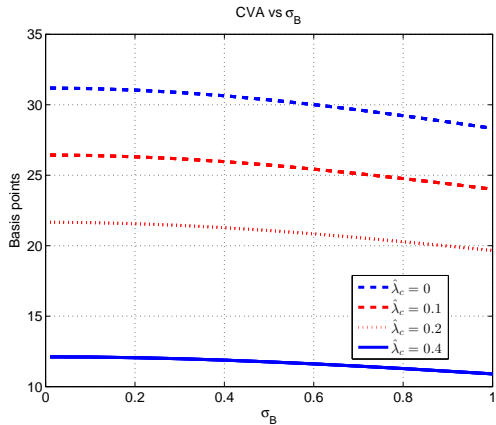},width=0.4\linewidth,clip=}
\end{tabular}
\caption{CVA adjustments with respect to the volatility $\sigma_B$.}
\label{fig:CVAvolB}
\end{figure}

Figure \ref{fig:CVAvolB} shows that the CVA slightly decreases when the credit risk volatility of the counterparty $B$ increases. When the default intensities are mainly driven by the common Poisson process, i.e. $\widehat{\lambda}_c$ is large, then (1) all CDS contracts become less in the money for $A$ and more in the money for $B$, and (2) a larger number of names anticipates $B$ in defaulting. All this reduces the positive on-default exposure of $A$ to $B$. The conclusion is that the CVA decreases, as also confirmed from the graph of Figure \ref{fig:CVAvolB}.

\begin{figure}
\centering
\begin{tabular}{cc}
\epsfig{file={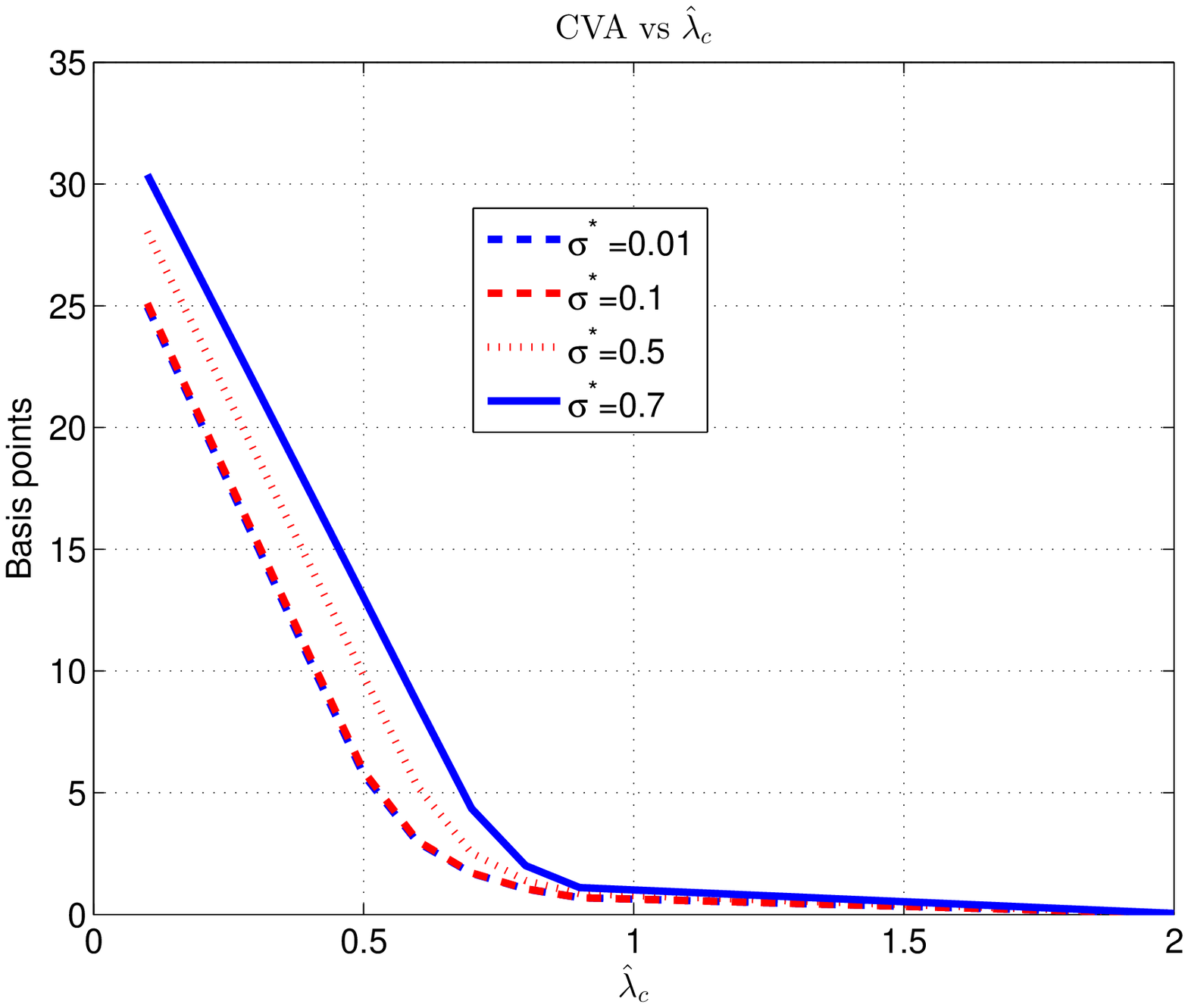},width=0.4\linewidth,clip=}
\epsfig{file={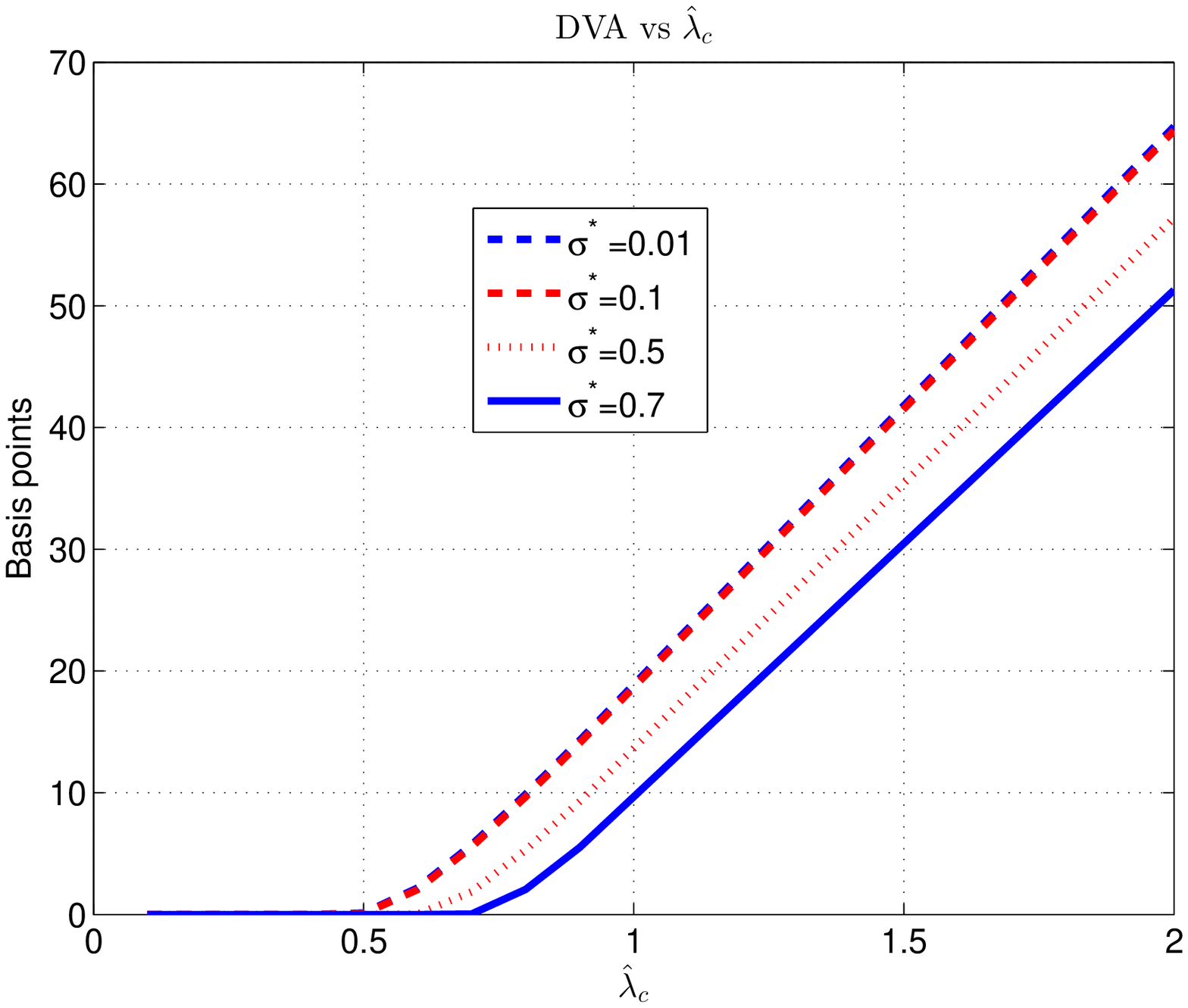},width=0.4\linewidth,clip=}
\end{tabular}
\caption{CVA and DVA adjustments with respect to the intensity $\widehat{\lambda}_c$.}
\label{fig:CVACI}
\end{figure}

Figure \ref{fig:CVACI} shows that when the underlying portfolio becomes riskier, bilateral counterparty risk inverts the sign. Indeed it changes from being negative (positive CVA and negligible DVA adjustments) to being positive (positive DVA and negligible CVA adjustments). Indeed, when the intensity of the common Poisson process is low, the portfolio default risk is small, and hence the counterparty A would always measure a positive on-default exposure to $B$. As jumps of the common Poisson process occur more frequently, the situation reverses and $B$ would measure a positive on-default exposure to $A$. Consequently, the CVA adjustments will become negligible, while the DVA adjustments increasingly higher.

\begin{figure}
\centering
\begin{tabular}{cc}
\epsfig{file={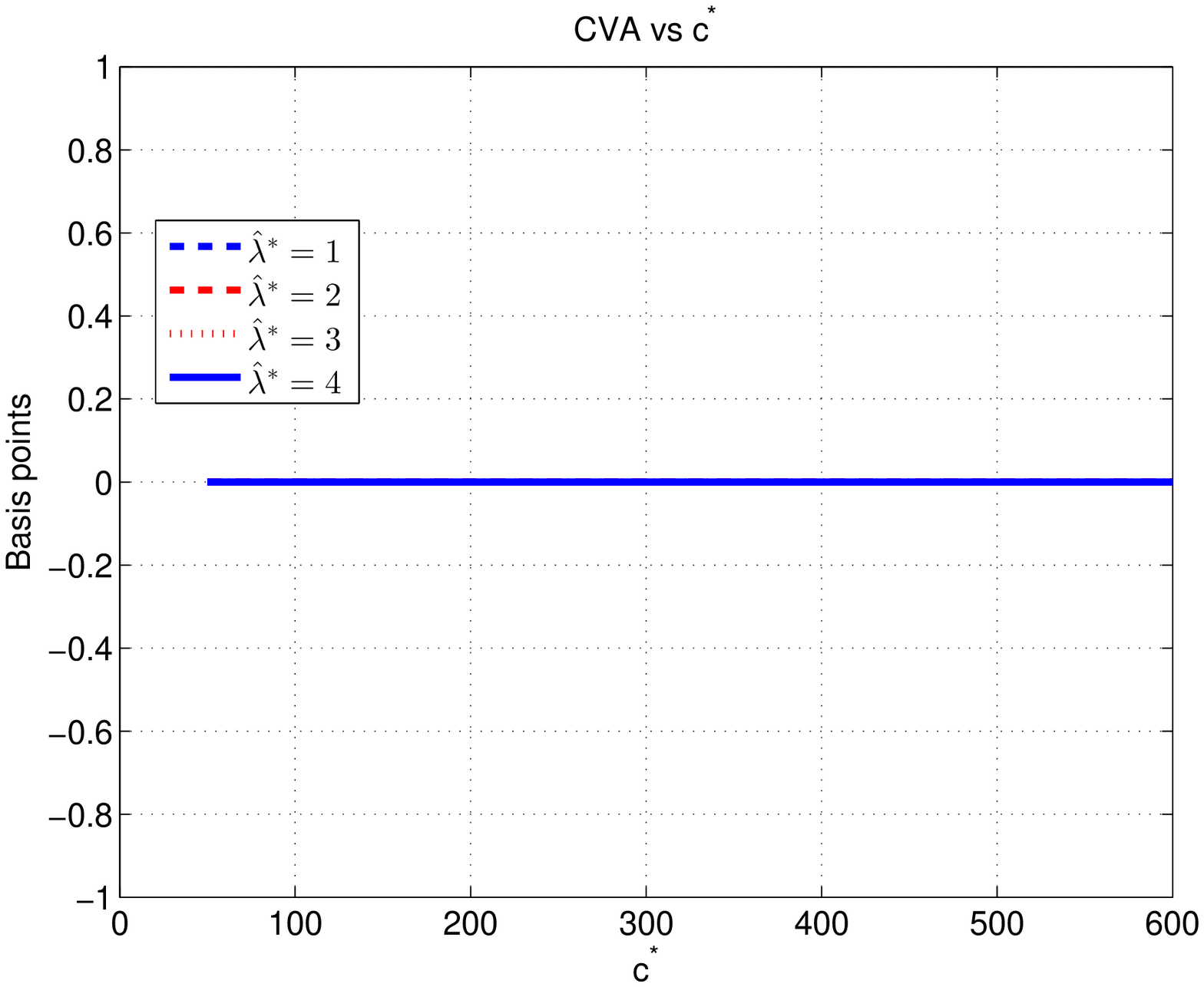},width=0.4\linewidth,clip=}
\epsfig{file={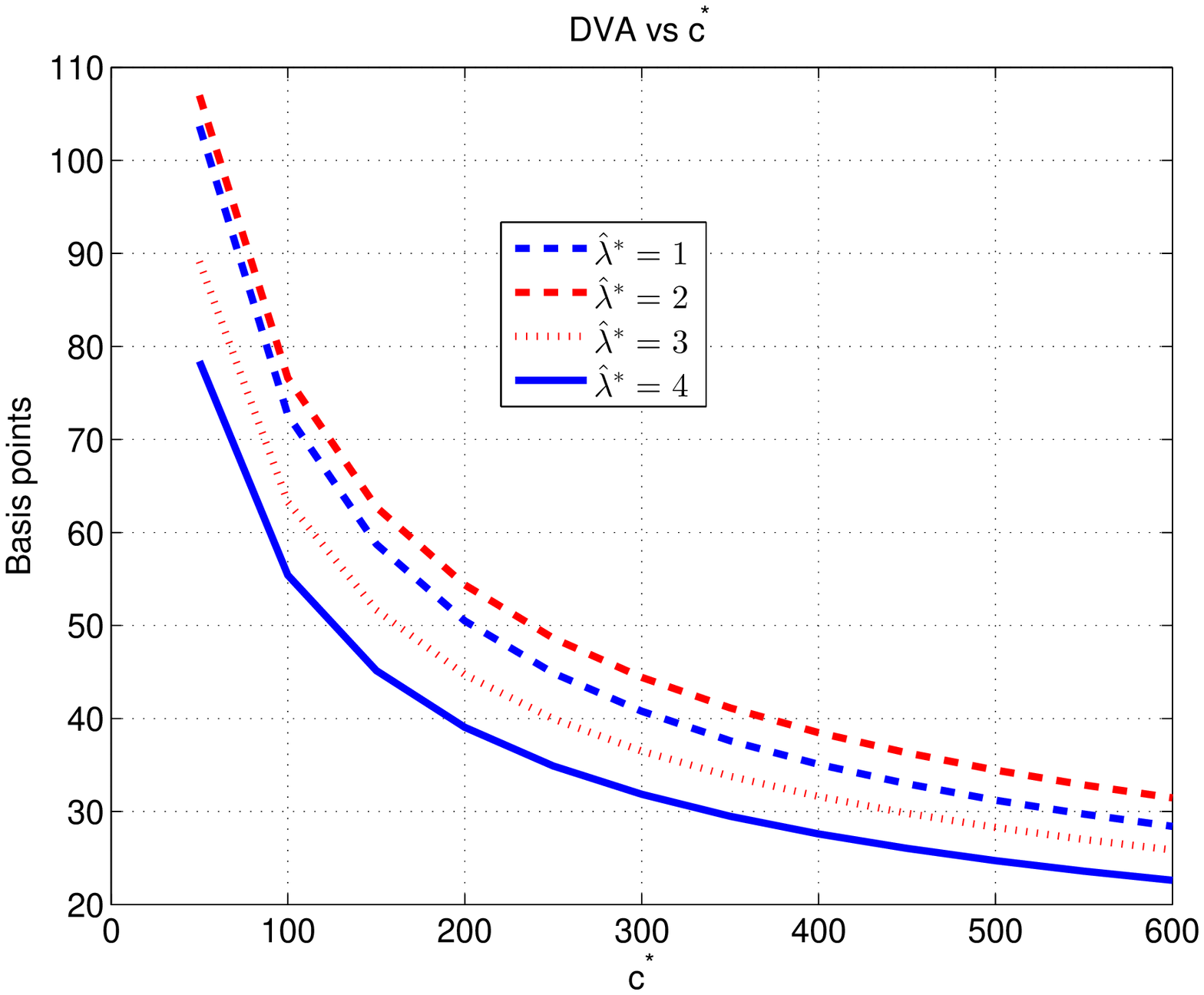},width=0.4\linewidth,clip=}
\end{tabular}
\caption{CVA and DVA adjustments with respect to the limiting jump size $c^*$.}
\label{fig:CVAvsc}
\end{figure}

Despite the long counterparty $A$ benefits from his own default when the credit risk level in the underlying portfolio increases (larger DVA adjustments), we next demonstrate that if the credit risk of the names becomes significantly higher than the one of either counterparty, both of them will measure a small exposure to the other. This is illustrated next, by considering a high risk CDS portfolio, with $x^* = 10$, $\alpha^*=5$ (the remaining parameters are left unchanged). We analyze the impact caused on the adjustments by increasingly larger jump sizes experienced by the intensities of the portfolio names. Clearly, the high default risk of the portfolio makes the on-default exposure of $A$ to $B$ negative, i.e. the portfolio is out-of-money for $A$, which is receiving too a low CDS premium. This immediately explains why the CVA in the left panel of Figure~\ref{fig:CVAvsc} is zero. Although a larger number of portfolio names default before both counterparties, when $A$ defaults his exposure to $B$ is highly negative due to the high default risk of the CDS portfolio. Hence, $A$ would still benefit from his own default, which result in a sizeable DVA adjustment. As $c^*$ increases, the intensities of the portfolio names will experience higher jumps, and thus a significantly larger fraction would default before $A$. As evidenced from Figure~\ref{fig:CVAvsc}, this reduces the size of DVA adjustments, especially if the idiosyncratic jumps occur at a high frequency.

\section{Conclusions} \label{sec:conclus}
We have developed a rigorous analytical framework for computing the bilateral CVA adjustment on a portfolio of credit default swaps. In the case when the portfolio consists of an asymptotically large number of credit default swaps, we have explicitly characterized the mark-to-market exposure. We have achieved that by means of a weak convergence analysis, relying on martingale arguments, showing that the aggregated intensity process can be recovered as the weak limit of a sequence of weighted empirical measure-valued processes. Using this result, we have provided a semi-closed-form expression for the BCVA adjustment under a conditionally independent default correlation model, where default intensities follow CEV processes with jumps. By further specializing the CEV to a CIR process enhanced with jumps, we have provided fully explicit expressions for the BCVA, by suitably combining our law of large numbers approximation formula for the exposure and the theory of affine processes. We have provided a detailed numerical analysis to measure the quality of the weak-limit approximation and a comparative statics analysis to interpret the financial meaning of the derived BCVA formula. We have found that our law of large numbers approximation for the portfolio exposure is accurate, regardless of the credit risk levels of the portfolio. From a financial perspective, the CVA adjustments in the large CDS portfolio are increasing in the credit risk volatility of the portfolio names, and highly sensitive to default correlation.


\appendix

\section{Proof of Lemma \ref{lem:positive-cve}} \label{app:positive-cve}

It can be easily checked that the drift and
volatility coefficients satisfy the Yamada \& Watanabe condition of
Proposition 2.13, Chapter 5 in \cite{KarShreve}. An application of such proposition
yields the existence of a unique non-explosive strong solution to the SDE
\eqref{eq:intensity-sde}.

Next we prove the positivity of this strong solution. Due to the
existence of only positive jumps in \eqref{eq:intensity-sde}, it is
enough to check that the intensity $\xi^{(k)}$ stays positive when
$c_k=d_k=0$. Let $\widetilde{\xi}^{(k)}$ be the corresponding (strong)
solution to SDE \eqref{eq:intensity-sde} with $c_k=d_k=0$.
Let $\ell^{a+}(M) = (\ell^{a+}_t(M); t\geq0)$ be the upper local
time process of any continuous semi-martingale $M = (M_t; t\geq0)$
concentrated on point $a\in\R$. Then the upper local time process
$\ell^{a+}(M)$ can be identified by
\begin{eqnarray*}
\ell^{a+}_t(M) =
\lim_{\varepsilon\downarrow0}\frac{1}{\varepsilon}\int_0^t{\bf1}_{\{a\leq
M_s<a+\varepsilon\}}\D\left<M,M\right>_s,\ \ \ \forall\ t\geq0.
\end{eqnarray*}
We next verify that the upper local time process
$\ell^{0+}(\widetilde{\xi}^{(k)})$ of the continuous semi-martingale
$\widetilde{\xi}^{(k)}$ concentrated on point $0$ is zero. When $\rho>\frac{1}{2}$, for all positive $t > 0$
and $\varepsilon > 0$, we have
\begin{eqnarray*}
\frac{1}{\varepsilon}\int_0^t{\bf1}_{\{0\leq
\widetilde{\xi}^{(k)}<\varepsilon\}}\D\left<\widetilde{\xi}^{(k)},\widetilde{\xi}^{(k)}\right>_s=\frac{\sigma_k^2}{\varepsilon}\int_0^t{\bf1}_{\{0\leq
\widetilde{\xi}^{(k)}<\varepsilon\}}(\widetilde{\xi}^{(k)})^{2\rho}\D s
\leq\sigma_k^2\varepsilon^{2\rho-1}t,
\end{eqnarray*}
which approaches zero as $\varepsilon\downarrow0$. This shows that
the upper local time process $\ell^{0+}(\widetilde{\xi}^{(k)})\equiv0$ when $\rho>\frac{1}{2}$. For
$\rho=\frac{1}{2}$, using the occupation time formula, we get
\begin{eqnarray*}
\int_{\R}\frac{1}{|a|}{\bf1}_{\{a\neq0\}}\ell^{a+}_t(\widetilde{\xi}^{(k)})\D a&=&\sigma_k^2\int_0^t\frac{1}{|\widetilde{\xi}^{(k)}|}{\bf1}_{\{|\widetilde{\xi}^{(k)}|>0\}}|\widetilde{\xi}^{(k)}|^{2\rho}\D
s\nonumber\\
&=&\sigma_k^2\int_0^t{\bf1}_{\{|\widetilde{\xi}^{(k)}|>0\}}\D s\leq\sigma_k^2t,\ \ \ \ \forall\ t\geq0.
\end{eqnarray*}
Note that $|a|^{-1}$ is not integrable in any neighborhood of $a =
0$. Then it must hold that $\ell^{0+}_t(\widetilde{\xi}^{(k)})=0$ for all
$t\geq0$. Using Tanaka's formula, it follows that
\begin{eqnarray*}
\Ex\left[(\widetilde{\xi}^{(k)}_{t\wedge\varsigma_m})_{-}\right]
&=&\Ex\left[(\xi^{(k)}_{0})_{-}\right]
-\Ex\left[\int_0^{t\wedge\varsigma_m}{\bf1}_{\{\widetilde{\xi}^{(k)}_s\leq0\}}\D\widetilde{\xi}^{(k)}_s\right]
+\frac{1}{2}\Ex\left[\ell_{t\wedge\varsigma_m}^{0+}(\widetilde{\xi}^{(k)})\right]\nonumber\\
&=&-\alpha_k\Ex\left[\int_0^{t\wedge\varsigma_m}{\bf1}_{\{\widetilde{\xi}^{(k)}_s\leq0\}}\D s\right]
+\kappa_k\Ex\left[\int_0^{t\wedge\varsigma_m}{\bf1}_{\{\widetilde{\xi}^{(k)}_s\leq0\}}\widetilde{\xi}^{(k)}_s\D
s\right]\nonumber\\
&\leq& 0,\ \ \ \ \forall\ t\geq0,
\end{eqnarray*}
where $\varsigma_m=\inf\{t>0;\ |\widetilde{\xi}^{(k)}|\geq m\}$ with $m\in\N$. This implies that $\widetilde{\xi}^{(k)}_{t\wedge\varsigma_m}\geq0$, $\Px$-a.s.
for each $m\in\N$. Letting $m\to+\infty$, we conclude that
$\widetilde{\xi}^{(k)}\geq0$ for all $t\geq0$.

By virtue of the Feller boundary classification criteria, we have
that the boundary $0$ is unattainable for $\widetilde{\xi}^{(k)}$ when
$\rho>\frac{1}{2}$. Thus the proof of the lemma is complete.
\hfill$\Box$

\section{Proofs Related to Weak Convergence Analysis}\label{app:weak-convergene}
\renewcommand\theequation{B.\arabic{equation}}
\setcounter{equation}{0}

\subsection{Moment Estimate of Intensities of $K$-Names} \label{app:momest}

Recall that the intensity process $\xi^{(k)}=(\xi^{(k)}_t;\ t\geq0)$ of the $k$-th name
follows the CEV process with jumps in \eqref{eq:intensity-sde}.

\begin{lemma}\label{lem:moment-intensity}
Let the assumption {\rm(\bf A2)} hold. Then for any $T>0$,
\begin{eqnarray}\label{eq:p-moment-average}
\sup_{0\leq t\leq T,\ K\in\N}\frac{1}{K}\sum_{k=1}^K\Ex\left[\left|\xi_{t}^{(k)}\right|^\beta\right]<+\infty,
\end{eqnarray}
where $1\leq \beta\leq 4$.
\end{lemma}

The proof procedure for the moment estimates \eqref{eq:p-moment-average} follows straightforward arguments.
First, we apply It\^o's formula, and then use H\"older inequality, BDG inequality and Gronwall Lemma. The full
details are omitted here.

\subsection{Proof of Lemma \ref{lem:limit-geneator}}\label{app:lem-generator}

It follows from the definition \eqref{eq:default-time} of default times, that for each $k\in\{1,2,\dots,K\}$
\begin{eqnarray}\label{eq:default-mart}
{\cal M}_t^{(k)}:=H_t^{(k)} - \int_0^t\overline{H}_s^{(k)}\xi_s^{(k)}\D s,\ \ \ \ \ \ t\geq0
\end{eqnarray}
is a $(\Px,\G_t^{(k)})$-martingale. Hence the third term on the r.h.s. of Eq.~\eqref{eq:ito} may be rewritten as
\begin{eqnarray*}
&&\frac{1}{K}\int_0^t\sum_{k=1}^Kf(p_k,(Y_1^{(k)},\widetilde{Y}_1^{(k)}),\xi_{s-}^{(k)})\D\overline{H}_s^{(k)}
=-\frac{1}{K}\int_0^t\sum_{k=1}^Kf(p_k,(Y_1^{(k)},\widetilde{Y}_1^{(k)}),\xi_{s-}^{(k)})\D{H}_s^{(k)}\nonumber\\
&&\qquad=-\frac{1}{K}\int_0^t\sum_{k=1}^Kf(p_k,(Y_1^{(k)},\widetilde{Y}_1^{(k)}),\xi_{s-}^{(k)})\D{\cal
M}_s^{(k)}-\frac{1}{K}\int_0^t\sum_{k=1}^K\xi_{s}^{(k)}f(p_k,(Y_1^{(k)},\widetilde{Y}_1^{(k)}),\xi_{s-}^{(k)})\overline{H}_s^{(k)}\D
s.
\end{eqnarray*}
Thus, we can conclude that there exists a {(local)} martingale
$\widehat{\cal M}^{(K)}=(\widehat{\cal M}_t^{(K)};\ t\geq0)$ such
that
\begin{eqnarray}\label{eq:martingale}
{\it\Phi}(\nu_t^{(K)}) &=& {\it\Phi}(\nu_0^{(K)}) + \sum_{m=1}^M\int_0^t\frac{\partial \varphi}{\partial x_m}(\nu_s^{(K)}({\bm f}))
\nu_s^{(K)}(\cL_{11}f_m)\D s + \widehat{\cal M}^{(K)}_t\nonumber\\
&&+\frac{1}{2K^2}\sum_{m,n=1}^M\int_0^t\frac{\partial^2\varphi}{\partial
x_m\partial x_{n}}(\nu_s^{(K)}({\bm f}))\nonumber\\
&&\qquad\times\left(\sum_{k=1}^K\sigma_k^2\frac{\partial f_m}{\partial x}(p_k,(Y_1^{(k)},\widetilde{Y}_1^{(k)}),\xi_s^{(k)})
\frac{\partial f_n}{\partial x}(p_k,(Y_1^{(k)},\widetilde{Y}_1^{(k)}),\xi_s^{(k)})(\xi_s^{(k)})^{2\rho}\overline{H}_s^{(k)}\right)\D s\nonumber\\
&&-\frac{1}{K} \sum_{m=1}^M \int_0^t \frac{\partial \varphi}{\partial x_m}(\nu_s^{(K)}({\bm f}))\left(\sum_{k=1}^K\xi_{s}^{(k)}
f_m(p_k,(Y_1^{(k)},\widetilde{Y}_1^{(k)}),\xi_{s-}^{(k)})\overline{H}_s^{(k)}\right)\D s\nonumber\\
&&+\sum_{k=1}^K\widehat{\lambda}_k\int_0^t\left[\varphi(\nu_s^{(K)}({\bm f})+{\bm J}_s^{(K,k)}(Y_1^{(k)},\widetilde{Y}_1^{(k)}))
-\varphi(\nu_s^{(K)}({\bm f}))\right]\D s\nonumber\\
&&+\widehat{\lambda}_c\int_0^t\left[\varphi(\nu_s^{(K)}({\bm
f})+{\bm J}_s^{(K,c)}({\bm Y}_1, {\widetilde{\bm  Y}}_1))-\varphi(\nu_s^{(K)}{\bm
f})\right]\D s,
\end{eqnarray}
where $t\geq0$. Notice that the third line of the above equation may be rewritten as
\[
-\sum_{m=1}^M\int_0^t\frac{\partial \varphi}{\partial x_m}(\nu_s^{(K)}({\bm f}))
\nu_{s}^{(K)}(\chi_0f_m)\D s,
\]
where {\Red $\chi_0 f(p,y,x)=xf(p,y,x)$} with $(p,y,x)\in\cO$.
Let ${\bm J}_{m,\cdot}^{(K,k)}$ (resp. ${\bm J}_{m,\cdot}^{(K,c)}$)
be the $m$-th component of ${\bm
J}_\cdot^{(K,k)}(Y_1^{(k)},\widetilde{Y}_1^{(k)})$ (resp.
${\bm J}_{\cdot}^{(K,c)}({\bm Y}_1,{\widetilde{\bm  Y}}_1 )$) with $m=1,2,\dots,M$.  Observe
that
\begin{eqnarray}\label{eq:jump1}
&&\varphi(\nu_s^{(K)}({\bm f})+{\bm
J}_s^{(K,k)}(Y_1^{(k)},\widetilde{Y}_1^{(k)}))-\varphi(\nu_s^{(K)}({\bm
f}))
\simeq\sum_{m=1}^M\frac{\partial \varphi}{\partial x_m}(\nu_s^{(K)}({\bm f})){\bm J}_{m,s}^{(K,k)}\nonumber\\
&&\qquad\simeq\sum_{m=1}^M\frac{\partial \varphi}{\partial
x_m}(\nu_s^{(K)}({\bm f})) \left(\frac{\partial f_m}{\partial
x}(p_k,(Y_1^{(k)},\widetilde{Y}_1^{(k)}),\xi_{s-}^{(k)})\frac{
d_k\widetilde{Y}_1^{(k)}}{K}\overline{H}_s^{(k)}\right),
\end{eqnarray}
and
\begin{eqnarray}\label{eq:jump2}
&&\varphi(\nu_s^{(K)}({\bm f})+{\bm J}_s^{(K,c)}({\bm Y}_1))-\varphi(\nu_s^{(K)}({\bm f}))
\simeq\sum_{m=1}^M\frac{\partial \varphi}{\partial x_m}(\nu_s^{(K)}({\bm f})){\bm J}_{m,s}^{(K,c)}\nonumber\\
&&\qquad\simeq\sum_{m=1}^M\frac{\partial \varphi}{\partial
x_m}(\nu_s^{(K)}({\bm f}))\left(\sum_{k=1}^K\frac{\partial
f_m}{\partial
x}(p_k,(Y_1^{(k)},\widetilde{Y}_1^{(k)}),\xi_{s-}^{(k)})\frac{c_kY_1^{(k)}}{K}\overline{H}_s^{(k)}\right).
\end{eqnarray}
{\Red Accordingly,} the fourth line of \eqref{eq:martingale} can be rewritten as
\begin{eqnarray*}
&&\sum_{m=1}^M\int_0^t\frac{\partial \varphi}{\partial x_m}(\nu_s^{(K)}({\bm f}))\left(\frac{1}{K}\sum_{k=1}^K\widehat{\lambda}_k
\frac{\partial f_m}{\partial x}(p_k,(Y_1^{(k)},\widetilde{Y}_1^{(k)}),\xi_{s-}^{(k)})d_k\widetilde{Y}_1^{(k)}\overline{H}_s^{(k)}\right)\D s\nonumber\\
&&\qquad\simeq\sum_{m=1}^M\int_0^t\frac{\partial \varphi}{\partial
x_m}(\nu_s^{(K)}({\bm f}))\nu_s^{(K)}(\cL_{21}f_m)\D s,
\end{eqnarray*}
and the fifth line of \eqref{eq:martingale} can be rewritten as
\begin{eqnarray*}
&&\widehat{\lambda}_c\sum_{m=1}^M\int_0^t\frac{\partial\varphi}{\partial x_m}(\nu_s^{(K)}({\bm f}))\left(\frac{1}{K}\sum_{k=1}^K
\frac{\partial f_m}{\partial x}(p_k,(Y_1^{(k)},\widetilde{Y}_1^{(k)}),\xi_{s-}^{(k)})c_kY_1^{(k)}\overline{H}_s^{(k)}\right)\D s\nonumber\\
&&\qquad\simeq\widehat{\lambda}_c\sum_{m=1}^M\int_0^t\frac{\partial\varphi}{\partial
x_m}(\nu_s^{(K)}({\bm f}))\nu_s^{(K)}(\cL_{22}f_m)\D s,
\end{eqnarray*}
where $a_K\simeq b_K$ means that $\lim_{K\too\infty}|a_K-b_K|=0$.

Finally, the second line
of~\eqref{eq:martingale} can be rewritten as
\begin{eqnarray}\label{eq:2nd-derivtive-phi}
\frac{1}{2K}\sum_{m,n=1}^M\int_0^t\frac{\partial^2\varphi}{\partial x_m\partial x_{n}}(\nu_s^{(K)}({\bm
f}))\nu_s^{(K)}(\chi_1(\cL_2f_m)\cdot\chi_1(\cL_2f_n))\D s,
\end{eqnarray}
where the operator $\chi_1f(p,y,x)=\sigma x^{\rho}f(p,y,x)$ and ${\cal L}_2f(p,y,x)=\frac{\partial f}{\partial x}(p,y,x)$.
We now prove that \eqref{eq:2nd-derivtive-phi} approaches zero when
$K\too\infty$. Indeed, let $\varsigma_a^{(K)}=\inf\{t\geq0;\
\bigvee_{k=1}^K|\xi_t^{(k)}|\geq a\}$ for $a>0$. Then, for each fixed
$a>0$,
\begin{eqnarray*}
\left|\frac{1}{2K}\sum_{m,n=1}^M\int_0^{t\wedge\varsigma_a^{(K)}}\frac{\partial^2\varphi}{\partial x_m\partial x_{n}}(\nu_s^{(K)}({\bm
f}))\nu_s^{(K)}(\chi_1(\cL_2f_m)\cdot\chi_1(\cL_2f_n))\D s\right|\leq \frac{C_a}{K}\too0,\ \ {\rm as}\ K\too\infty.
\end{eqnarray*}
Letting $a\rightarrow\infty$, we conclude that the quantity in
\eqref{eq:2nd-derivtive-phi} approaches zero as $K\too+\infty$ since
$\varsigma_a^{(K)}\too+\infty$. Thus we complete the proof of the lemma. \hfill$\Box$\\

\subsection{Proof of Lemma \ref{lem:ccc}}

Let $0\leq t\leq T$. Recall that the decomposition of $\nu_t^{(K)}(f)$ for any $f\in C^{\infty}(\cO)$ admits the form:
\begin{eqnarray}\label{eq:decom-<f,nu>0}
\nu_t^{(K)}(f)=\nu_0^{(K)}(f) + A_t^{(K)} + \widehat{A}_t^{(K)} + B_t^{(K)} + \widehat{B}_t^{(K)},
\end{eqnarray}
where we have defined
\begin{eqnarray}\label{eq:decom-<f,nu>}
A_t^{(K)} &=& \int_0^t\nu_s^{(K)}(\cL_{11}f)\D s,\nonumber\\
\widehat{A}_t^{(K)}&=&\frac{1}{K}\int_0^t\sum_{k=1}^Kf(p_k,(Y_1^{(k)},\widetilde{Y}_1^{(k)}),\xi_{s-}^{(k)})\D\overline{H}_s^{(k)},\nonumber\\
B_t^{(K)}
&=&\frac{1}{K}\int_0^t\sum_{k=1}^K\left[\sigma_k\frac{\partial
f}{\partial x}
(p_k,(Y_1^{(k)},\widetilde{Y}_1^{(k)}),\xi_s^{(k)})\overline{H}_s^{(k)}(\xi_s^{(k)})^{\rho} \D W_s^{(k)}\right]\\
\widehat{B}_t^{(K)}
&=&\frac{1}{K}\int_0^t\sum_{k=1}^K\left([f(p_k,(Y_1^{(k)},\widetilde{Y}_1^{(k)}),\xi_{s-}^{(k)}+
d_k\widetilde{Y}_1^{(k)})
-f(p_k,(Y_1^{(k)},\widetilde{Y}_1^{(k)}),\xi_{s-}^{(k)})]\overline{H}_s^{(k)}\D
\widehat{N}_s^{(k)}\right)\nonumber\\
&&+
 \frac{1}{K}\int_0^t\left(\sum_{k=1}^K[f(p_k,(Y_1^{(k)},\widetilde{Y}_1^{(k)}),\xi_{s-}^{(k)}+
c_kY_1^{(k)})
-f(p_k,(Y_1^{(k)},\widetilde{Y}_1^{(k)}),\xi_{s-}^{(k)})]\overline{H}_s^{(k)}\right)\D
\widehat{N}_s^{(c)}.\nonumber
\end{eqnarray}
Then, for any $T>0$, we have
\begin{eqnarray}\label{Eq:decomsup-f}
\sup_{0\leq t\leq T}\left|\nu_t^{(K)}(f)\right|\leq\sup_{0\leq t\leq T}\left|A_t^{(K)}\right|
+\sup_{0\leq t\leq T}\left|\widehat{A}_t^{(K)}\right|+\sup_{0\leq t\leq T}\left|B_t^{(K)}\right|
+\sup_{0\leq t\leq T}\left|\widehat{B}_t^{(K)}\right|.
\end{eqnarray}
Next we estimate the expectation of each term on the r.h.s. of the above equation. First, by the assumption ({\bf A2}), we have
\begin{eqnarray*}
&&\Ex\left[\sup_{0\leq t\leq T}\left|A_t^{(K)}\right|\right]\nonumber\\
&&\ \leq\frac{1}{K}\sum_{k=1}^K\Ex\left[\int_0^T\left|\frac{1}{2}\sigma_k^2(\xi_s^{(k)})^{2\rho}
\frac{\partial^2 f}{\partial
x^2}(p_k,(Y_1^{(k)},\widetilde{Y}_1^{(k)}),\xi_s^{(k)})
+(\alpha_k-\kappa_k\xi_{s}^{(k)})\frac{\partial f}{\partial x}(p_k,(Y_1^{(k)},\widetilde{Y}_1^{(k)}),\xi_s^{(k)})\right|\D s\right]\nonumber\\
&&\ \leq\frac{C_p^2}{2}\left\|\frac{\partial^2 f}{\partial
x^2}\right\|\int_0^T\left(\frac{1}{K}\sum_{k=1}^K\Ex\left[(\xi_s^{(k)})^{2\rho}\right]\right)\D
s +C_p\left\|\frac{\partial f}{\partial
x}\right\|\int_0^T\left(\frac{1}{K}\sum_{k=1}^K\Ex\left[\xi_s^{(k)}\right]\right)\D
s+C_p T\left\|\frac{\partial f}{\partial x}\right\|,
\end{eqnarray*}
where, for a given function $f\in C^{\infty}(\cO)$, $\|f\|$ denotes
the supremum norm, i.e. $\|f\| = \sup_{(p,y,x)\in\cO} |f(p,y,x)|$.
The same definition of supremum norm applies to
$\|\frac{\partial f}{\partial x}\|$ and $\|\frac{\partial^2
f}{\partial x^2}\|$. The constant $C_p>0$ is chosen to be $C_p
=\max_{k\in\{1,\dots,K\}}\{\alpha_k,\kappa_k,\sigma_k,c_k,d_k,\widehat{\lambda}_k,m_k^{Y}, m_k^{\tilde{Y}}\}$,
and is finite by assumption ({\bf A2}).

We can bound the second term on the r.h.s. of Eq.~\eqref{Eq:decomsup-f} as
\begin{eqnarray*}
\Ex\left[\sup_{0\leq t\leq
T}\left|\widehat{A}_t^{(K)}\right|\right]\leq
\frac{1}{K}\Ex\left[\int_0^T\sum_{k=1}^K\left|f(p_k,(Y_1^{(k)},\widetilde{Y}_1^{(k)}),\xi_{s-}^{(k)})\right|\D{H}_s^{(k)}\right]
\leq\left\|f\right\|\Ex\left[\frac{1}{K}\sum_{k=1}^K{H}_T^{(k)}\right]\leq\|f\|,
\end{eqnarray*}
where we have used the fact that
$\frac{1}{K}\sum_{k=1}^K{H}_T^{(k)}\leq 1$ for all $K\in\N$. Using the Burkholder-Davis-Gundy's inequality, we can bound the third term on the r.h.s. of~\eqref{Eq:decomsup-f} as
\begin{eqnarray*}
\Ex\left[\sup_{0\leq t\leq T}\left|B_t^{(K)}\right|\right]&\leq&\frac{1}{K}
\Ex\left[\int_0^T\sum_{k=1}^K\sigma_k^2\left|\frac{\partial f}{\partial x}(p_k,(Y_1^{(k)},\widetilde{Y}_1^{(k)}),\xi_s^{(k)})\right|^2\overline{H}_s^{(k)}(\xi_s^{(k)})^{2\rho}\D s\right]^{\frac{1}{2}}\nonumber\\
&\leq&{C_p}\left\|\frac{\partial f}{\partial x}\right\|\Ex\left[\int_0^T\frac{1}{K}\sum_{k=1}^K(\xi_s^{(k)})^{2\rho}\D s\right]^{1/2}\nonumber\\
&\leq&\frac{C_p}{2}\left\|\frac{\partial f}{\partial x}\right\|\left[\int_0^T\frac{1}{K}\sum_{k=1}^K\Ex\left[(\xi_s^{(k)})^{2\rho}\right]\D s+\frac{1}{K}\right]\nonumber\\
&\leq&\frac{C_p}{2}\left\|\frac{\partial f}{\partial x}\right\|\left[\int_0^T\frac{1}{K}\sum_{k=1}^K\Ex\left[(\xi_s^{(k)})^{2\rho}\right]\D s+1\right].
\end{eqnarray*}
Finally, we have
\begin{eqnarray*}
\Ex\left[\sup_{0\leq t\leq
T}\left|\widehat{B}_t^{(K)}\right|\right]
&\leq&
\frac{1}{K}\Ex\left[\int_0^T\sum_{k=1}^K\left|f(p_k,(Y_1^{(k)},\widetilde{Y}_1^{(k)}),\xi_{s-}^{(k)}+d_k
\widetilde{Y}_1^{(k)})-f(p_k,(Y_1^{(k)},\widetilde{Y}_1^{(k)}),\xi_{s-}^{(k)})\right|\overline{H}_s^{(k)}\D \widehat{N}_s^{(k)}\right]\nonumber\\
&&+\frac{1}{K}\Ex\left[\int_0^T\sum_{k=1}^K\left|f(p_k,(Y_1^{(k)},\widetilde{Y}_1^{(k)}),\xi_{s-}^{(k)}+c_k
Y_1^{(k)})-f(p_k,(Y_1^{(k)},\widetilde{Y}_1^{(k)}),\xi_{s-}^{(k)})\right|\overline{H}_s^{(k)}\D \widehat{N}_s^{(c)}\right]\nonumber\\
&\leq&\left\|\frac{\partial f}{\partial
x}\right\|\frac{1}{K}\sum_{k=1}^K(\lambda_k(c_k\vee d_k)m_k^Y) \leq
C_p^2(C_p+\widehat{\lambda}_c)\left\|\frac{\partial f}{\partial
x}\right\|,
\end{eqnarray*}
where we have used the mean-value theorem in the last inequality.

Note that $\Ex[\nu_0^{(K)}(f)]\leq \|f\|$. Using
\eqref{eq:p-moment-average} in Lemma \ref{lem:moment-intensity}, {
we can find a constant $C=C(T,\|f\|,\|\frac{\partial f}{\partial x}\|,\|\frac{\partial^2 f}{\partial x^2}\|)>0$ such that
\begin{eqnarray*}
\sup_{K\in\N}\Ex\left[\sup_{0\leq t\leq T}\left|\nu_t^{(K)}(f)\right|\right]<C<+\infty.
\end{eqnarray*}
From Chebyshev's inequality, it follows that \eqref{lem:ccc-eq} holds. \hfill$\Box$

\subsection{Proof of Lemma \ref{lem:cond-b}}

From the decomposition \eqref{eq:decom-<f,nu>0}, it follows that
\begin{eqnarray}\label{eq:diff-docm}
(\nu_{t+u}^{(K)}-\nu_{t}^{(K)})(f)&=&A_{t+u}^{(K)}-A_{t}^{(K)}+\widehat{A}_{t+u}^{(K)}-\widehat{A}_{t}^{(K)}
+B_{t+u}^{(K)}-B_{t}^{(K)}\nonumber\\
&&+\widehat{\cal M}_{t+u}^{(K)}-\widehat{\cal M}_{t}^{(K)} + P_{t+u}^{(K)} - P^{(K)}_t,
\end{eqnarray}
where $A_t^{(K)},\widehat{A}_{t}^{(K)},B_{t+u}^{(K)}$ are given by
\eqref{eq:decom-<f,nu>} and we have defined \begin{eqnarray*}
\widehat{\cal
M}_{t}^{(K)}&=&\frac{1}{K}\sum_{k=1}^K\int_0^t\int_{\R_+}[f(p_k,y,\xi_{s-}^{(k)}+d_ky_2)-f(p_k,y,\xi_{s-}^{(k)})]
\ \overline{H}_s^{(k)}\widetilde{N}^{(k)}(\D s,\D y_2)\\
&&+\frac{1}{K}\sum_{k=1}^K\int_0^t\int_{\R_+}[f(p_k,y,\xi_{s-}^{(k)}+c_ky_1)-f(p_k,y,\xi_{s-}^{(k)})]
\ \overline{H}_s^{(k)}\widetilde{N}^{(c)}(\D s,\D y_1),\\
P^{(K)}_t&=&\frac{1}{K}\sum_{k=1}^K\widehat{\lambda}_k\int_0^t\int_{\R_+}[f(p_k,y,\xi_{s-}^{(k)}+d_ky_2)-f(p_k,y,\xi_{s-}^{(k)})]\
\overline{H}_s^{(k)}F^{(k)}_{\widetilde{Y}}(\D y_2)\D s\\
&&+\frac{1}{K}\sum_{k=1}^K\widehat{\lambda}_c\int_0^t\int_{\R_+}[f(p_k,y,\xi_{s-}^{(k)}+c_ky_1)-f(p_k,y,\xi_{s-}^{(k)})]\
\overline{H}_s^{(k)}F^{(k)}_{Y}(\D y_1)\D s.
\end{eqnarray*}
Here, for $(t,y)=(t,y_1,y_2)\in\R_+^3$, $\widetilde{N}^{(k)}(\D t,\D
y_2)$ and $\widetilde{N}^{(c)}(\D t,\D y_1)$ denote the compensated
Poisson random measures associated, respectively, to the
{systematic} compound Poisson process
$\sum_{i=1}^{\widehat{N}_{\cdot}^{(c)}}Y_i^{(k)}$ and to the
{idiosyncratic one} given by
$\sum_{\ell=1}^{\widehat{N}_{\cdot}^{(k)}}\widetilde{Y}_\ell^{(k)}$.
Moreover, the measures $F_Y^{(k)}(\D y_1)$ and
$F_{\widetilde{Y}}^{(k)}(\D y_2)$ are the distributions of the jump
amplitude $Y_1^{(k)}$ and $\widetilde{Y}_1^{(k)}$, respectively.
Then
\begin{eqnarray*}
h^2\left(\nu_{t+u}^{(K)}(f),\nu_{t}^{(K)}(f)\right)&\leq& 8\bigg[\left|A_{t+u}^{(K)}-A_{t}^{(K)}\right|+\left|\widehat{A}_{t+u}^{(K)}-\widehat{A}_{t}^{(K)}\right|
+\left|P_{t+u}^{(K)}-P_{t}^{(K)}\right|\nonumber\\
&&+\left|{B}_{t+u}^{(K)}-B_{t}^{(K)}\right|^2+\left|\widehat{\cal M}_{t+u}^{(K)}-\widehat{\cal M}_{t}^{(K)} \right|^2\bigg].
\end{eqnarray*}
First, we have, for $0\leq u\leq\delta$,
\begin{eqnarray*}
\left|A_{t+u}^{(K)}-A_{t}^{(K)}\right|&\leq&
\frac{C_p^2}{2}\left\|\frac{\partial^2 f}{\partial x^2}\right\|\int_t^{t+u}\left(\frac{1}{K}\sum_{k=1}^K(\xi_s^{(k)})^{2\rho}\right)\D s
+C_p\left\|\frac{\partial f}{\partial x}\right\|\int_{{t}}^{t+u}\left(\frac{1}{K}\sum_{k=1}^K\xi_s^{(k)}\right)\D s
+C_p\left\|\frac{\partial f}{\partial x}\right\|u\nonumber\\
&\leq&\frac{C_p^2}{4}\left\|\frac{\partial^2 f}{\partial x^2}\right\|\delta^{\frac{1}{4}}\left[1+\int_0^{T}\left(\frac{1}{K}\sum_{k=1}^K(\xi_s^{(k)})^{4\rho}\right)\D s\right]+C_p\left\|\frac{\partial f}{\partial x}\right\|\delta\nonumber\\
&&+\frac{C_p}{2}\left\|\frac{\partial f}{\partial x}\right\|\delta^{\frac{1}{4}}\left[1+\int_0^{T}\left(\frac{1}{K}\sum_{k=1}^K(\xi_s^{(k)})^{2}\right)\D s\right]=:H_K^{1}(\delta).
\end{eqnarray*}
Next, we have
\begin{eqnarray*}
\left|\widehat{A}_{t+u}^{(K)}-\widehat{A}_{t}^{(K)}\right|
\leq\frac{1}{K}\sum_{k=1}^K\int_t^{t+u}\left\|f\right\|\D{H}_s^{(k)}=\left\|f\right\|\frac{1}{K}\sum_{k=1}^K(H_{t+u}^{(k)}-H_t^{(k)}).
\end{eqnarray*}
Note that the difference $H_{t+u}^{(k)}-H_t^{(k)}$ admits the decomposition:
\begin{eqnarray*}
H_{t+u}^{(k)}-H_t^{(k)}={\cal M}_{t+u}^{(k)}-{\cal M}_{t}^{(k)} +\int_t^{t+u}\overline{H}^{(k)}_s\xi_s^{(n)}\D s,
\end{eqnarray*}
where the martingale ${\cal M}^{(k)}=({\cal M}_{t}^{(k)};\ t\geq0)$ is defined by \eqref{eq:default-mart}.
Then it holds that
\begin{eqnarray*}
\Ex\left[\left|\widehat{A}_{t+u}^{(K)}-\widehat{A}_{t}^{(K)}\right|\Big|\bigvee_{k=1}^K\G_t^{(k)}\right]
&\leq&\left\|f\right\|\frac{1}{K}\sum_{k=1}^K\Ex\left[{\cal M}_{t+u}^{(k)}-{\cal M}_{t}^{(k)} +\int_t^{t+u}\overline{H}^{(k)}_s\xi_s^{(k)}\D s\bigg|\bigvee_{k=1}^K\G_t^{(k)}\right]\nonumber\\
&\leq&\left\|f\right\|\Ex\left[\int_t^{t+u}\frac{1}{K}\sum_{k=1}^K\xi_s^{(k)}\D s\bigg|\bigvee_{k=1}^K\G_t^{(k)}\right]\nonumber\\
&\leq&\Ex\left[\frac{\|f\|}{2}\delta^{\frac{1}{4}}\left(1+\int_0^T\frac{1}{K}\sum_{k=1}^K(\xi_s^{(k)})^2\D s\right)\bigg|\bigvee_{k=1}^K{\G}_t^{(k)}\right]\nonumber\\
&=:&\Ex\left[H_K^{2}(\delta)\Big|\bigvee_{k=1}^K\G_t^{(k)}\right].
\end{eqnarray*}
For the third term on the r.h.s. of \eqref{eq:diff-docm}, we have
\begin{eqnarray*}
\Ex\left[\left|B_{t+u}^{(K)}-B_t^{(K)}\right|^2\Big|\bigvee_{k=1}^K\G_t^{(k)}\right]&=&
{\Ex\left[\left|B_{t+u}^{(K)}\right|^2-\left|B_t^{(K)}\right|^2\Big|\bigvee_{k=1}^K\G_t^{(k)}\right]}\nonumber\\
&\leq&C_p^2\left\|\frac{\partial f}{\partial x}\right\|^2\Ex\left[\int_t^{t+u}\frac{1}{K}\sum_{k=1}^K(\xi_s^{k})^{2\rho}\D s\bigg|\bigvee_{k=1}^K\G_t^{(k)}\right]\nonumber\\
&\leq&\Ex\left[\frac{C_p^2}{2}\left\|\frac{\partial f}{\partial x}\right\|^2\delta^{\frac{1}{4}}\left(1+\int_0^{T}\frac{1}{K}\sum_{k=1}^K(\xi_s^{k})^{4\rho}\D s\right)\bigg|\bigvee_{k=1}^K\G_t^{(k)}\right]\nonumber\\
&=:&\Ex\left[H_K^{3}(\delta)\Big|\bigvee_{k=1}^K\G_t^{(k)}\right].
\end{eqnarray*}
Next, we consider the fourth term on the r.h.s. of
\eqref{eq:diff-docm}. Using the equality $(x+y)^2\leq
2(x^2+y^2)$, the martingale property of $(\widehat{\cal
M}^{(K)}_{t};\ t\geq0)$, the mean-value theorem and the assumption
{\bf(A2)}, we have
\begin{eqnarray*}
&&\Ex\left[\left|\widehat{\cal
M}^{(K)}_{t+u}-\widehat{\cal
M}^{(K)}_{t}\right|^2\Big|\bigvee_{k=1}^K\G_t^{(k)}\right]
\nonumber\\
&\leq&2\Ex\left[\frac{1}{K^2}\sum_{k=1}^K\int_t^{t+u}\int_{\R_+}\left|f(p_k,y,\xi_{s-}^{(k)}+d_ky_2)-f(p_k,y,\xi_{s-}^{(k)})\right|^2
\overline{H}_s^{(k)}\widehat{N}^{(k)}(\D s,\D y_2)\bigg|\bigvee_{k=1}^K\G_t^{(k)}\right]\nonumber\\
&&+2\Ex\left[\frac{1}{K^2}\int_t^{t+u}\int_{\R_+}\left(\sum_{k=1}^K\left|f(p_k,y,\xi_{s-}^{(k)}+c_ky_1)-f(p_k,y,\xi_{s-}^{(k)})\right|
\ \overline{H}_s^{(k)}\right)^2\widehat{N}^{(c)}(\D s,\D y_1)\bigg|\bigvee_{k=1}^K\G_t^{(k)}\right]\nonumber\\
&\leq& 2\Ex\left[\frac{1}{K^2}\left\|\frac{\partial
f}{\partial x}\right\|^2\sum_{k=1}^K\widehat{\lambda}_kd_k^2\int_t^{t+u}\int_{\R_+}y_2^2{F}_{\widetilde{Y}}^{(k)}(\D y_2)\D s\bigg|\bigvee_{k=1}^K\G_t^{(k)}\right]\nonumber\\
&&+2\Ex\left[\frac{1}{K^2}\left\|\frac{\partial f}{\partial
x}\right\|^2\widehat{\lambda}_c\int_t^{t+u}\int_{\R_+}\left(\sum_{k=1}^Kc_k\overline{H}_s^{(k)}\right)^2y_1^2F_Y^{(k)}(\D
y_1)\D s\bigg|\bigvee_{k=1}^K\G_t^{(k)}\right]
\nonumber\\
&\leq& 2 C_p^2(C_p+\widehat{\lambda}_c)\left\|\frac{\partial
f}{\partial x}\right\|^2\delta=:H_K^{4}(\delta),
\end{eqnarray*}
where $\widehat{N}^{(k)}(\D t,\D y_2)$ and $\widehat{N}^{(c)}(\D
t,\D y_2)$ denote the Poisson random measures associated,
respectively, to the {systematic} compound Poisson process
$\sum_{i=1}^{\widehat{N}_{\cdot}^{(c)}}Y_i^{(k)}$ and to the
{idiosyncratic one} given by
$\sum_{\ell=1}^{\widehat{N}_{\cdot}^{(k)}}\widetilde{Y}_\ell^{(k)}$.
Finally, we have
\begin{eqnarray*}
\left|P_{t+u}^{(K)}-P_t^{(K)}\right|&\leq&C_p(C_p+\widehat{\lambda}_c)\left\|\frac{\partial
f}{\partial x}\right\|
\frac{1}{K}\sum_{k=1}^K\int_t^{t+u}\int_{\R_+}y_1F_{Y}^{(k)}(\D y_1)\D s\nonumber\\
&&+2C_p^2\left\|\frac{\partial f}{\partial x}\right\|\frac{1}{K}\sum_{k=1}^K\int_t^{t+u}\int_{\R_+}y_2F_{\widetilde{Y}}^{(k)}(\D y_2)\D s\nonumber\\
&=&C_p(3C_p+\widehat{\lambda}_c)\left\|\frac{\partial f}{\partial x}\right\|\frac{1}{K}\sum_{k=1}^K\int_t^{t+u}
\left(\Ex[Y_1^{(k)}]+\Ex[\widetilde{Y}_1^{(k)}]\right)\D s\nonumber\\
&\leq& 2C_p^2(3C_p+\widehat{\lambda}_c)\left\|\frac{\partial
f}{\partial x}\right\|\delta=:H_K^{5}(\delta).
\end{eqnarray*}
Note that $h^2(\nu_{t}^{(K)}(f),\nu_{t-v}^{(K)}(f))\leq 1$. Let
$H_K(\delta)=\sum_{n=1}^5H_K^{n}(\delta)$. It satisfies
$\lim_{\delta\too 0}\sup_{K\in\N}\Ex[H_K(\delta)]=0$ and
\eqref{eq:lem-b} holds, due to the above estimates and
\eqref{eq:p-moment-average} from Lemma \ref{lem:moment-intensity}.
\hfill$\Box$

\section{Proofs related to Section \ref{sec:BCVAformula}} \label{app:basics}
\renewcommand\theequation{C.\arabic{equation}}
\setcounter{equation}{0}


\begin{lemma}\label{lem:cond-indepedence}
The default times $\tau_1,\dots,\tau_K$, $\tau_A$ and $\tau_B$ are
conditionally independent. Namely, for any
$t_1,\dots,t_K,t_A,t_B\geq0$ and any
$T\geq\max\{t_1,\dots,t_K,t_A,t_B\}$, it holds that
\begin{eqnarray}\label{eq:lem-cond-independence}
\Px\left(\tau_1>t_1,\dots,\tau_K>t_K,\tau_A>t_A,\tau_B>t_B\Big|\F_{T}^{(K,A,B)}\right)
=\prod_{j\in\{1,\dots,A,B\}}\exp\left(-\int_{0}^{t_j}\xi_s^{(j)}\D s\right).
\end{eqnarray}
\end{lemma}

\noindent{\it Proof.}\quad The proof is straightforward and follows immediately from the discussion in Section 9.1.1 of \cite{bielecki01}.
\hfill$\Box$\\

\noindent{\bf Proof of Theorem \ref{thm:cva1}.}\quad First, we define the conditional cumulative distribution function associated to the default times
$(\tau_X^*,\tau_A,\tau_B)$. For $\min\{t^*,t_A,t_B\}>t$, define
\begin{eqnarray}
P(t;t^*,t_A,t_B):=\Px\left(\tau_X^*\leq t^*,\tau_A\leq t_A,\tau_B\leq t_B |\G_t^{(K,A,B)}\right).
\end{eqnarray}
Then
\begin{eqnarray}\label{eq:dis-expression}
P(t;t^*,t_A,t_B)
&=&1-\Px \left(\tau_A> t_A|\G_t^{(K,A,B)} \right)-\Px\left(\tau_B>
t_B|\G_t^{(K,A,B)}\right)\nonumber\\
&&+\Px\left(\tau_A> t_A,\tau_B> t_B|\G_t^{(K,A,B)} \right)\nonumber\\
&&-\left[\Px\left(\tau_X^*>t^*|\G_t^{(K,A,B)} \right)-\Px \left(\tau_A>t_A,\tau_X^*>t^*|\G_t^{(K,A,B)} \right)\right]\nonumber\\
&&+\left[\Px\left(\tau_B>t_B,\tau_X^*>t^*|\G_t^{(K,A,B)} \right)-\Px\left(\tau_A>t_A,\tau_B>t_B,\tau_X^*>t^*|\G_t^{(K,A,B)}\right)\right].
\end{eqnarray}
Using Lemma \ref{lemma:joint-survive}, and the definition of the limit default time $\tau_X^*$ given in Section \ref{sec:approx}, on the event $\{\tau_X^*>t,\tau_A>t_A,\tau_B>t_B\}$, we have
\begin{eqnarray}\label{eq:joint-csp}
\Px\left(\tau_X^*>t^*,\tau_A> t_A,\tau_B> t_B |\G_t^{(K,A,B)}\right)
=\widehat{F}(t,t^*)\Ex\left[\exp\left(-\int_t^{t_A}\xi_s^{(A)}\D s-\int_t^{t_B}\xi_s^{(B)}\D s\right)\Big|\F_t^{(K,A,B)}\right].
\end{eqnarray}
By virtue of \eqref{eq:dis-expression}, it follows that
\begin{eqnarray}\label{eq:density}
\frac{\partial P(t;t^*,t_A,t_B)}{\partial t_B}=(1-\widehat{F}(t,t^*))\frac{\partial P(t;\infty,t_A,t_B)}{\partial t_B},
\end{eqnarray}
where
\begin{eqnarray}\label{eq:density-infty}
\frac{\partial P(t;\infty,t_A,t_B)}{\partial t_B}&:=&\frac{\partial
P(t;t^*,t_A,t_B)}{\partial
t_B}\Big|_{t^*=\infty}
=\Ex\left[\exp\left(-\int_t^{t_B}\xi_s^{(B)}\D
s\right)\xi_{t_B}^{(B)}\Big|\F_t^{(K,A,B)}\right]\nonumber\\
&&-\Ex\left[\exp\left(-\int_t^{t_A}\xi_s^{(A)}\D
s-\int_t^{t_B}\xi_s^{(B)}\D
s\right)\xi_{t_B}^{(B)}\Big|\F_t^{(K,A,B)}\right].
\end{eqnarray}
Hence we have that
\begin{eqnarray}\label{eq:cva-0}
B^{(K,*)}(t,T)&=&\int_{t}^T\int_{t}^{\infty}\int_{t}^{ \infty}\Ex\left[{\bf1}_{\{t_B\leq
t_A\}}{\bf1}_{\{t_B<t^*\}}D(t,t_B)\varepsilon_+^{(K,*)}(t_B,T)\frac{\partial^3P(t;t^*,t_A,t_B)}{\partial
t^*\partial t_A\partial t_B}\Big|\F_t^{(K,A,B)}\right]\D
t^*\D t_A\D t_B\nonumber\\
&=&\int_{t}^T\int_{t}^\infty
\Ex\left[{\bf1}_{\{t_B<t^*\}}D(t,t_B)\varepsilon_+^{(K,*)}(t_B,T)\frac{\partial^2P(t;t^*,t_A,t_B)}{\partial
t^*\partial t_B}\Big|_{t_A=t_B}^{\infty}\Big|\F_t^{(K,A,B)}\right]\D
t^*\D t_B\nonumber\\
&=&\int_{t}^T\int_{t}^\infty
\Ex\left[{\bf1}_{\{t_B<t^*\}}D(t,t_B)\varepsilon_+^{(K,*)}(t_B,T)\frac{\partial^2P(t;t^*,\infty,t_B)}{\partial
t^*\partial t_B}\Big|\F_t^{(K,A,B)}\right]\D t^*\D t_B\nonumber\\
&&-\int_{t}^T\int_{t}^\infty
\Ex\left[{\bf1}_{\{t_B<t^*\}}D(t,t_B)\varepsilon_+^{(K,*)}(t_B,T)\frac{\partial^2P(t;t^*,t_B,t_B)}{\partial
t^*\partial t_B}\Big|\F_t^{(K,A,B)}\right]\D
t^*\D t_B\nonumber\\
&=&\int_{t}^TD(t,t_B)\varepsilon_+^{(K,*)}(t_B,T)\Ex\left[\frac{\partial
P(t;\infty,\infty,t_B)}{\partial t_B}-\frac{\partial
P(t;t_B,\infty,t_B)}{\partial t_B}\Big|\F_t^{(K,A,B)}\right]\D t_B\nonumber\\
&&-\int_{t}^TD(t,t_B)\varepsilon_+^{(K,*)}(t_B,T)\Ex\left[\frac{\partial
P(t;\infty,t_B,t_B)}{\partial t_B}-\frac{\partial
P(t;t_B,t_B,t_B)}{\partial t_B}\Big|\F_t^{(K,A,B)}\right]\D t_B\nonumber\\
&=&\int_{t}^TD(t,t_B)\varepsilon_+^{(K,*)}(t_B,T)\widehat{F}(t,t_B)\Ex\left[\frac{\partial
P(t;\infty,\infty,t_B)}{\partial t_B}-\frac{\partial
P(t;\infty,t_B,t_B)}{\partial t_B}\Big|\F_t^{(K,A,B)}\right]\D t_B,
\end{eqnarray}
where we have used \eqref{eq:density} to obtain the last equality in \eqref{eq:cva-0}. Using \eqref{eq:density-infty}, we have
\begin{eqnarray}\label{eq:cva-1}
B^{(K,*)}(t,T)&=&
\int_t^T\Ex\left[D(t,t_B)\varepsilon_{+}^{(K,*)}(t_B,T)\widehat{F}(t,t_B)
\exp\left(-\int_t^{t_B}\xi_s^{(A)}+\xi_s^{(B)}\D s\right)\xi_{t_B}^{(B)}\bigg|\F_t^{(K,A,B)}\right]\D t_B\nonumber\\
&&-\int_t^T\Ex\left[D(t,t_B)\varepsilon_{+}^{(K,*)}(t_B,T)\widehat{F}(t,t_B)
\exp\left(-\int_t^{\infty}\xi_s^{(A)}\D s - \int_t^{t_B}\xi_s^{(B)}\D
s\right)\xi_{t_B}^{(B)}\bigg|\F_t^{(K,A,B)}\right]\D t_B.\nonumber\\
\end{eqnarray}
For $t\leq t_B\leq T$ and $(x_A,x_B)\in\R_+^2$, define the function
\begin{eqnarray}\label{eq:hatH1}
\widehat{H}_1(t_B-t,x_A,x_B):=\Ex\left[\exp\left(-\int_t^{\infty}\xi_s^{(A)}\D
s-\int_t^{t_B}\xi_s^{(B)}\D
s\right)\xi_{t_B}^{(B)}\bigg|\xi_t^{(A)}=x_A,\xi_t^{(B)}=x_B\right].
\end{eqnarray}

Using the definition of the function $H_1$ given in \eqref{eq:H-hatH12} along with Eq.~\eqref{eq:hatH1}, we obtain that $B^{(K,*)}(t,T)$ is given by
\begin{eqnarray}\label{eq:cva-rhs1-0}
B^{(K,*)}(t,T)&:=&\Ex\left[{\bf1}_{\{t<\tau_B \leq \min(\tau_A,T) \}}{\bf1}_{\{\tau_B<\tau_X^*\}}D(t,\tau_B)\varepsilon_{+}^{(K,*)}(\tau_B,T)\Big|\G_t^{(K,A,B)}\right]\nonumber\\
&=&\int_t^TD(t,t_B)\varepsilon_{+}^{(K,*)}(t_B,T)\widehat{F}(t,t_B)\nonumber\\
&&\qquad\times\left(H_1(t_B-t,\xi_t^{(A)},\xi_t^{(B)})-\widehat{H}_1(t_B-t,\xi_t^{(A)},\xi_t^{(B)})\right)\D
t_B.
\end{eqnarray}

Next, we prove that the function $\widehat{H}_1$ defined in
\eqref{eq:hatH1} vanishes. To this purpose, let
$\xi^{(A,noj)}=(\xi_t^{(A,noj)};\ t\geq0)$ be the CEV
process satisfying the SDE:
\begin{eqnarray}\label{eq:cevano-jump}
\D\xi_t^{(A,noj)} = -\kappa_A \xi_t^{(A,noj)}\D t + \sigma_A
(\xi_t^{(A,noj)})^{\widehat{\rho}}\D W_t^{(A)},\ \ \ \ \
\xi_t^{(A,noj)}=x_A>0,
\end{eqnarray}
where $\kappa_A,\sigma_A>0$ and the elasticity factor
$\widehat{\rho}\in[\frac{1}{2},1)$ are the parameters specified in
\eqref{eq:default-intensity-counterparty}. Then we have that
\begin{eqnarray*}
0\leq \Ex_{t_B,x_A}\left[\exp\left(-\int_{t_B}^{\infty}\xi_s^{(A)}\D
s\right)\right]\leq\Ex_{t_B,x_A}\left[\exp\left(-\int_{t_B}^{\infty}\xi_s^{(A,noj)}\D
s\right)\right]=:M(t_B,x_A),
\end{eqnarray*}
where $\Ex_{t_B,x_A}[\ \cdot\ ]$ represents the expectation
conditional on the underlying state process  being equal to $x_A$ at
time $t_B$}. Hereafter, we use $\Ex_{x_A}[\ \cdot\ ]:=\Ex_{0,x_A}[\
\cdot\ ]$.

\renewcommand\theequation{C.\arabic{equation}}

We want to verify that $M(t_B,x_A)=0$ for fixed $t_B,x_A>0$. Using the
Markov property of $\xi^{(A,noj)}$,
\begin{eqnarray}\label{eq:math-expec2}
M(t_B,x_A) = M(x_A):=
\Ex_{x_A}\left[\exp\left(-\int_{0}^{\infty}\xi_s^{(A,noj)}\D
s\right)\right],\ \ \ \ x_A>0.
\end{eqnarray}
Let $BESQ_{(\delta,x_A)}=(BESQ_{(\delta,x_A)}(t);\ t\geq0)$ denote a
squared Bessel process with dimension $\delta>0$. This is a
particular CIR process satisfying the SDE:
\begin{eqnarray}\label{eq:besq-sde}
\D BESQ_{(\delta,x_A)}(t) = \delta \D t +
2\sqrt{BESQ_{(\delta,x_A)}(t)}\D W_t^{(A)},\ \ \ \ \
BESQ_{(\delta,x_A)}(0)=x_A.
\end{eqnarray}
From Proposition 2.3 in \cite{AtlanLeblanc}, it follows that
\begin{eqnarray}\label{eq:cev-besq}
\xi_t^{(A,noj)} =
e^{-\kappa_At}\left(BESQ_{\delta,x_A^{1/p}}(a(t))\right)^{p},\ \ \ \
\ t\geq0,
\end{eqnarray}
where $p=\frac{1}{2(1-\widehat{\rho})}>1$,
$\delta=\frac{2\widehat{\rho}-1}{\widehat{\rho}-1}$, and the
time-changed function $a(t)$ is defined as
\begin{eqnarray}\label{eq:time-changed-fcn}
a(t) =
\frac{(1-\widehat{\rho})\sigma_A^2}{2\kappa_A}\left(e^{2(1-\widehat{\rho})\kappa_A
t}-1\right)=\frac{1}{l_A}\left(e^{(\kappa_A/p)t}-1\right).
\end{eqnarray}
Here $l_A=\frac{2\kappa_A}{(1-\widehat{\rho})\sigma_A^2}$. Then
\begin{eqnarray}\label{eq:math-expec-estimate}
M(x_A) =
\Ex_{x_A}\left\{\exp\left[-\int_{0}^{\infty}e^{-\kappa_As}\left(BESQ_{\delta,x_A^{1/p}}(a(s))\right)^{p}\D
s\right]\right\}.
\end{eqnarray}
Set the time variable $v=a(s)$. Then
$s=a^{-1}(v)=\frac{p}{\kappa_A}\log\{l_Av+1\}$. Observing that $a(0)=0$, we obtain that
\begin{eqnarray}\label{eq:math-expec-estimate1}
M(x_A) =
\Ex_{x_A}\left\{\exp\left[-\frac{pl_A}{\kappa_A}\int_{0}^{\infty}\frac{1}{(l_Av+1)^{p+1}}\left(BESQ_{\delta,x_A^{1/p}}(v)\right)^{p}\D
v\right]\right\}.
\end{eqnarray}
For any $T>0$, define
\begin{eqnarray}\label{eq:math-expec-estimate2}
M_T(x_A) &:=&
\Ex_{x_A}\left\{\exp\left[-\frac{pl_A}{\kappa_A}\int_{0}^{T}\frac{1}{(l_Av+1)^{p+1}}\left(BESQ_{\delta,x_A^{1/p}}(v)\right)^{p}\D
v\right]\right\}\nonumber\\
&\leq&\Ex_{x_A}\left\{\exp\left[-\frac{pl_A}{\kappa_A}\frac{1}{(l_AT+1)^{p+1}}\int_{0}^{T}\left(BESQ_{\delta,x_A^{1/p}}(v)\right)^{p}\D
v\right]\right\}\nonumber\\
&=&g(T)\Ex_{x_A}\left\{\exp\left[-\int_{0}^{T}\left(BESQ_{\delta,x_A^{1/p}}(v)\right)^{p}\D
v\right]\right\},
\end{eqnarray}
where the function
\[
g(T)=\exp\left(\frac{pl_A}{\kappa_A}\frac{1}{(l_AT+1)^{p+1}}\right),\
\ {\rm and\ hence}\ \lim_{T\too\infty}g(T)=1.
\]

Next, we prove the following limit:
\begin{eqnarray}\label{eq:limit1}
\lim_{T\too\infty}\Ex_{x_A}\left\{\exp\left[-\int_{0}^{T}\left(BESQ_{\delta,x_A^{1/p}}(v)\right)^{p}\D
v\right]\right\}=0.
\end{eqnarray}
Let $\nu=\frac{\delta-2}{2}$ and hence $\nu=-p$.
In terms of \eqref{eq:besq-sde}, we have
\begin{eqnarray}\label{eq:besq-sde2}
\D BESQ_{(\delta,x_A)}(t) = 2(\nu+1) \D t +
2\sqrt{BESQ_{(\delta,x_A)}(t)}\D W_t^{(A)},\ \ \ \ \
BESQ_{(\delta,x_A)}(0)=x_A.
\end{eqnarray}
Notice that $\nu<0$ and $p\geq 1$. Using Lemma 2.1 in
\cite{Cetin}, we have that, for any $\alpha>0$, the process
\begin{eqnarray}\label{eq:martingale}
M_t^{(u)}:=u(\sqrt{X_t})(X_t)^{p}\exp\left(-\frac{\alpha}{2}\int_0^t(X_s)^p\D
s\right),\ \ \ \ \ t\geq0
\end{eqnarray}
is a (local) martingale, where $X=(X_t;\ t\geq0)$ denotes any
squared Bessel process starting at $x>0$ with the above dimension
$\delta>0$ and the function $u(\cdot)$ satisfies the ODE:
\begin{eqnarray}\label{eq:ode}
x^2u{''}(x) + xu'(x) -u(x)\left(p^2+\alpha x^{2(p+1)}\right)=0,\ \ \
x>0.
\end{eqnarray}
Then for the stopping time $\tau_R:=\inf\{t\geq0;\ X_t\geq R\}$ with
$R\geq x$, $x=x_A^{1/p}$ and $X_t=BESQ_{(\delta,x_A^{1/p})}(t)$, it
holds that
\begin{eqnarray*}
\Ex_{x_A}\left[M_{t\wedge\tau_R}\right]=u(x_A^{1/(2p)})x_A^{1/2},\ \
\ \ \ \ {\rm for\ all}\ t>0.
\end{eqnarray*}
Letting $t\too\infty$, it follows that
\begin{eqnarray}\label{eq:int-fcn}
\Ex_{x_A}\left[\exp\left(-\frac{\alpha}{2}\int_0^{\tau_R}(X_s)^p\D
s\right)\right] =
\frac{u(x_A^{1/(2p)})x_A^{1/2}}{u(\sqrt{R})R^{p/2}}.
\end{eqnarray}
Since we are considering the case $R\geq x_A^{1/p}$, where
$x_A^{1/p}$ is the starting value of the squared Bessel process
$BESQ_{(\delta,x_A^{1/p})}$, we must have that $R\to
u(\sqrt{R})R^{p/2}$ is an increasing function. This is the case
because when $R$ increases, $\tau_R$ would be increasing hence
$\Ex_{x_A}\left[\exp\left(-\frac{\alpha}{2}\int_0^{\tau_R}(X_s)^p\D
s\right)\right]$ is decreasing w.r.t. $R$. Using Eq.~(2.8) in
\cite{Cetin}, we deduce that
\begin{eqnarray}\label{eq:int-fcn2}
\Ex_{x_A}\left[\exp\left(-\frac{\alpha}{2}\int_0^{\tau_R}(X_s)^p\D
s\right)\right] =
\frac{u_0(x_A^{1/(2p)})x_A^{1/2}}{u_0(\sqrt{R})R^{p/2}},
\end{eqnarray}
where the function $u_0(\cdot)$ is defined as
\[
u_0(x)
=I_{\frac{p}{p+1}}\left(\frac{1}{p+1}\sqrt{\alpha}x^{p+1}\right),\ \
\ \ x>0.
\]
Here $I_{b}(\cdot)$ represents the modified Bessel function of the
first kind with $b>-\frac{1}{2}$, defined by
\begin{eqnarray}
I_b(x) =
\frac{(x/2)^{b}}{\Gamma(b+\frac{1}{2})\Gamma(\frac{1}{2})}\int_{-1}^{1}e^{-xt}(1-t^2)^{b-\frac{1}{2}}\D
t,\ \ \ \ x>0.
\end{eqnarray}
Using the fact that $\lim_{x\too\infty}I_b(x)=+\infty$ and
taking $R\too\infty$ in \eqref{eq:int-fcn2}, it follows that for any
$\alpha>0$,
\begin{eqnarray}\label{eq:int-fcn3}
\Ex_{x_A}\left[\exp\left(-\frac{\alpha}{2}\int_0^{\infty}(X_s)^p\D
s\right)\right] = 0.
\end{eqnarray}
Accordingly, the limit equality \eqref{eq:limit1} is proven by taking
the parameter $\alpha=2$ in \eqref{eq:int-fcn3} and hence
$M(t_B,x_A)=0$. This results in
$\Ex_{t_B,x_A}\left[\exp\left(-\int_{t_B}^{\infty}\xi_s^{(A)}\D
s\right)\right]=0$. For $t\leq t_B\leq T$, using the tower property,
it follows that
\begin{eqnarray}\label{eq:cva-3}
&&\widehat{H}_1(t_B-t,\xi_t^{(A)},\xi_t^{(B)})=\Ex\left[\exp\left(-\int_t^{\infty}\xi_s^{(A)}\D
s - \int_t^{t_B}\xi_s^{(B)}\D s\right)\xi_{t_B}^{(B)}\Bigg|\F_t^{(K,A,B)}\right]\nonumber\\
&&\qquad=\Ex\left\{\Ex\left[\exp\left(-\int_t^{\infty}\xi_s^{(A)}\D
s - \int_t^{t_B}\xi_s^{(B)}\D
s\right)\xi_{t_B}^{(B)}\bigg|\F_{t_B}^{(K,A,B)}\right]\Bigg|\F_t^{(K,A,B)}\right\}\nonumber\\
&&\qquad=\Ex\left\{\exp\left(-\int_t^{t_B}(\xi_s^{(A)}+\xi_{s}^{(B)})\D
s\right)\xi_{t_B}^{(B)}\Ex\left[\exp\left(-\int_{t_B}^{\infty}\xi_s^{(A)}\D
s\right) \bigg|\F_{t_B}^{(K,A,B)}\right]\Bigg|\F_{t}^{(K,A,B)}\right\}\nonumber\\
&&\qquad=0.
\end{eqnarray}
This yields Eq.~\eqref{eq:cva-rhs1}. Similarly for the second term on the r.h.s. of
\eqref{eq:cvagen}, we have, on the event
$\{\tau_X^*\wedge\tau_A\wedge\tau_B>t\}$,
\begin{eqnarray*}
A^{(K,*)}(t,T)&=&
\int_t^TD(t,t_A)\varepsilon_{-}^{(K,*)}(t_A,T)\widehat{F}(t,t_A)\nonumber\\
&&\qquad\times\left(H_2(t_A-t,\xi_t^{(A)},\xi_t^{(B)})-\widehat{H}_2(t_A-t,\xi_t^{(A)},\xi_t^{(B)})\right)\D
t_A,
\end{eqnarray*}
where the function $H_2$ is defined as in \eqref{eq:H-hatH12} and
the function
\begin{eqnarray*}
\widehat{H}_2(t_A-t,x_A,x_B):=\Ex\left[\exp\left(-\int_t^{t_A}\xi_s^{(A)}\D
s-\int_t^{\infty}\xi_s^{(B)}\D
s\right)\xi_{t_B}^{(A)}\bigg|\xi_t^{(A)}=x_A,\xi_t^{(B)}=x_B\right].
\end{eqnarray*}
Using a symmetric argument to the one used to show that
$\widehat{H}_1(t_B-t,x_A,x_B)=0$, it follows that
$\widehat{H}_2(t_A-t,x_A,x_B)=0$. Hence, we obtain that
$A^{(K,*)}(t,T)$ is given by \eqref{eq:cva-rhs2}. This
completes the proof of the theorem yielding the law of large number approximation formula for
 the BCVA \eqref{eq:cvagen}.
\hfill$\Box$\\

\section{Solutions to Riccati Equations} \label{app:closeforms}
\renewcommand\theequation{D.\arabic{equation}}
\setcounter{equation}{0}

\begin{lemma}\label{appen:Riccati1}
The explicit solution to the following Riccati equation:
\begin{eqnarray}\label{eq:solu-B}
B'(u) &=& -\kappa B(u) + \frac{1}{2}\sigma^2
B^2(u) -1\nonumber\\
B(0) &=& 0
\end{eqnarray}
is given by
\begin{eqnarray}\label{eq:solu-express-B}
B(\kappa,\sigma;u) = -\frac{2(e^{\varpi u}-1)}{2\varpi +
(\kappa+\varpi)(e^{\varpi u}-1)},\ \ \  \ \ \ 0\leq u\leq T.
\end{eqnarray}
where $\kappa>0$, $\sigma>0$ and $\varpi=\sqrt{\kappa^2+2\sigma^2}$.
\end{lemma}

\begin{lemma}\label{lem:solution-R1}
Let the real number $b\neq0$. Then the explicit solution of the following Riccati equation:
\begin{eqnarray}\label{eq:lem-eqn-B1}
\beta'(u) &=& -\kappa \beta (u) + \frac{1}{2}\sigma^2
\beta^2(u) -1\nonumber\\
\beta(0) &=& b
\end{eqnarray}
is given by
\begin{eqnarray}\label{eq:lem-sol-B1}
\beta(\kappa,\sigma,b;u) = B(\kappa,\sigma;u) +
e^{\phi(u)}\frac{1}{\frac{1}{b}-\frac{\sigma^2}{2}\int_0^u
e^{\phi(v)}\D v},
\end{eqnarray}
where the function $B(\kappa,\sigma;u)$ is given by \eqref{eq:solu-express-B}, and
\[
\phi(u)=\sigma^2\int_0^u B(\kappa,\sigma;v)\D v-\kappa u,\ \ \ \ \ 0\leq u\leq T.
\]
Moreover, we have
\begin{eqnarray}\label{appen:intephi}
\int_0^u e^{\phi(v)}\D v&=&\int_0^ue^{\varpi
v}\left(\frac{2\varpi}{(\varpi-\kappa)+e^{\varpi
v}(\kappa+\varpi)}\right)^2\D v\nonumber\\
&=&\frac{2}{\kappa + \varpi {\rm Coth}\left(\frac{\varpi
u}{2}\right)},
\end{eqnarray}
where ${\rm Coth}(u)=\frac{\cosh(u)}{\sinh(u)}$ gives the hyperbolic
cotangent of $u$.
\end{lemma}

\noindent{\it Proof.}\quad For $b\neq0$, we define the following
function
\begin{eqnarray}\label{eq:f}
f(u) = B(\kappa,\sigma;u) +
e^{\phi(u)}\frac{1}{C-\frac{\sigma^2}{2}\int_0^u e^{\phi(v)}\D v},
\end{eqnarray}
where $C$ is an unspecified real constant. Then
\begin{eqnarray*}
f'(u) &=& B'(\kappa,\sigma;u) +
\phi'(u)(f(u)-B(\kappa,\sigma;u))+\frac{\sigma^2}{2}(f(u)-B(\kappa,\sigma;u))^2\nonumber\\
&=& B'(\kappa,\sigma;u) + \kappa
B(\kappa,\sigma;u)-\frac{\sigma^2}{2}B^2(\kappa,\sigma;u)-\kappa
f(u)+\frac{\sigma^2}{2}f^2(u)\nonumber\\
&=&-1-\kappa f(u)+\frac{\sigma^2}{2}f^2(u).
\end{eqnarray*}
This yields that the function given by \eqref{eq:f} is the general
solution to the above Riccati equation. Taking the initial condition
$\beta(0)=b$ into account, we have the constant $C=\frac{1}{b}$ in
\eqref{eq:f}, since $B(\kappa,\sigma;0)=0$. \hfill$\Box$

\begin{lemma}\label{lem:solution-R2}
Let $b\in\R$ and $a_{\ell}>0$. Then the explicit solution to the following
Riccati equation:
\begin{eqnarray}\label{eq:lem-eqn-B1}
\widehat{\beta}'(u) &=& -\kappa \widehat{\beta} (u) +
\frac{1}{2}\sigma^2
\widehat{\beta}^2(u) - a_\ell\nonumber\\
\widehat{\beta}(0) &=& b
\end{eqnarray}
is given by
\begin{eqnarray}\label{eq:lem-sol-B2}
\widehat{\beta}(\kappa,\sigma,a_\ell,b;u) =a_\ell\cdot
\beta\left(\kappa,\sigma\sqrt{a_\ell},\frac{b}{a_{\ell}};u\right),
\end{eqnarray}
where the function $\beta(\kappa,\sigma,b;u)$ is given by
\eqref{eq:lem-sol-B1}.
\end{lemma}

\noindent{\it Proof.}\quad Let $g(u)=\frac{\widehat{\beta}(u)}{a_{\ell}}$.
Then the function $g(u)$ satisfies the Riccati equation
\eqref{eq:lem-eqn-B1} with coefficient $\sigma$ and initial value
$b$ replaced by $\sigma\sqrt{a_{\ell}}$ and $\frac{b}{a_{\ell}}$
respectively. Thus $g(u) =
\beta\left(\kappa,\sigma\sqrt{a_\ell},\frac{b}{a_{\ell}};u\right)$ and hence the
solution \eqref{eq:lem-sol-B2} follows. \hfill$\Box$\\

\begin{lemma}\label{lem:pide-affine}
Assume the default intensities $\xi^{(A)}$ and $\xi^{(B)}$ of the two counterparties to be
CIR processes (i.e., the elasticity factor $\widehat{\rho}=\frac{1}{2}$ in \eqref{eq:default-intensity-counterparty}).
Define the conditional expectations:
\begin{eqnarray}\label{eq:me}
Q_{t,T}g(x_A,x_B) :=
\Ex\left[\exp\left(-\int_{t}^{T}\ell(\xi_s^{(A)},\xi_s^{(B)})\D
s\right)g(\xi_T^{(A)},\xi_T^{(B)})\ \Big|\ \xi_t^{(A)}=x_A,\
\xi_t^{(B)}=x_B\right],
\end{eqnarray}
where $\ell(x_A,x_B)$ and $g(x_A,x_B)$ are two measurable functions satisfying the form specified as in the following lemma.
Assume that the functions $\ell(x_A,x_B)$ and $g(x_A,x_B)$ are of the following forms on $(x_A,x_B)\in\R_+^2$,
\begin{eqnarray}\label{eq:ell-g}
\ell(x_A,x_B)&=&a_{\ell}x_A + b_{\ell}x_B + c_{\ell},\nonumber\\
g(x_A,x_B)&=&(a_g + b_g x_A+ c_g x_B)e^{d_g + e_g x_A + f_g x_B},
\end{eqnarray}
where $d_g,e_g,f_g$ and $a_{i},b_{i},c_{i}$ for
$i\in\{\ell,g\}$ are real constants. Then we have
\begin{eqnarray}\label{solution:pide-affine}
Q_{t,T}g(x_A,x_B) &=& \left[\theta_{AB}(T-t)+\theta_{A}(T-t)x_A +\theta_{B}(T-t)x_B\right]\nonumber\\
&&\times e^{\beta_{AB}(T-t) + \beta_A(T-t)x_A + \beta_B(T-t)x_B},\ \
\ \ 0\leq t\leq T,
\end{eqnarray}
where the unspecified functions in \eqref{solution:pide-affine} satisfy
the following generalized Riccati equations:
\begin{eqnarray}\label{Riccati:betaA-B-AB}
R(a_\ell,b_\ell,c_\ell):\quad\left\{\begin{array}{l}
-\beta_A'(u)-\kappa_A\beta_A(u)+\frac{1}{2}\sigma_A^2\beta_A^2(u)-a_{\ell}=0,\\
-\beta_B'(u)-\kappa_B\beta_B(u)+\frac{1}{2}\sigma_B^2\beta_B^2(u)-b_{\ell}=0,\\
\alpha_A\beta_A(u)+\alpha_B\beta_B(u)-\lambda-c_{\ell}+\widehat{\lambda}_c{\it\Phi}(c_A\beta_A(u),c_B\beta_B(u))\\
\qquad+\widehat{\lambda}_A\widetilde{{\it\Phi}}(d_A\beta_A(u),0)+\widehat{\lambda}_B\widetilde{{\it\Phi}}(0,d_B\beta_B(u))=\beta_{AB}'(u),
\end{array}\right.
\end{eqnarray}
and
\begin{eqnarray}\label{Riccati:thetaA-B-AB}
\left\{\begin{array}{l}
-\theta'_A(u) - \kappa_A \theta_A(u)+ \sigma^2_A
\theta_A(u) \beta_A(u)=0,\\
-\theta'_B(u) - \kappa_B \theta_B(u)+ \sigma^2_B \theta_B(u) \beta_B(u)=0,\\
\alpha_A \theta_A(u) + \alpha_B \theta_B(u)+
\widehat{\lambda}_cc_A\theta_A(u)\frac{\partial {\it \Phi}(c_A
\beta_A(u), c_B \beta_B(u))}{\partial \theta_A}\\
\qquad + \widehat{\lambda}_cc_B\theta_B(u)\frac{\partial
{\it\Phi}(c_A \beta_A(u), c_B \beta_B(u))}{\partial \theta_B} +
 \widehat{\lambda}_Ad_A\theta_A(u) \frac{\partial
\widetilde{{\it\Phi}}(d_A\beta_A(u), 0)}{\partial \theta_A}\\
\qquad + \widehat{\lambda}_Bd_B\theta_B(u) \frac{\partial
\widetilde{{\it\Phi}}(0, d_B \beta_B(u))}{\partial \theta_B} =
\theta'_{AB}(u).
\end{array}\right.
\end{eqnarray}
Here the function ${\it\Phi}(\theta_A,\theta_B)$ denotes the moment generating
function of the bivariate random variable $(Y_1^{(A)},Y_1^{(B)})$
defined by
\begin{eqnarray}\label{eq:mgfY}
{\it\Phi}(\theta_A,\theta_B)=\int_{\R_+^2}e^{\theta_A y_A+\theta_B
y_B}F_{AB}(\D y_A,\D y_B),\ \ \ \ \ \theta_A,\theta_B\leq0.
\end{eqnarray}
Similarly,
$\widetilde{{\it\Phi}}(\theta_A,\theta_B)$ denotes the moment
generating function of the bivariate random variable
$(\widetilde{Y}_1^{(A)},\widetilde{Y}_1^{(B)})$ defined by
\begin{eqnarray}\label{eq:mgftildeY}
\widetilde{{\it\Phi}}(\theta_A,\theta_B)=\int_{\R_+^2}e^{\theta_A
\widetilde{y}_A+\theta_B \widetilde{y}_B}\widetilde{F}_{AB}(\D
\widetilde{y}_A,\D \widetilde{y}_B),\ \ \ \ \ \theta_A,\theta_B\leq0,
\end{eqnarray}
Here, $F_{AB}(\D y_A,\D y_B)$ and $\widetilde{F}_{AB}(\D
\widetilde{y}_A,\D \widetilde{y}_B)$ are the joint
distribution functions of $(Y_1^{(A)},Y_1^{(B)})$, and
$(\widetilde{Y}_1^{(A)},\widetilde{Y}_1^{(B)})$ respectively.

In addition, $\lambda = \widehat{\lambda}_A +
\widehat{\lambda}_B + \widehat{\lambda}_c$,  with
$\widehat{\lambda}_A$ and $\widehat{\lambda}_B$ being the individual
intensities associated to counterparties $A$ and $B$, while
$\widehat{\lambda}_c$ is the intensity of the common Poisson
process. The initial conditions of the unspecified functions in
\eqref{solution:pide-affine} are given by
\begin{eqnarray}\label{eq:Ric-ini-conds}
\theta_{AB}(0)=a_g,\ \theta_A(0)=b_g,\ \theta_B(0)=c_g,\
\beta_{AB}(0)=d_g,\ \beta_A(0)=e_g,\ \beta_B(0)=f_g.
\end{eqnarray}
\end{lemma}

\noindent{\it Proof.}\quad Applying the Feynman-Kac formula to \eqref{eq:me}, it follows that
the function $Q_{t,T}g(x_A,x_B)$ satisfies on $(x_A,x_B)\in\R_+^2$,
\begin{eqnarray}\label{eq:Fey-kac-eqn}
\left(\frac{\partial }{\partial t} + \cL\right)f(t,x_A,x_B) &=&
\ell(x_A,x_B)f(t,x_A,x_B)\nonumber\\
f(T,x_A,x_B)&=&{g(x_A,x_B)},
\end{eqnarray}
where the integro-differential operator $\cL$ acting on $h\in
C^2(\R_+^2)$ is given by
\begin{eqnarray}\label{eq:idoperaotr}
\cL h(x_A,x_B) &=& \frac{1}{2}\sigma_A^2x_A\frac{\partial^2h}{\partial x_A^2}(x_A,x_B) + \frac{1}{2}\sigma_B^2x_B\frac{\partial^2h}{\partial x_B^2}(x_A,x_B) \nonumber\\
&&+(\alpha_A-\kappa_Ax_A)\frac{\partial h}{\partial x_A}(x_A,x_B) + (\alpha_B-\kappa_Bx_B)\frac{\partial h}{\partial x_B}(x_A,x_B)-\lambda h(x_A,x_B)\nonumber\\
&&+\widehat{\lambda}_A\int_{\R_+}h(x_A+
d_A\widetilde{y}_A,x_B)\widetilde{F}_{A}(\D\widetilde{y}_A)
+\widehat{\lambda}_B\int_{\R_+}h(x_A,x_B+ d_B\widetilde{y}_B)\widetilde{F}_{B}(\D\widetilde{y}_B)\nonumber\\
&&+\widehat{\lambda}_c\int_{\R_+^2}h(x_A+c_Ay_A,x_B+c_By_B)F_{AB}(\D
y_A,\D y_B).
\end{eqnarray}
Plugging the solution form \eqref{solution:pide-affine} into the PIDE
\eqref{eq:Fey-kac-eqn}, we obtain
\begin{eqnarray*}
\frac{\partial f}{\partial t}(t,x_A,x_B)&=&-f(t,x_A,x_B)\left[\beta_{AB}'(T-t)+\beta_A'(T-t)x_A + \beta_B'(T-t)x_B\right] \\
& & -(\theta'_{AB}(T-t) + \theta'_A(T-t) x_A + \theta'_B(T-t) x_B) e^{\beta_{AB}(T-t) + \beta_A(T-t)x_A + \beta_B(T-t)x_B}  \\
\frac{\partial f}{\partial x_A}(t,x_A,x_B)&=&f(t,x_A,x_B)\beta_A(T-t) + \theta_A(T-t) e^{\beta_{AB}(T-t) + \beta_A(T-t)x_A + \beta_B(T-t)x_B} \\
\frac{\partial f}{\partial x_B}(t,x_A,x_B)&=&f(t,x_A,x_B)\beta_B(T-t) + \theta_B(T-t) e^{\beta_{AB}(T-t) + \beta_A(T-t)x_A + \beta_B(T-t)x_B} \\
\frac{\partial^2 f}{\partial x_A^2}(t,x_A,x_B)&=&f(t,x_A,x_B)\beta_A^2(T-t) + 2 \theta_A(T-t) \beta_A(T-t) e^{\beta_{AB}(T-t) + \beta_A(T-t)x_A + \beta_B(T-t)x_B} \\
\frac{\partial^2 f}{\partial x_B^2}(t,x_A,x_B)&=&f(t,x_A,x_B)\beta_B^2(T-t) + 2 \theta_B(T-t) \beta_B(T-t) e^{\beta_{AB}(T-t) + \beta_A(T-t)x_A + \beta_B(T-t)x_B}
\end{eqnarray*}
and
\begin{eqnarray*}
& & \int_{\R_+}f(t,x_A+
d_A\widetilde{y}_A,x_B)\widetilde{F}_{A}(\D\widetilde{y}_A)
= f(t,x_A,x_B)\int_{\R_+}e^{\beta_A(T-t)
d_A\widetilde{y}_A}\widetilde{F}_{A}(\D\widetilde{y}_A) \\
& & + e^{\beta_{AB}(T-t) + \beta_A(T-t)x_A + \beta_B(T-t)x_B}
\int_{\R_+} \theta_A(T-t) d_A\widetilde{y}_A e^{\beta_A(T-t)
d_A\widetilde{y}_A} \widetilde{F}_{A}(\D\widetilde{y}_A) \\
& & \int_{\R_+}f(t,x_A,x_B+
d_B\widetilde{y}_B)\widetilde{F}_{B}(\D\widetilde{y}_B) =
f(t,x_A,x_B)\int_{\R_+}e^{\beta_B(T-t)
d_B\widetilde{y}_B}\widetilde{F}_{B}(\D\widetilde{y}_B) \\
& & + e^{\beta_{AB}(T-t) + \beta_A(T-t)x_A + \beta_B(T-t)x_B}
\int_{\R_+} \theta_B(T-t) d_B\widetilde{y}_B e^{\beta_B(T-t)
d_B\widetilde{y}_B} \widetilde{F}_{B}(\D\widetilde{y}_B) \\
& & \int_{\R_+^2}f(t,x_A+c_Ay_A,x_B+c_By_B)F_{AB}(\D y_A,\D y_B) =
f(t,x_A,x_B)\nonumber\\
&&\qquad\times\int_{\R_+^2}e^{\beta_A(T-t)c_Ay_A
+\beta_B(T-t)c_B y_B} F_{AB}(\D y_A,\D y_B)  \\
& &\qquad + e^{\beta_{AB}(T-t) + \beta_{A}(T-t) x_A + \beta_{B}(T-t) x_B} \\
& &\qquad\quad \times\int_{\R_+^2} \left( \theta_A(T-t) c_A y_A +
\theta_B(T-t) c_B y_B  \right) e^{\beta_A(T-t) c_A y_A +
\beta_B(T-t) c_B y_B} F_{AB}(\D y_A,\D y_B).
\end{eqnarray*}
In order for \eqref{eq:Fey-kac-eqn} to hold, we need that for all $u\in[0,T]$ and
$(x_A,x_B)\in\R_+^2$, the following two equalities are satisfied
\begin{eqnarray}\label{eq:Ricati}
&&x_A\left[-\beta_A'(u)-\kappa_A\beta_A(u)+\frac{1}{2}\sigma_A^2\beta_A^2(u)-a_{\ell}\right]
+x_B\left[-\beta_B'(u)-\kappa_B\beta_B(u)+\frac{1}{2}\sigma_B^2\beta_B^2(u)-b_{\ell}\right]\nonumber\\
&&\qquad+\alpha_A\beta_A(u)+\alpha_B\beta_B(u)-\beta_{AB}'(u)-\lambda-c_{\ell}+\widehat{\lambda}_c{\it\Phi}(c_A\beta_A(u),c_B\beta_B(u))\nonumber\\
&&\qquad+\widehat{\lambda}_A\widetilde{{\it\Phi}}(d_A\beta_A(u),0)+\widehat{\lambda}_B\widetilde{{\it\Phi}}(0,d_B\beta_B(u))=0,
\end{eqnarray}
and
\begin{eqnarray}\label{eq:Ricati1}
& & + x_A \left[-\theta'_A(u) - \kappa_A \theta_A(u)+ \sigma^2_A
\theta_A(u) \beta_A(u)  \right]
+ x_B \left[-\theta'_B(u) - \kappa_B \theta_B(u)+ \sigma^2_B \theta_B(u) \beta_B(u)  \right] \nonumber\\
&&-\theta'_{AB}(u) +  \alpha_A \theta_A(u) + \alpha_B \theta_B(u) \nonumber\\
& & + \widehat{\lambda}_cc_A\theta_A(u)\frac{\partial {\it \Phi}(c_A
\beta_A(u), c_B \beta_B(u))}{\partial \theta_A}
+ \widehat{\lambda}_cc_B\theta_B(u)\frac{\partial {\it\Phi}(c_A \beta_A(u), c_B \beta_B(u))}{\partial \theta_B} \nonumber\\
& & + \widehat{\lambda}_Ad_A\theta_A(u) \frac{\partial
\widetilde{{\it\Phi}}(d_A\beta_A(u), 0)}{\partial \theta_A} +
\widehat{\lambda}_Bd_B\theta_B(u) \frac{\partial
\widetilde{{\it\Phi}}(0, d_B \beta_B(u))}{\partial \theta_B} = 0,
\end{eqnarray}
Hence the unspecified functions in \eqref{solution:pide-affine} satisfy the Riccati equations \eqref{Riccati:betaA-B-AB} and \eqref{Riccati:thetaA-B-AB}. From the terminal condition in \eqref{eq:Fey-kac-eqn}, we further
have the initial conditions given by \eqref{eq:Ric-ini-conds}.
This completes the proof of the lemma. \hfill$\Box$

The following results give the explicit solutions to the generalized Riccati equations \eqref{Riccati:betaA-B-AB}
and \eqref{Riccati:thetaA-B-AB}.

\begin{lemma}\label{lem:solution-genRiccatibeta}
Assume that the initial values $\beta_A(0)=e_g\leq0$ and $\beta_B(0)=f_g\leq0$. If the constants $a_\ell,\ b_\ell>0$, then the generalized Riccati
equation \eqref{Riccati:betaA-B-AB} admits the explicit solution given by
\begin{eqnarray}\label{solu:betaA-B}
\beta_A(u)&=&a_\ell\cdot\beta\left(\kappa_A,\sigma_A\sqrt{a_\ell},\frac{e_g}{a_{\ell}};u\right),\nonumber\\
\beta_B(u)&=&b_\ell\cdot\beta\left(\kappa_B,\sigma_B\sqrt{b_\ell},\frac{f_g}{b_{\ell}};u\right),
\end{eqnarray}
and
\begin{eqnarray}\label{eq:betaAB}
\beta_{AB}(u) &=&
\int_0^u\bigg[\alpha_A\beta_A(v)+\alpha_B\beta_B(v)+\widehat{\lambda}_c{\it\Phi}(c_A\beta_A(v),c_B\beta_B(v))\nonumber\\
&&\qquad\quad+\widehat{\lambda}_A\widetilde{\it\Phi}(d_A\beta_A(v),0)+\widehat{\lambda}_B\widetilde{\it\Phi}(0,d_B\beta_B(v))\bigg]\D
v +d_g - (\lambda+c_{\ell})u,
\end{eqnarray}
where the function $\beta\left(\kappa,\sigma,b;u\right)$ is given by
\eqref{eq:lem-sol-B1} and ${\it\Phi}(\theta_A,\theta_B)$,
$\widetilde{\it\Phi}(\theta_A,\theta_B)$ are the moment generating
functions defined in~\eqref{eq:mgfY} and \eqref{eq:mgftildeY} with
$\theta_A,\theta_B\leq0$.
\end{lemma}

\noindent{\it Proof.}\quad The solutions $(\beta_A(u),\beta_B(u))$
given in \eqref{solu:betaA-B} can be obtained by an immediate
application of Lemma \ref{lem:solution-R2}. Note that the bivariate
random variables $(Y_1^{(A)},Y_1^{(B)})$ and
$(\widetilde{Y}_1^{(A)},\widetilde{Y}_1^{(B)})$ associated to the
jumps of the counterparties, are assumed to take values on $\R_+^2$.
Then the corresponding moment generating function
${\it\Phi}(\theta_A,\theta_B)$ exists if $\theta_A,\theta_B\leq0$.
From Lemma \ref{lem:solution-R1}, it follows that the solution
\eqref{eq:lem-sol-B2} given by
\begin{eqnarray*}
\beta(\kappa,\sigma,b;u) = B(\kappa,\sigma;u) +
e^{\phi(u)}\frac{1}{\frac{1}{b}-\frac{\sigma^2}{2}\int_0^u
e^{\phi(v)}\D v}\leq0,\ \ \ \ \ \forall\ 0\leq u\leq T,
\end{eqnarray*}
provided the initial value $\beta(\kappa,\sigma,b;0)=b\leq0$. This
is because $B(\kappa,\sigma;u)\leq0$ for all $0\leq u\leq T$, by
\eqref{eq:solu-express-B}. Hence if the initial values
$\beta_A(0)=e_g\leq0$ and $\beta_B(0)=f_g\leq0$ in
\eqref{solu:betaA-B}, then
${\it\Phi}(c_A\beta_A(u),c_B\beta_B(u))$,
$\widetilde{\it\Phi}(d_A\beta_A(u),0)$ and
$\widetilde{\it\Phi}(0,d_B\beta_B(u))$ exist since
$c_A,c_B,d_A,d_B>0$, and can be computed using
\eqref{eq:mgfY} and \eqref{eq:mgftildeY}. Hence, we can derive the
solution $\beta_{AB}(u)$ to the third equation in
\eqref{Riccati:betaA-B-AB}, which is given by \eqref{eq:betaAB}.
\hfill$\Box$

Based on the above explicit solution to the generalized Riccati
equation \eqref{Riccati:betaA-B-AB}, we immediately have
\begin{lemma}\label{lem:solution-genRiccatitheta}
The generalized Riccati
equation \eqref{Riccati:thetaA-B-AB} admits the explicit solution given by
\begin{eqnarray}\label{solu:thetaA-B}
\theta_A(u)&=&b_g\exp\left(-\kappa_Au +
\sigma_A^2\int_0^u\beta_A(v)\D v\right),\nonumber\\
\theta_B(u)&=&c_g\exp\left(-\kappa_Bu +
\sigma_B^2\int_0^u\beta_B(v)\D v\right),
\end{eqnarray}
and
\begin{eqnarray}\label{eq:thetaAB}
\theta_{AB}(u) &=&a_g + \int_0^u\Bigg[\alpha_A \theta_A(v) +
\alpha_B \theta_B(v)+
\widehat{\lambda}_cc_A\theta_A(v)\frac{\partial {\it \Phi}(c_A
\beta_A(v), c_B \beta_B(v))}{\partial \theta_A}\nonumber\\
&&+ \widehat{\lambda}_cc_B\theta_B(v)\frac{\partial {\it\Phi}(c_A
\beta_A(v), c_B \beta_B(v))}{\partial \theta_B} +
\widehat{\lambda}_Ad_A\theta_A(v) \frac{\partial
\widetilde{\it\Phi}(d_A\beta_A(v), 0)}{\partial \theta_A}\nonumber\\
&&+ \widehat{\lambda}_Bd_B\theta_B(v)
\frac{\partial\widetilde{\it\Phi}(0, d_B \beta_B(v))}{\partial
\theta_B}\Bigg]\D v,
\end{eqnarray}
where $0\leq u\leq T$.
\end{lemma}

\section{Proof of Proposition \ref{lem:Ftsexpl}} \label{app:lemma-Ftsexpl}
\renewcommand\theequation{E.\arabic{equation}}
\setcounter{equation}{0}

Recall from Eq.~\eqref{eq:FLS1} that, for $0\leq t\leq s\leq T$,
\begin{eqnarray*}
\widehat{F}(t,s)=\Ex\left[\int_{\cO}\Ex\left[\exp\left(-\int_t^{{s}}X_u({\bm p})\D u\right)\right]q(\D p)\eta(\D y)\phi_0(\D
x)\right],
\end{eqnarray*}
where the `type' parameter ${\bm p}=(p,y,x)\in\cO$ with
$p=(\alpha,\kappa,\sigma,c,d,\widehat{\lambda})\in\cO_p$.
The limit process $X({\bm p})=(X_t({\bm p});\ t\geq0)$ is a shifted
square root diffusion process given by
\begin{eqnarray}\label{eq:limit-Xp-cir}
X_t({\bm p}) = x + \int_0^t \left[D({\bm p}) + \alpha - \kappa
X_u({\bm p})  \right] \D u + \sigma \int_0^t (X_u({\bm p}))^{1/2} \D
W_u,
\end{eqnarray}
where the drift $D({\bm p})$ is given by \eqref{eq:drift}, i.e.,
$D({\bm p})=dy_2\widehat{\lambda}+cy_1\widehat{\lambda}_c$.

Note that the limit process $X({\bm p})$ is an affine process. Using
Lemma \ref{lem:pide-affine}, we have
\[
\Ex\left[\exp\left(-\int_t^sX_u({\bm p})\D u\right)\bigg|X_t({\bm
p})=x\right]=\exp\left(A_{\bm p}(s-t) + B_{\bm p}(s-t) x\right),\ \
\  0\leq t\leq s,
\]
where the functions $A_{\bm p}(u)$ and $B_{\bm p}(u)$ satisfy the
following system of Riccati equations
\begin{eqnarray}\label{Riccati:simulation}
\left\{\begin{array}{l}
-A'_{\bm p}(u) + (D({\bm p})+\alpha)B_{\bm p}(u) =0,\\
-B_{\bm p}'(u)-\kappa B_{\bm p}(u) + \frac{1}{2}\sigma^2B_{\bm
p}^2(u)-1=0,
\end{array}\right.
\end{eqnarray}
with the following initial conditions
\begin{eqnarray}\label{Riccati:simulation-intial-conds}
A_{\bm p}(0) = B_{\bm p}(0) = 0.
\end{eqnarray}

From Lemma \ref{appen:Riccati1}, it follows that  the solution to
the second equation of the Riccati system \eqref{Riccati:simulation} is
given by

\begin{eqnarray*}
B_{\bm p}(u) = -\frac{2 \left(e^{\varpi u } - 1\right)}{2 \varpi +
\left( \kappa+\varpi\right) \left(e^{\varpi u } -1 \right) },\ \ \ \ 0\leq u\leq s,
\end{eqnarray*}
where $\varpi = \sqrt{\kappa^2 + 2 \sigma^2}$.  Using the first equation of the Riccati
system \eqref{Riccati:simulation} and the initial conditions \eqref{Riccati:simulation-intial-conds}, it follows that
\begin{eqnarray*}
e^{A_{\bm p}(s)}=\exp\left[\left(\alpha + D({\bm p})\right)\int_0^s B_{\bm p}(u)\D u\right],
\end{eqnarray*}
where
\begin{eqnarray*}
\int_0^s B_{\bm p}(u) \D u = 
\frac{2T}{\varpi-\kappa}+\frac{4}{\varpi^2-\kappa^2}\log\left[\frac{2
\varpi}{(1+e^{\varpi T})\varpi+(e^{\varpi T}-1)\kappa}\right].
\end{eqnarray*}
Hence, we obtain
\begin{eqnarray*}\label{eq:integrcomp}
&&\int_{\cO}\Ex\left[\exp\left(-\int_t^{s}X_u({\bm p})\D u\right)\right]q(\D p)\eta(\D y)\phi_0(\D x)\nonumber\\
&&\quad=\int_{\cO}\exp\left(A_{\bm p}(s-t) + B_{\bm p}(s-t) x\right) q(\D p) \eta(\D y)\phi_0(\D x)\nonumber\\
&&\quad = e^{x^* B_{p^*}(s-t)} \int_{\R_+^2}
e^{A_{(p^*,y_1,y_2)}(s-t)} \delta(y_1-Y)\D
y_1\delta(y_2-\widetilde{Y})\D y_2\nonumber\\
&&\quad=e^{x^* B_{p^*}(s-t)+{A_{(p^*,Y,\widetilde{Y})}(s-t)}}\nonumber\\
&&\quad = e^{x^* B_{p^*}(s-t)} \exp\left(\left[\alpha^*+
d^*\widehat{\lambda}^*\widetilde{Y} +
c^*\widehat{\lambda}_cY\right]\int_0^{s-t} B_{p^*}(u)\D u\right).
\end{eqnarray*}
Using the independence of the exponential random variables
$Y$ and $\widetilde{Y}$, we have
\begin{eqnarray}\label{hatF:simulation}
\widehat{F}(t,s) &=& e^{x^* B_{p^*}(s-t)}
\Ex\left[\exp\left(\left[\alpha^*+
d^*\widehat{\lambda}^*\widetilde{Y} +
c^*\widehat{\lambda}_cY\right]\int_0^{s-t} B_{p^*}(u)\D u\right)\right]\nonumber\\
&=&\exp\left(x^* B_{p^*}(s-t) + \alpha^*\int_0^{s-t} B_{p^*}(u)\D u \right)\nonumber\\
&&\times\Ex\left[\exp\left(\widetilde{Y}d^*\widehat{\lambda}^*\int_0^{s-t} B_{p^*}(u)\D u\right)\right]
\Ex\left[\exp\left(Yc^*\widehat{\lambda}_c\int_0^{s-t} B_{p^*}(u)\D u\right)\right]\nonumber\\
&=&\exp\left(x^* B_{p^*}(s-t) + \alpha^*\int_0^{s-t} B_{p^*}(u)\D u
\right)\frac{\gamma_1}{\gamma_1-c^*\widehat{\lambda}_c\int_0^{s-t}
B_{p^*}(u)\D u}\nonumber\\
&&\times\frac{\gamma_2}{\gamma_2-d^*\widehat{\lambda}^*\int_0^{s-t}
B_{p^*}(u)\D u},
\end{eqnarray}
since $\int_0^u B_{p^*}(z)\D z<0$ for all $u>0$. Hence, the proof of the Lemma is complete.
\hfill$\Box$

\section{Proof of Proposition \ref{prop:HhatH12}} \label{app:prop-HhatH12}
\renewcommand\theequation{F.\arabic{equation}}
\setcounter{equation}{0}

For $t\leq t_B\leq T$ and $(x_A,x_B)\in\R_+^2$, using \eqref{eq:me}, we obtain
\begin{eqnarray}\label{eq:cva-2}
H_1(t_B-t,x_A,x_B)&:=&\Ex\left[\exp\left(-\int_t^{t_B} (\xi_s^{(A)}+\xi_s^{(B)}) \D s\right)\xi_{t_B}^{(B)}\bigg|\xi_t^{(A)}=x_A,\xi_t^{(B)}=x_B\right]\nonumber\\
&=&\left[h_1(t_B-t)+h_A(t_B-t)\xi_{t}^{(A)}+h_B(t_B-t)\xi_{t}^{(B)}\right]\nonumber\\
&&\times\exp\left(\widehat{h}_1(t_B-t) +
\widehat{h}_A(t_B-t)\xi_t^{(A)}+
\widehat{h}_B(t_B-t)\xi_t^{(B)}\right),
\end{eqnarray}
where the functions $(\widehat{h}_1(u),\widehat{h}_A(u),\widehat{h}_B(u))$ satisfy the generalized Riccati equation
$R(1,1,0)$ given by \eqref{Riccati:betaA-B-AB}, while the functions  $({h}_1(u),{h}_A(u),{h}_B(u))$ satisfy the
generalized Riccati equation given by \eqref{Riccati:thetaA-B-AB}. The initial conditions are given by
\begin{eqnarray}\label{eq:initial-conds-left}
h_1(0)=h_A(0)=\widehat{h}_1(0)=\widehat{h}_A(0)=\widehat{h}_B(0)=0,\ {\rm and}\ h_B(0)=1.
\end{eqnarray}
Solving the corresponding Riccati equations via Lemma \ref{lem:solution-genRiccatibeta} and Lemma \ref{lem:solution-genRiccatitheta} respectively, we have the solutions $(\widehat{h}_1(u),\widehat{h}_A(u),\widehat{h}_B(u))$ and $({h}_1(u),{h}_A(u),{h}_B(u))$ are \eqref{eq:solu-hat-h} and \eqref{eq:solu-h} respectively.

By virtue of Lemma \ref{lem:pide-affine}, we have,  for $t\leq t_A\leq T$,
\begin{eqnarray}\label{eq:H2}
H_2(t_A-t,x_A,x_B)&=&\left[w_1(t_A-t)+w_A(t_A-t)x_A+w_B(t_A-t)x_B\right]\nonumber\\
&&\times\exp\Big(\widehat{w}_1(t_A-t) +\widehat{w}_A(t_A-t)x_A+\widehat{w}_B(t_A-t)x_B\Big),
\end{eqnarray}
where the functions $(\widehat{w}_1(u),\widehat{w}_A(u),\widehat{w}_B(u))$ satisfy the generalized Riccati equation
$R(1,1,0)$ given by \eqref{Riccati:betaA-B-AB}, while the functions  $({w}_1(u),{w}_A(u),{w}_B(u))$ satisfy the
generalized Riccati equation given by \eqref{Riccati:thetaA-B-AB}. The initial conditions are given by
\begin{eqnarray}\label{eq:initial-conds-left1}
w_1(0)=w_B(0)=\widehat{w}_1(0)=\widehat{w}_A(0)=\widehat{w}_B(0)=0,\ {\rm and}\ w_A(0)=1.
\end{eqnarray}
Solving the corresponding Riccati equations by using Lemma
\ref{lem:solution-genRiccatibeta} and Lemma
\ref{lem:solution-genRiccatitheta} respectively, we have the
solutions $(\widehat{w}_1(u),\widehat{w}_A(u),\widehat{w}_B(u))$ and
$({w}_1(u),{w}_A(u),{w}_B(u))$ are given by \eqref{eq:solu-hat-w}
and \eqref{eq:solu-w} respectively. Hence, the proof of the
proposition is complete. \hfill$\Box$


\end{document}